\documentclass[aps,pra,superscriptaddress,showpacs,twocolumn,longbibliography,10pt]{revtex4-1}

\usepackage{graphicx}
\usepackage{amsmath}
\usepackage{amssymb}
\usepackage{dsfont}
\usepackage{xspace}
\usepackage{color}
\usepackage[usenames,dvipsnames]{xcolor}
\usepackage[colorlinks=true,linkcolor=blue,urlcolor=blue,citecolor=blue]{hyperref}
\usepackage[normalem]{ulem}
\usepackage{setspace}
\usepackage{xfrac}
\usepackage{simplewick}
\usepackage{mathtools}

\DeclareMathOperator{\Tr}{\mbox{Tr}}
\DeclareMathOperator{\tr}{\mbox{tr}}
\DeclareMathOperator{\re}{\mbox{Re}}

\newcommand{\eqw}[1]{(\ref{#1})}
\newcommand{\eq}[1]{Eq.\thinspace{}(\ref{#1})}

\newcommand{\be}{\begin{eqnarray}}
\newcommand{\ee}{\end{eqnarray}}

\newcommand{\fc}[1]{({#1})}
\newcommand{\figc}[2]{Fig.\thinspace{}\ref{#1}\thinspace{}\fc{#2}}

\def\bra#1{\mathinner{\langle{#1}|}}
\def\ket#1{\mathinner{|{#1}\rangle}}
\newcommand{\braket}[2]{\langle #1 | #2 \rangle}

\definecolor{dgreen}{rgb}{0.0, 0.5, 0.0}

\newcommand{\veck}{\mathbf k}

\newcommand{\vecp}{\mathbf p}
\newcommand{\vecq}{\mathbf q}

\newcommand{\vecr}{\mathbf r}
\newcommand{\etal}{\textit{et~al.}~}

\newcommand{\rtext}[1]{\textcolor{black}{{{#1}}}}

\begin{document}

\title{Universal many-body response of heavy impurities coupled to a Fermi sea\rtext{:\\a review of recent progress}}

\author{Richard Schmidt}
\affiliation{Department of Physics, Harvard University, Cambridge MA 02138, USA}
\affiliation{ITAMP, Harvard-Smithsonian Center for Astrophysics, Cambridge, MA 02138, USA}

\author{Michael Knap}
\affiliation{Department of Physics, Walter Schottky Institute, and Institute for Advanced Study, Technical University Munich, 85748 Garching, Germany}

\author{Dmitri A. Ivanov}
\affiliation{Institute for Theoretical Physics, ETH Z\"urich, 8093 Z\"urich, Switzerland}
\affiliation{Department of Physics, University of Z\"urich, 8057 Z\"urich, Switzerland}

\author{Jhih-Shih You}
\affiliation{Department of Physics, Harvard University, Cambridge MA 02138, USA}

\author{Marko Cetina}
\affiliation{Institut f\"ur Quantenoptik und Quanteninformation (IQOQI), \"Osterreichische Akademie der Wissenschaften, 6020 Innsbruck, Austria}
\affiliation{Institut f\"ur Experimentalphysik, Universit\"at Innsbruck, 6020 Innsbruck, Austria}
\affiliation{Joint Quantum Institute, Joint Center for Quantum Information and Computer Science, and Department of Physics, University of Maryland, College Park, MD 20742}

\author{Eugene Demler}
\affiliation{Department of Physics, Harvard University, Cambridge MA 02138, USA}

\date{\today}

\begin{abstract}
In this report we discuss the dynamical response of heavy quantum impurities immersed in a Fermi gas at zero and at finite temperature. \rtext{Studying} both the frequency and the time domain \rtext{allows one to} identify interaction regimes that are characterized by  distinct  many-body dynamics.  From \rtext{this} theoretical study a picture emerges in which impurity dynamics is universal on essentially all time scales, and where the high-frequency few-body response is related to the long-time dynamics of the Anderson orthogonality catastrophe by Tan relations.  Our theoretical description relies on different and complementary  approaches: \rtext{functional determinants give an exact numerical solution for time- and frequency-resolved responses, bosonization provides accurate analytical expressions at low temperatures, and the theory of Toeplitz determinants allows one to analytically predict response up to high temperatures.} Using these approaches we predict the thermal decoherence rate of the fermionic system and prove that within the considered model the fastest rate of long-time decoherence is given by $\gamma=\pi k_BT/4$. We show that  Feshbach resonances in cold atomic systems give access to new interaction regimes where quantum effects can prevail even in the thermal regime of many-body dynamics.   The key signature of this phenomenon is a crossover between different exponential decay rates of the real-time Ramsey signal. \rtext{It is shown} that the physics of the orthogonality catastrophe \rtext{is experimentally observable up to temperatures $T/T_F\lesssim 0.2$ where it leaves its fingerprint in a power-law temperature dependence of thermal spectral weight and we review how this phenomenon is related to the  physics of heavy ions in liquid $^3$He and the formation of Fermi polarons.} \rtext{The presented} results  are in excellent agreement with  recent experiments on LiK mixtures, and  we predict several new phenomena that can be tested using currently available experimental technology.
\end{abstract}

\maketitle

\tableofcontents

\section{Introduction}

Exactly solvable models are a rare commodity in  many-body physics \cite{mahan_many_2000}. Yet they provide a unique basis to develop a profound understanding of physical systems and \rtext{can serve as}  a benchmark for approximate theoretical approaches. Examples include the Ising model \cite{Onsager1944},  one-dimensional systems \cite{Bethe1931,giamarchi_quantum_2004}, the Kitaev model \cite{Kitaev2003,Kitaev2006}, or the Dicke model \cite{dicke_coherence_1954} which became paradigms of condensed matter  physics and quantum optics.

Heavy impurities interacting with a Fermi gas are another paradigmatic example that is exactly solvable and yet  retains the complexity of an interacting many-body system  exhibiting rich physics \cite{weiss1999,rosch_quantum-coherent_1999}. Most prominently it features the orthogonality catastrophe which defies a simple perturbative description and signals the absence of quasiparticles even at weak interactions,  as first investigated by Anderson \cite{anderson_infrared_1967}. 

The Anderson orthogonality catastrophe manifests itself not only in ground state properties but also in the non-equilibrium dynamics. At long times, the power-law decay of coherence (for a definition see Eq.~\eqref{eq:ramsey}  and Refs.~\cite{nozieres_singularities_1969,nozieres_recoil_94,dambrumenil_fermi_2005}) is one of its key manifestations  which universally depends only on the \rtext{scattering} phase shift close to the Fermi surface. In contrast, the description of the short-time dynamics, being testament of the short-distance physics,  suffers typically from the microscopic unknown. In particular in conventional solid-state systems physics depends on the details of chemical bonding, the core-hole potentials, as well as their screening and relaxation,  which leads to highly non-universal short-time dynamics \cite{Schweigert2007,shah2013ultrafast}.

Here we show that ultracold atoms provide a system where the physics of impurities is universal on essentially all time scales.  This special property of cold atoms has its origin in their diluteness and ultra-low temperatures \cite{bloch_many-body_2008}, which renders even the few-body physics   universal \cite{Braaten2006}. Specifically, we propose the realization of an  Anderson-Fano  model \cite{mahan_many_2000} which is fully tunable by the use of Feshbach resonances \cite{chin_feshbach_2010}. \rtext{We introduce and review a functional determinant approach \cite{levitov_electron_1996,klich_03,schoenhammer07,Ivanov2013}, that allows us not only to provide the exact numerical solution of its full non-equilibrium quench dynamics, but also to derive analytical expressions} for the short- and long-time dynamics so that a complete analytical understanding of this paradigm model  arises.

From this solution, an overarching picture of universality emerges, reaching from short to long time scales. 
The dynamics at short times is determined by universal tails of the impurity high-frequency response \cite{Braaten2010,Nishida2012}. Those tails are connected with the long-time dynamics by  exact operator identities known as the Tan relations \cite{Tan2008a,Tan2008b,Tan2008c,Braaten2008}. For the long-time, respectively low-frequency, response of the system we derive analytical expressions  valid at arbitrary temperatures by a combination of the theory of Toeplitz determinants and bosonization. We find that the physics  of impurities can be universally  described in terms of excitation branches which represent collections of relevant many-body states. By using this approach, even the non-equilibrium dynamics beyond the standard Luttinger liquid paradigm \cite{giamarchi_quantum_2004,gogolin_bosonization_2004} finds an analytical description.

\begin{figure}[t] 
  \centering 
  \includegraphics[width=\linewidth]{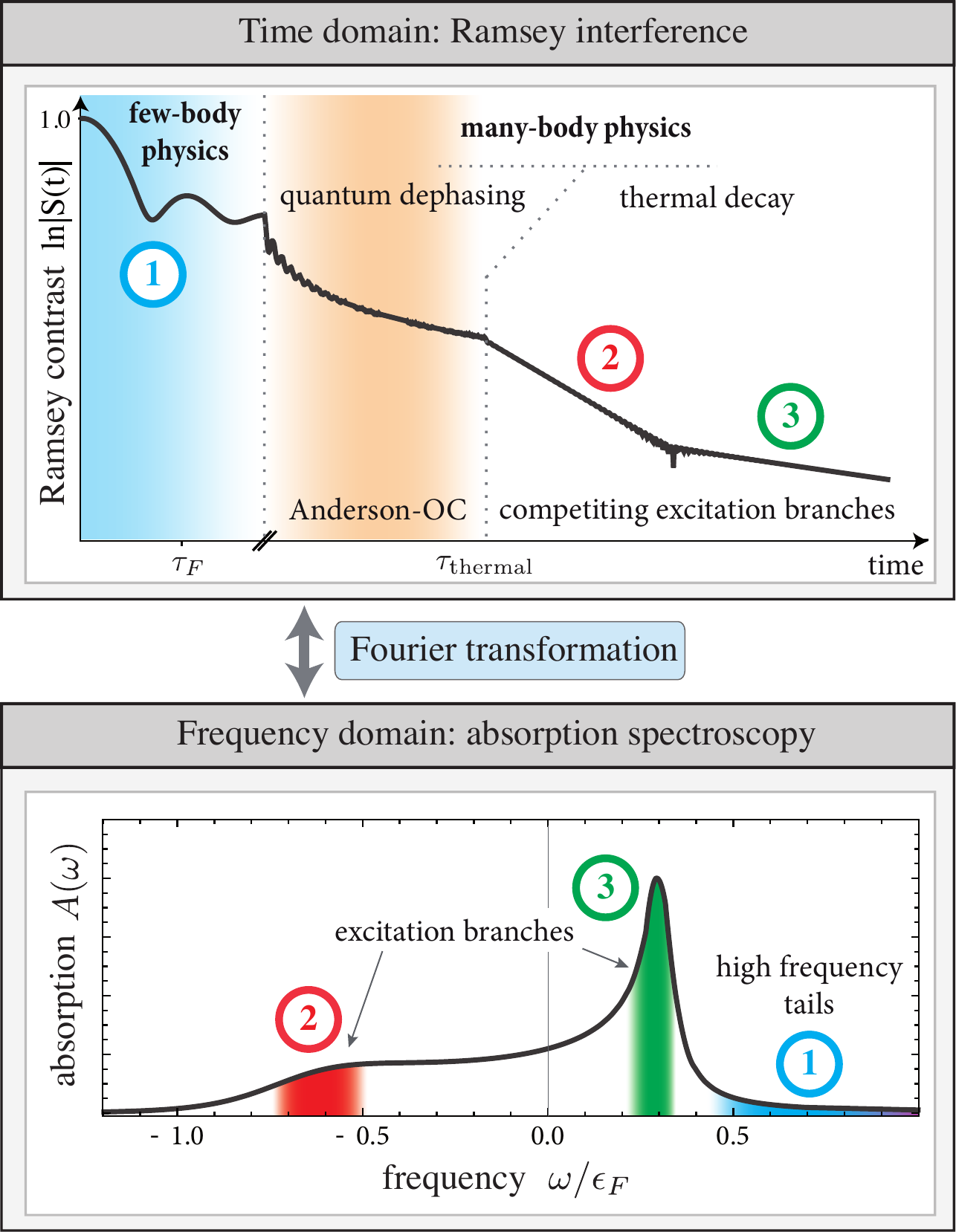}\vspace{-3mm}
  \caption{ \textbf{Illustration of  universal impurity dynamics.} \rtext{The upper panel shows a sketch of the dephasing dynamics of a fermionic bath interacting with a single localized impurity as function of time. The corresponding many-body overlap $S(t)$, defined in Eq.~\eqref{eq:ramsey}, see also Eq.~\eqref{eq:MixedRF}, can be measured in Ramsey spectroscopy. It is related by Fourier transformation to the absorption spectrum $A(\omega)$ of the impurity which hence contains equivalent physical information in the frequency domain (lower panel).  The short-time  dynamics, which is reflected in high-frequency tails of the absorption response, is dominated by few-body physics. At  times exceeding the Fermi time scale $\tau_F=\hbar/\epsilon_F$, with the Fermi energy $\epsilon_F$,  many-body physics becomes relevant and the dynamics of the orthogonality catastrophe is manifest in a power-law decay of coherence. Beyond the thermal time scale $\tau_\text{thermal}=\hbar/k_B T$, exponential decay takes over. Even in this long-time regime, quantum effects can prevail. They lead to competing exponential decay rates that signal the superposition of various excitation branches visible in the absorption spectrum.}}
  \label{fig:intro}
\end{figure}

\rtext{Due to the low temperatures and diluteness of cold atomic gases, only short-range s-wave interactions are relevant. These} can be tuned \cite{chin_feshbach_2010} and  universal regimes reached that are not readily accessible in conventional solid state systems \cite{Sidler2016}. We identify various interaction regimes  exhibiting distinct decoherence dynamics and spectral properties that can be related  to  dominant excitation branches. The main results are summarized in Fig.~\ref{fig:intro}. \rtext{Here we show an illustration of the time- and  frequency-resolved response of the system following a quench of the impurity-Fermi-gas interaction. Experimentally the real-time dephasing signal $S(t)$ can be measured using an interferometric Ramsey scheme, while the frequency-resolved response $A(\omega)$  is accessible in absorption spectroscopy. Both signals are related by Fourier transformation and hence provide complementary probes of quantum many-body dynamics.}

\rtext{ One finds various regimes of universal real-time dynamics.} At ultrashort times dephasing is dominated by few-body physics \rtext{leading to universal high-frequency tails  in the absorption response} \cite{Braaten2010,Nishida2012,cetina_2016,parish2016}. Intermediate time scales exhibit strong oscillations signifying the dressing of the impurity by excitations from the full depth of the  Fermi sea. These oscillations are a robust and universal feature of short-range, strongly interacting impurity Fermi systems \rtext{and they govern not only the dynamics of infinitely heavy impurities, but also appear in the case of impurities of finite mass where the dressing by bath excitations leads to the formation of polaronic quasiparticles \cite{cetina_2016,parish2016}}. 

At longer times, the dynamics of heavy impurities is governed by the power-law dephasing of the Anderson orthogonality catastrophe which is universally dependent only on the phase shift at the Fermi surface. Even longer times are dominated by exponential decay due to finite temperature. Quite surprisingly, \rtext{we find that even at times substantially exceeding the thermal time scale $\hbar/k_BT$, the quantum nature of excitation branches  still persists. This is reflected in new features of competing dynamics where the robustness of superpositions of excitations branches leads to a crossover between characteristic exponential decays of coherence at very long times. }

Remarkably, although there is a enormous  scale separation between the many-body regime at long times and the intrinsic few-body short-time dynamics,  both regimes are connected by the Tan relations \cite{Tan2008a,Tan2008b,Tan2008c,Braaten2008}. We  show that based on these exact relations one can  relate the long-time phase evolution of the many-body wave function to the high-frequency tail of the absorption spectrum. 

\paragraph*{\textbf{Progress towards studying real-time impurity dynamics.}---}
\rtext{Recent years have seen an extensive interest in the Fermi polaron problem where one considers a single impurity immersed in a Fermi gas. Quite generally, the interaction between the impurity and the Fermi gas leads to the dressing of the impurity by excitations in the Fermi sea. When the mass of the impurity is finite, this dressing by excitations remains quite moderate. In consequence, the new many-body ground state of the system -- the `dressed impurity' -- retains some resemblance to its non-interacting counterpart and one speaks of the formation of a well-defined quasiparticle: the Fermi polaron. In contrast, an infinitely heavy impurity is subject to a much stronger dressing by bath excitations. Here the dressing becomes even so extensive that the system is left in a state orthogonal to its original non-interacting state. Hence the dressed impurity completely looses its quasiparticle nature -- its quasiparticle weight $Z=0$ vanishes -- which signifies the hallmark of the Anderson orthogonality catastrophe \cite{anderson_infrared_1967}.}

\rtext{The case of impurities of finite mass has been considered first}  in the context of the phase diagram of spin-imbalanced Fermi gases \cite{Chevy2006,Lobo2006,Pilati2008,Comescot2007,Punk2009} where the Fermi polaron problem represents the extreme limit of spin imbalance. Special attention was given to the transition from a polaronic to a molecular ground state \rtext{\cite{prok2008,Prokofiev2008b,Punk2009,Mora2009,Bruun2010,Parish2011,Mathy2011,Zoellner2011,Levinsen2013,Massignan2014}}, which serves as a benchmark for theoretical approaches ranging from variational wave functions \cite{Chevy2006,Punk2009,Mora2009,Trefzger2012}, \rtext{diagrammatic resummation \cite{Comescot2007,Massignan2008,Comescot2008,Combescot2009,Schmidt2012b}, $1/N$ expansions \cite{Nikolic2007}, Monte Carlo calculations \cite{Lobo2006,Pilati2008},} and functional renormalization group \cite{schmidt_excitation_2011} to  diagrammatic Monte Carlo calculations \cite{prok2008,Prokofiev2008b,Kroiss2014,Kroiss2015,goulko2016}. Using radio-frequency spectroscopy, the ground state properties of the Fermi polaron, including the polaron to molecule transition, were first observed by Schirotzek \textit{et al.} \cite{schirotzek_observation_2009}; for an experimental study in two dimensions we refer to Ref.~\cite{Zhang2012}.

Shortly after these observations it was theoretically predicted that the impurity excitation spectrum contains a rich and interesting structure also above the ground state. In particular studies of the impurity spectral function revealed a metastable excitation at positive energies on the `repulsive' side of the Feshbach resonance (i.e. at positive scattering length $a>0$), separated by a large gap from the ground state \cite{Cui2010,schmidt_excitation_2011,massignan_repulsive_2011}. Since this `repulsive polaron' excitation can again be viewed as the extreme limit of a spin-imbalanced Fermi gas, its properties are of significance for studies of the repulsive Fermi gas and the question of a phase transition to itinerant Stoner ferromagnetism \cite{Jo2009,Barth2011,Pekker2011,Weld2011,Sanner2012,Valtolina2016}. For a detailed discussion we refer the reader to the excellent review by Massignan \textit{et al.} \cite{Massignan2014}. The  repulsive polaron  was experimentally observed in three dimensions for the first time by Kohstall \textit{et al.} \cite{kohstall_metastability_2012} and also in two-dimensional Fermi gases by Koschorreck \textit{et al.} \cite{koschorreck_attractive_2012} following its theoretical prediction \cite{Schmidt2012b,Ngampruetikorn2012}. For a recent experimental study of  repulsive polarons in a $^6$Li Fermi gas using radio-freqency spectroscopy we refer to \cite{Scazza2016}.  

While the ground state and zero-temperature properties of Fermi polarons have received much theoretical attention, only  recently the study of real-time dynamics of Fermi polarons came into reach of experimental techniques. First experimental steps towards the study of real-time dynamics were taken by Cetina \textit{et al.} \cite{cetina_decoherence_2015}, where the long-time impurity decoherence dynamics following an interaction quench was studied (the results of this work will be discussed in detail in Section \ref{sec_decoh}). Very recently,  interferometric Ramsey techniques were  used to experimentally observe the real-time formation of Fermi polarons for the first time  \cite{cetina_2016}. In the work \cite{cetina_2016}, the functional determinant approach, presented and reviewed in the present report, had been employed for a detailed description of the observed dephasing dynamics (for a detailed discussion of a variational approach to the problem see~Ref.~\cite{parish2016}). \rtext{With these recent developments  the stage it set for the experimental study of impurities in the heavy-mass limit where strong fluctuations lead to intriguing many-body dynamics accompanied by the complete disintegration of  Fermi polarons.  } 

\rtext{It is this real-time dynamics of heavy impurities immersed in a Fermi gas that is} at the center of this report. \rtext{While developing the theoretical description, we will make  connections} to known results on the excitation spectrum of Fermi polarons wherever applicable. In this respect, this report serves not only to introduce new approaches to the dynamics of heavy impurities in Fermi gases as well as to highlight new directions in the study of such systems, but also to make the connection to previous theoretical work on ground and equilibrium properties of Fermi polarons.

\paragraph*{\textbf{Outline.}---}The structure of this report builds on the observation that  many aspects of the dynamics of impurities can  be studied equivalently in the frequency or time domain. Experimentally, the time-resolved impurity Green's function $S(t)$ can be measured using Ramsey interference, while its Fourier transform to the frequency domain $A(\omega)$ is accessible in absorption spectroscopy \cite{goold_orthogonality_2011,knap2012,cetina_2016}. While both signals contain in principle the same information, their measurement can present different challenges to experiments. Also, their separate theoretical analysis gives  insight to non-equilibrium impurity dynamics from different perspectives. Beyond that, time domain methods also allow one to study  observables, such as the spin-echo signal, which have no analogue in frequency space and hence yield  information about the many-body system not  accessible by frequency-resolved methods. Following this route --- after we introduce the two-channel scattering model describing the scattering in the vicinity of a Feshbach resonance and show its equivalence to the Fano-Anderson model in Section \ref{model} ---  we  introduce the dynamic response functions in the time and frequency domain in Sec.~\ref{response}. 

The radio-frequency absorption response of the system is discussed in detail in Sec.~\ref{rfsection}.  We  analyze its universal properties and develop a simple interpretation of spectral features in terms of a  single-particle picture. This sets the basis to identify universal excitation branches pertinent for our discussion of the time-resolved response in Section \ref{sec_decoh}. Here we first provide a numerically exact  solution of the non-equilibrium quench dynamics followed by the discussion of the relevant excitation branches.  The identification of these branches allows us to derive analytical formulas for the universal asymptotic long-time  dynamics based on bosonization and the theory of  Toeplitz determinants. Finally, in Section \ref{sec:shorttime}, we introduce the Tan relations which relate the long-time phase evolution of the impurity Green's function to  high-frequency response. We summarize our findings and discuss future perspectives in Sec.~\ref{outlook}.

\section{Anderson-Fano model with ultracold atoms \label{model}}

\begin{figure}[t] 
  \centering 
  \includegraphics[width=\linewidth]{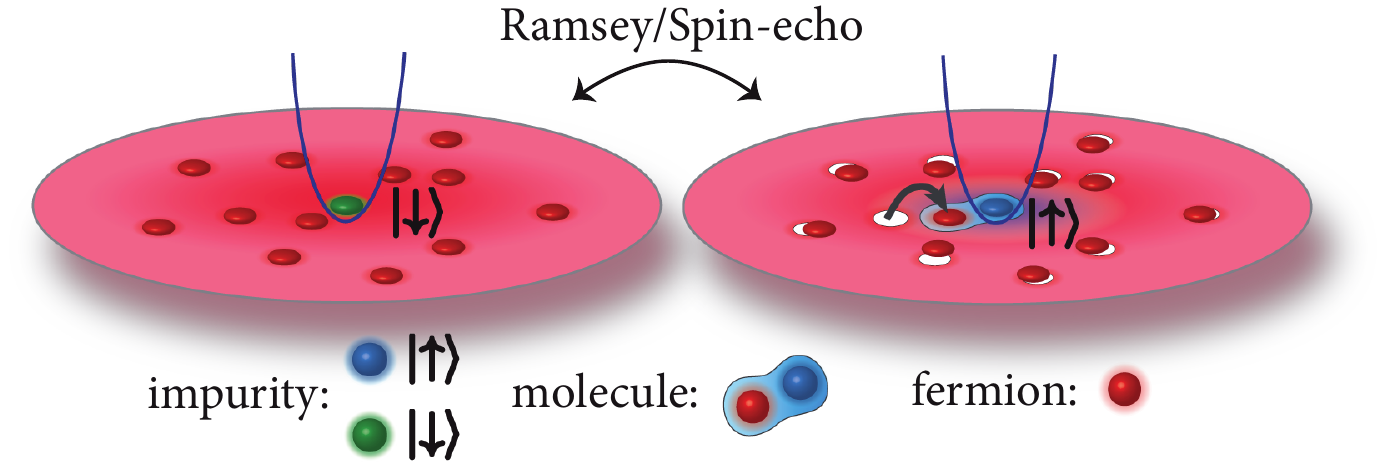}\vspace{-3mm}
  \caption{ \textbf{Schematic representation of the model.} Left: \rtext{an impurity atom in the $\ket{\downarrow}$ hyperfine state (green sphere) is decoupled from atoms in the Fermi sea (red spheres). Right: the impurity atom in the $\ket{\uparrow}$ hyperfine state (blue sphere) interacts with the Fermi sea atoms with a strength that is tunable by a Feshbach resonance. The impurity is either of infinite mass or, as illustrated, trapped in an additional, strong optical potential. When the impurity interacts with the Fermi gas it can bind with an atom from the Fermi sea and form a molecule.} The internal hyperfine spin state is  addressed using radio-frequency (RF) pulses in RF absorption spectroscopy or Ramsey and spin-echo interferometry.}
  \label{fig:setup}
\end{figure}

We study a low density of impurity atoms immersed into a Fermi gas of atoms of mass $m$. As illustrated in Fig.~\ref{fig:setup}, two hyperfine states of the impurity atoms, which we refer to as $\ket{\downarrow}$ and $\ket{\uparrow}$, respectively, participate in the dynamics. These states are chosen such that only one of them interacts with the Fermi gas while the other does not. In the following, we consider impurities of an infinite mass which are localized in space. Experimentally, this can be achieved by  using atomic species with a different polarizability so that  only the impurities are trapped by an optical lattice or microtraps  while the fermions in the bath remain mobile \cite{greiner_quantum_2002,bloch_many-body_2008,Lamporesi2010,Serwane2011,catani_quantum_2012,wenz_few_2013}.

\subsection{{Feshbach resonances}}
In ultracold atoms, the scattering of the bath atoms with the impurity is described by the s-wave scattering amplitude
  \begin{equation}\label{scattampl}
f(k)=\frac{1}{k\cot \delta_k-ik}\approx\frac{1}{-1/a+\frac{1}{2} r_e k^2-ik}
\end{equation}
where $k$ is the momentum of the incoming atom, and $\delta_k$ is the s-wave scattering phase shift. The second expression in Eq.~\eqref{scattampl} represents the effective range expansion of the phase shift which is valid for small scattering momenta $k$. Here $a$ is the scattering length, and $r_e$ is the so-called effective range. In the expansion also higher-order terms exist which are, however,  typically negligible for ultracold atoms. Therefore, the two parameters $a$ and $r_e$ provide an accurate and universal description of the scattering physics \cite{Braaten2006}.

It is one of the great appeals of ultracold gases that the scattering length $a$, and hence the interaction strength, can be tuned almost at will using Feshbach resonances \cite{chin_feshbach_2010}. Here one makes use of the coupling of the atoms in an open scattering channel to a molecular state in a closed channel, which is energetically accessible only by virtual processes. Due to its magnetic moment the energy $\epsilon_m(B)$ of the closed-channel molecule can be tuned with respect to the open channel by an external magnetic field $B$.

Depending on the relative detuning $\epsilon_m(B)$ from the scattering threshold at zero energy, the scattering length can be manipulated and at optimal detuning it diverges, $a\to\infty$, defining the Feshbach resonance. Close to the resonance, \rtext{where the non-resonant (background) scattering in the open channel can be neglected}, the magnetic-field dependence of the scattering length can be parametrized as \cite{chin_feshbach_2010}
\begin{equation}\label{aBDep}
a(B)\approx-\frac{\hbar^2}{2\mu_\text{red} r^* \delta\mu\,\, (B-B_0)}.
\end{equation}
Here \rtext{the reduced mass} $\mu_\text{red}=m M /(m+M)$ becomes $\mu_\text{red}=m$ for the case of an impurity of mass $M=\infty$. Moreover, $\delta\mu$ is the differential magnetic moment of the closed-channel molecule, and $B_0$ denotes the magnetic field strength at which the scattering length diverges. Most importantly, Eq.~\eqref{aBDep} defines the range parameter \rtext{$r^*>0$} which determines the character of the Feshbach resonance.

To a good approximation, $r^*$ is related to the effective range as \cite{chin_feshbach_2010}
\begin{equation}
r_e\approx -2 r^*.
\end{equation} 
Therefore the range parameter $r^*$ provides the second parameter $r_e$ required to characterize the scattering properties of ultracold atoms to high accuracy. In the following we will use both $r^*$ and $r_e$ interchangeably to describe the momentum dependence of the scattering phase shift $\delta_k$. Note that unlike the scattering length $a$, the range parameter $r^*$ can typically not be tuned in a practical way \cite{Fedichev1996,Bohn1996,Bohn1999}, but is fixed by microscopic molecular details. 

\begin{figure*}[t] 
  \centering 
  \includegraphics[width=\linewidth]{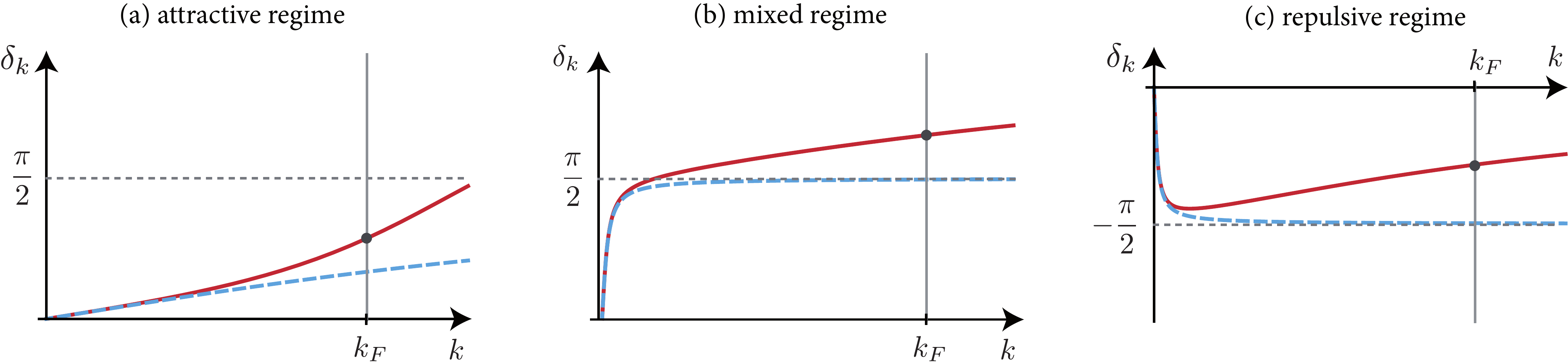}
    \caption{ \textbf{Interaction regimes and scattering phase shift.} \rtext{The figures show the momentum dependence of the scattering phase shift in the three interaction regimes. The various regimes are characterized by the scattering phase shift $\delta_F\equiv\delta_{k=k_F}$ evaluated at the Fermi momentum $k_F$. The attractive regime (a) is characterized by a negative scattering length $a<0$ and a positive phase shift $\delta_F$ that does not exceed $\pi/2$. In the mixed regime (b) the scattering length remains negative while the positive phase shift at the Fermi surface $\delta_F$ exceeds $\pi/2$. The repulsive regime (c) borrows its name from the fact that here single-particle scattering states are shifted upwards in energy (see the detailed discussion in Section~\ref{rfsection}). In this regime the scattering length $a>0$ and a bound state exists in the single-particle spectrum. The existence of the bound state leads to a jump of the phase shift by $\pi$ and we adapt the convention of  choosing $\delta_k<0$. Blue dashed lines correspond to a contact-interaction model with $k_F r^*=0$ and red lines correspond to $k_F r^*=0.8$. In the contact interaction model the mixed regime cannot be realized. The individual plots are shown for scattering lengths (a) $k_F a = -0.5$, (b) $k_F a = -100$, (c) $k_F a = 100$. } }
  \label{fig.phaseshift}
\end{figure*}

\subsection{{Interaction regimes}}
The range $r^*$  influences the phase shift $\delta_k$ in Eq.~\eqref{scattampl} and thus can have a profound effect on the many-body physics.  While for contact interactions, where $r^*=0$, the phase shift is bounded by $|\delta_k|\leq\pi/2$, for finite $r^*$ it can exceed $\pi/2$, see Fig.~\ref{fig.phaseshift}. 
\rtext{This allows one to realize interaction regimes that feature universal physics and that are not accessible in simple contact-interaction models.} 
 \rtext{The interaction regime} are identified by  phase shift  $\delta_F\equiv\delta_{k=k_F}$ evaluated at the Fermi momentum $k_F$.  One can distinguish three interaction regimes, cf.~Fig.~\ref{fig.phaseshift}:
\begin{description}
\item[(a)] \textbf{`attractive regime'}: here $a<0$  and  phase shift at the Fermi surface $\delta_F<\pi/2$. In this regime the phase shift is positive for all energies, $\delta(E)\equiv\delta_{k=\sqrt{2m E}}>0$. \rtext{As discussed in Section \ref{rfsection} the  term `attractive' is derived from the fact that single-particle scattering states are shifted  to lower energies.}
\item[(b)] 
\textbf{\rtext{`mixed regime'}}: 
here $a<0$ and $\delta_F>\pi/2$. In this regime the phase shift is again positive for all energies, $\delta(E)>0$.  \rtext{Within the two-channel model introduced below it is realized for negative inverse dimensionless scattering lengths $-k_F r^*<1/k_F a <0$. }
\item[(c)] \textbf{`repulsive regime'}: here $a>0$ which in our convention for the scattering phase implies $\delta(E)<0$ for all energies.  \rtext{As discussed in Section \ref{rfsection} in this case single-particle scattering states are shifted to higher energies motivating the  term `repulsive' for this regime.}
\end{description}
\rtext{While the} \rtext{mixed} regime (b) cannot be reached in a system with  contact interactions\rtext{, it is accessible} in the Anderson-Fano model \rtext{introduced below}. Alternatively, the \rtext{mixed} regime can also be realized using pure open-channel scattering potentials. Here, according to Levinson's theorem \cite{friedrich2013}, a phase shift $\delta_F>\pi/2$ is  possible when more than one bound state is supported by the interaction potential. \rtext{Note that the term `mixed' is derived from the observation (discussed in Section \ref{rfsection}) that the many-body dynamics in this regime shares properties of both the attractive and repulsive regimes. }

Following the classification of the various interaction regimes, one can use the value of  $k_Fr^*$ for a  many-body characterization of Feshbach resonances. When $k_F r^*\ll 1$, one speaks of  so-called `broad' Feshbach resonance \rtext{\cite{bloch_many-body_2008}}. For those, the physics is universally parametrized by  $k_F a$ alone, and Fermi gases close to such a  resonance can be described by simple contact interaction models. For broad Feshbach resonances, the realization of the \rtext{mixed} regime requires very large, negative values of $k_Fa$. 

In contrast, for so-called `narrow' Feshbach resonances, where $k_F r^*\gtrsim 1$, the \rtext{mixed} regime can be realized more easily. In this regime unique dynamics beyond the simple paradigm of contact interactions  appears, and one has to employ models beyond contact interactions for their description.

\subsection{{Anderson-Fano model}}

In the following we develop a theory which describes the dynamical response of  a Fermi gas coupled to an \rtext{immobile} impurity close to a Feshbach resonance of arbitrary `width' as determined by  $k_Fr^*$. Theoretically, Feshbach resonances can be described with high accuracy by a two-channel model, where the interaction between  atoms is mediated by the exchange of a closed-channel molecule \cite{pethick2002,chin_feshbach_2010}. 

Specifically, for a bath of fermions interacting with a localized impurity, the system is described by the two-channel Hamiltonian
\begin{eqnarray}\label{H2ch}
H_\text{2-ch}&=& - \int_\vecr \hat\psi^\dag(\vecr) \frac{\hbar^2\nabla^2 }{2m}\hat\psi(\vecr) +\epsilon_m \hat \phi^\dagger \phi\nonumber\\
&+& g \int_\vecr \chi(\vecr) [\hat \phi^\dagger \hat \psi_\sigma \hat \psi(\vecr) + \text{h.c.}]
\end{eqnarray}
where the first term describes the free fermions of mass $m$ with \rtext{creation operators $\hat \psi^\dagger(\vecr)$}. The second term represents the closed-channel molecule \rtext{created by $\hat \phi^\dagger$} with energy  $\epsilon_m(B)$ detuned from the scattering threshold.  Furthermore, \rtext{the impurity, which is localized at $\vecr=0$, is created by the operator $\hat \psi^\dagger_\sigma$ in  the atomic spin state $\sigma$ in which it interacts with the Fermi sea};  c.f.~Fig.~\ref{fig:setup} where we have chosen $\sigma=\uparrow$.

\rtext{Like the impurity, the molecule is immobile and hence the corresponding creation operators carry no coordinate dependence. Since their mass is infinite both molecule and impurity also have no kinetic energy term.} Finally, the  interaction of the impurity with the bath of fermions is described by the third term, where the impurity $\hat \psi_\sigma$ and a host atom $\hat \psi$ are converted into the molecular state $\hat \phi$. Here the form factor $\chi(\vecr)$ determines the shape of the atom-to-molecule coupling and we choose $\chi(\vecr)={e^{-r/\rho}}/{4\pi \rho^2 r}$ \rtext{where the range $\rho$ is determined by the van-der-Waals length \cite{schmidt_efimov_2012,Langmack2017}. }

For an impurity of infinite mass the molecule creation and impurity annihilation operator can be combined to a new operator $\hat m^\dag = \hat \phi^\dag  \hat \psi_\sigma$ which allows us to map the two-channel model onto the Fano-Anderson model 
\begin{equation}
\hat H = \epsilon_m \hat m^\dag\hat m +\sum_\veck \epsilon_\veck \hat c_\veck^\dagger \hat c_\veck + \frac{g}{\sqrt{V}} \sum_\veck \chi_\veck [\hat m^\dag \hat c_\veck + \text{h.c.}].
\label{eq:h}
\end{equation}
\rtext{Here the operators $\hat c^\dagger_\veck$ are the  creation operators of fermions with dispersion relation}  $\epsilon_\veck=\hbar^2\veck^2/2m$ and $\chi_\veck$ is the Fourier transform of the form factor $\chi(\vecr)$.  The Fano-Anderson model has been introduced in solid state physics to describe localized magnetic states in metals~\cite{anderson_localized_1961,Fano1961}. It reappears here in the new context where it can be experimentally studied with the high control of ultracold atoms with  tunable model parameters. 

The parameters $\epsilon_m$, $g$, and $\rho$ of the model \eqw{eq:h} can be deduced from the scattering properties. Calculating the scattering phase shift from the model  \eqw{eq:h} we can relate those parameters  to  experimentally accessible quantities which are the scattering length $a$,   the  range parameter $r^*$, and the van der Waals length $l_\text{vdW}$, which determines the range of the underlying atomic interaction potentials  \cite{goral_adiabatic_2004,szymanska_conventional_2005}. Using the abbreviation $\bar a=4\pi/\Gamma(1/4)^2\ l_\text{vdW}\approx 0.956\ l_\text{vdW}$ \cite{chin_feshbach_2010}, one finds (for details see Appendix \ref{sec:solModel} and \cite{schmidt_efimov_2012,cetina_2016})
\begin{equation}
g^2 = \frac{\hbar^4 \pi}{\mu_\text{red}^2 r^*},\quad
 \epsilon_m= \frac{\hbar^2}{2\mu_\text{red}r^*}\Big(\frac{1}{\bar a}-\frac{1}{a}\Big),\quad
 \rho = \bar a/2.
 \label{eq:parameters}
\end{equation}
\rtext{Here we  keep the effective mass $\mu_\text{red}$ explicit as these equations apply to arbitrary mass ratios between impurity and bath atoms; for the case of an immobile impurity  $\mu_\text{red}=m$. Note that the} first equation shows that the range parameter $r^*$ controls the strength of the coupling of the atoms  to the closed-channel molecule.

The following many-body calculation requires to find the single-particle solutions of Hamiltonian \eqw{eq:h}. Those are obtained from the ansatz $\ket{\Psi}=\alpha_m\ket{m}+\ket{\psi}$ which takes into account explicitly the molecular state $\ket{m}$ and  solves the Schr\"odinger equation $\hat H \ket{\Psi}=E\ket{\Psi}$ \cite{Fano1961}. Here $\alpha_m$ is a constant that determines the occupation of the closed-channel molecule $\ket{m}$, and $\psi(r) =\braket{r}{\psi}=  A\ {\sin (kr +\delta_k)}/{r}+ B\ \chi(r)$ represents the open-channel scattering wave function. \rtext{For the bound state we use the ansatz  $A\ {\sinh (\kappa (r-R))}/r+ B\ \chi(r)$ for the radial wave function where $\kappa$ is the binding wave vector and $R$ is the size of a spherical box $R$}. From the solution of the Schr\"odinger equation  we  evaluate the unknown coefficients $A,\ B,\ \alpha_m$. Next, we calculate the single-particle eigenenergies, determined through $k$ and $\kappa$, from the boundary conditions of atoms being confined in the spherical box of size $R$. For  details we refer to Appendix~\ref{sec:solModel}.

\section{Dynamic many-body responses\label{response}}
Information about a many-body system can be obtained from response measurements in both the frequency and  time domain. In ultracold atomic systems experimental tools exist to address both domains with high precision.  While radio-frequency (RF) spectroscopy gives access to the spectrum of a many-body Hamiltonian,  Ramsey or spin-echo interferometry reveals information about the time-evolution of the many-body wave function.  Using functional determinants one can solve numerically  the response of the model \eqw{eq:h} in both the frequency and time domain at arbitrary temperature without  approximations. In this section we introduce some of the responses typically studied in cold atom experiments. These will then be investigated in more detail in the following sections.

\subsection{Radio-frequency spectroscopy}

Information about the spectrum, and in particular the ground state of the Anderson-Fano model \eqref{eq:h}, can be obtained experimentally from RF spectroscopy. In such an experiment the impurity is prepared in an initial spin state, $\ket{\sigma}$, where $\sigma \in \{\uparrow,\downarrow\}$. Using a weak RF signal, the spin is then driven into a final state $\ket{\bar \sigma}$, orthogonal to $\ket{\sigma}$. Theoretically, the RF signal can be modeled by a monochromatic perturbation $\sim e^{i\omega t} \ket{\bar\sigma}\bra{\sigma}+\text{h.c.}$ of frequency $\omega$. In linear response theory the absorption is given by Fermi's golden rule ($\hbar=1$)
\begin{eqnarray}\label{Awresolved}
A(\omega)=2\pi \Omega^2 \sum_{i,f}w_i|\bra{f}\hat W\ket{i}|^2 \delta[\omega-(E_f-E_i)],
\end{eqnarray}
where the transition operator $\hat W=\ket{\bar\sigma}\bra{\sigma}+\text{h.c.}$ acts only on the spin state of the impurity.  Furthermore, $\Omega$ is the Rabi frequency which determines the power of the applied RF field. In the following we set $\Omega=1$. The sum in Eq.~\eqref{Awresolved} extends over complete sets of initial $\ket{i}$ and final many-body states $\ket{f}$ with energies $E_i$ and $E_f$. The weights $w_i$ are determined by the initial state density matrix as $w_i=\bra{i}\hat\rho_i\ket{i}$. 

The measured RF signal depends on the specific initial and final states chosen and in the following we will focus on two scenarios. In the \textit{`standard RF'} scheme, the system is driven from the spin state, in which the impurity interacts with the Fermi sea, to a non-interacting spin state. In contrast, in the \textit{`reverse RF'} procedure the impurity is initially in a non-interacting state and then driven to an interacting one \cite{schmidt_excitation_2011}. In this section we formally define the two schemes, while their distinct responses are discussed in detail in Section \ref{rfsection}.

\paragraph*{\textbf{`Standard' RF spectroscopy}.---}
In the standard RF scheme, the impurity is prepared in the state $\ket{\uparrow}$ in which it interacts with the Fermi sea.  At $T=0$ the fermions are initially in the many-body ground state \rtext{$\ket{\psi_\text{GS}}$ of the Hamiltonian \eqw{eq:h}, i.e. $\hat H\ket{\psi_\text{GS}}=E_\text{GS}\ket{\psi_\text{GS}}$}, where they experience the impurity as a scattering center at $\vecr=0$. 
The initial state, including the impurity state, is then given by \rtext{$\ket{i}=\ket{\uparrow}\otimes\ket{\psi_\text{GS}}$}. At finite temperature the fermionic initial `state' is  determined by the thermal density matrix \rtext{$\hat \rho_\text{GS}\equiv e^{-\beta( \hat H-\mu \hat N)}/Z_\text{GS}$}, with $\beta=1/k_B T$ the inverse temperature with $k_B$ the Boltzmann constant, which we set to one in the following, $\hat N$ is the fermion number, $\mu$ their chemical potential, \rtext{$Z_\text{GS}$} the partition sum, and $\hat H$ is given by Eq.~\eqref{eq:h}.

In the final  state  the impurity is in the spin state $\ket{\downarrow}$ which is non-interacting with the Fermi gas. Since then the scattering center is absent, the system is described by the Hamiltonian $\hat H_0$, defined by Eq.~\eqref{eq:h} with $g=0$. The sum over the systems' final states can be written in terms of the states $\ket{f}=\ket{\downarrow}\otimes\ket{\psi_n}$ where $\ket{\psi_n}$ denotes the complete set of many-body fermionic eigenstates of $\hat H_0$, $\hat H_0\ket{\psi_n}=E_n^0\ket{\psi_n}$, i.e. $E_f\equiv E_n^0$ in Eq.~\eqref{Awresolved}. Using these definitions and transforming Eq.~\eqref{Awresolved} to the time domain \rtext{(for details see App.~\ref{SAAppendix})}, one arrives at the expression for the `standard RF absorption spectrum'
\begin{equation}\label{eq:StandardRF}
R(\omega)=2\text{Re}\int_0^\infty dt e^{i\omega t} \Tr [e^{i \hat H t}e^{-i \hat H_0 t}\hat \rho_\text{GS}],
\end{equation}
which solely contains operators in the fermionic Hilbert space.  

\rtext{The standard RF response of an impurity immersed in a Fermi gas has been measured by Schirotzek \textit{et al.} \cite{schirotzek_observation_2009}. In this case the impurity was mobile and the ground state  of the system $\ket{\psi_\text{GS}}\to\ket{\psi_\text{pol}}$ is a well-defined quasiparticle, the Fermi polaron. The Fermi polaron is a well-defined quasiparticle since its state has a finite overlap, the so-called quasiparticle weight $Z$, with the non-interacting ground state $\ket{\psi_0}$ of the system, i.e. $Z=|\braket{\psi_0}{\psi_\text{pol}}|^2>0  $. This is different for an infinitely heavy impurity at zero temperature. In this case the overlap of the interacting ground state $\ket{\psi_\text{GS}}$ with the non-interacting ground state vanishes, i.e. $Z=|\braket{\psi_0}{\psi_\text{GS}}|^2\equiv 0 $. Hence no quasiparticle exists in this case, giving rise to the term `orthogonality catastrophe' (OC) \cite{anderson_infrared_1967}. The relation between the Anderson OC and Fermi polarons will be discussed in more detail in the following Section \ref{rfsection}.}

\paragraph*{\textbf{`Reverse' RF spectroscopy}.---}
In the `reverse' RF scheme, sometimes also called the `spin-injection' scheme \cite{Cheuk2012}, the role of interactions are reversed. Here the impurity in its initial atomic state $\ket{\downarrow}$  is not interacting with the Fermi gas. Hence, at $T=0$, the fermions build a perfect free Fermi sea $\ket{\text{FS}}$ by filling up all single-particle eigenstates of the non-interacting Hamiltonian $\hat H_0$ up to the Fermi energy $\epsilon_F$. The initial state is then given by $\ket{i}=\ket{\downarrow}\otimes\ket{\text{FS}}$. At finite temperature $T$, $\ket{\text{FS}}$ is replaced by the thermal density matrix $\hat \rho_{\text{FS}}=e^{-\beta(\hat H_0-\mu\hat N)}/Z_{\text{FS}}$ of a free Fermi gas.

The impurity is then driven into its final state $\ket{\uparrow}$ in which the impurity is interacting with the gas. In that state, the dynamics of the fermions is described by the Hamiltonian $\hat H$ where the scattering center at $\vecr=0$ is present; i.e. $g>0$ in Eq.~\eqref{eq:h}.  The final states in Eq.~\eqref{Awresolved} are then given by $\ket{f}=\ket{\uparrow}\otimes\ket{\psi_\alpha}$, where the fermionic states are defined by $\hat H  \ket{\psi_\alpha}= E_\alpha \ket{\psi_\alpha}$ \footnote{Note that we set the energy splitting between hyperfine levels to zero. }.
Basic manipulations \rtext{(for details see App.~\ref{SAAppendix})} lead to the reverse RF response 
\begin{equation}\label{eq:InverseRF}
A(\omega)=2\text{Re}\int_0^\infty dt e^{i\omega t} \Tr [e^{i \hat H_0 t}e^{-i \hat H t}\hat \rho_\text{\text{FS}}].
\end{equation}
The reverse RF scheme  had been implemented  for the observation of the full impurity spectral function by Kohstall \textit{et al.} \cite{kohstall_metastability_2012}, and found multiple applications in the observation of polaronic physics with ultracold atoms \cite{kohstall_metastability_2012,Zhang2012,cetina_2016,jorgensen2016,hu2016,Scazza2016,SchmidtDemKillian2016,SchmidtDemKillian2017}.

\subsection{Real-time responses}

Information about the many-body dynamics can also be obtained from real-time observables. Here we give two examples for real-time responses, namely Ramsey and spin-echo interferometry~\cite{knap2012,Mark2015}.

\begin{figure*}[t] 
  \centering 
  \includegraphics[width=\linewidth]{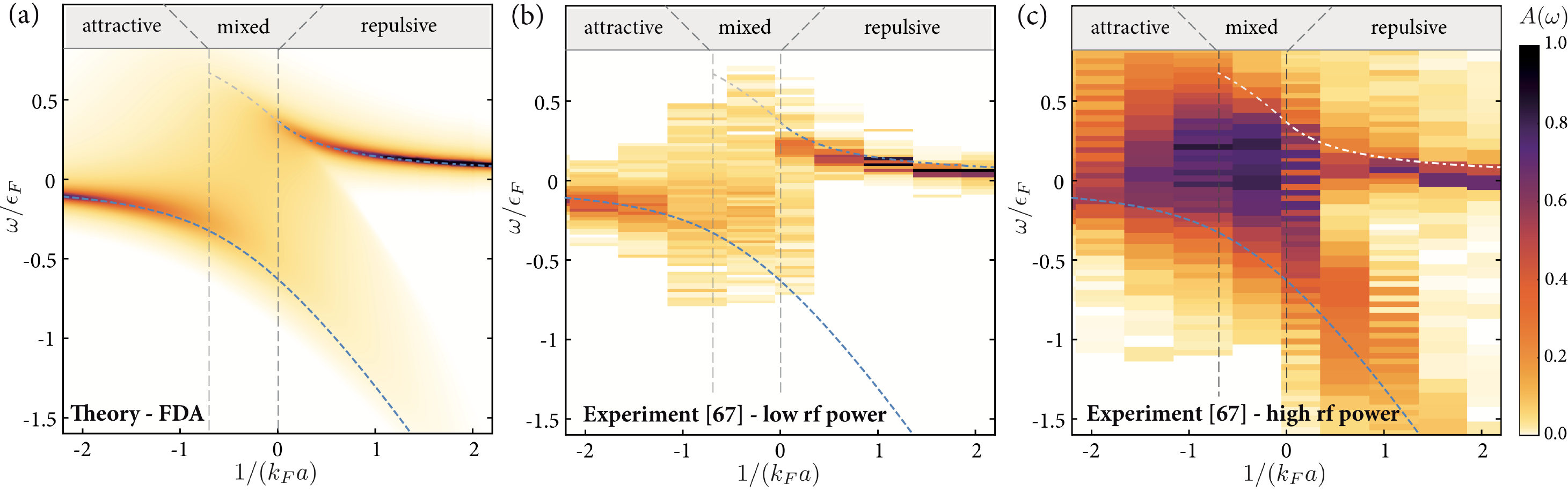}
    \caption{ \textbf{Finite temperature RF absorption spectra -- theory and experiment \cite{kohstall_metastability_2012}.}  (a) Reverse RF absorption spectrum $A(\omega)$ as function of frequency $\omega/\epsilon_F$ and inverse interaction strength $1/k_F a$  in the `attractive', `\rtext{mixed}', and `repulsive' interaction regime as obtained from an FDA calculation (throughout this work  absorption spectra $A(\omega)$ are shown in units of $\epsilon_F$). The temperature $T/T_F=0.16$ and range parameter $k_Fr^*=0.71$ are chosen as in the experiment by Kohstall \textit{et al.} \cite{kohstall_metastability_2012}.  Their experimental results are shown for comparison: in (b) for a weak RF drive and (c) for a strong RF drive which saturated the signal and hence was  beyond linear response.  The dashed and dot-dashed lines, shown in (a-c) correspond to the \rtext{energy of the onset of the attractive and repulsive excitation branch.} They are  calculated from Fumi's theorem \cite{Fumi1955} which relates the sum over phase shifts to the ground state energy of the system, see also Refs.~\cite{mahan_many_2000,Comescot2007,massignan_repulsive_2011} \rtext{and App.~\ref{FumiApp}.} }
  \label{fig.densityrfspec}
\end{figure*}

\paragraph*{\textbf{Ramsey interferometry}.---}
In many-body Ramsey interferometry, the impurity is initially prepared in a non-interacting spin state $\ket{\downarrow}$. Then using a $\pi/2$ rotation a superposition of the two impurity states, $(\ket{\downarrow}+\ket{\uparrow})/\sqrt{2}$ is prepared, where in the state $\ket{\uparrow}$ the impurity interacts with the Fermi gas. After a finite interaction time $t$ a second $\pi/2$ pulse with variable phase $\varphi$ is applied. Measuring $\hat \sigma_z$ then yields  the Ramsey signal $S(t)$ \cite{knap2012,cetina_decoherence_2015}. \rtext{Throughout this work we assume infinitely fast $\pi/2$ rotations that leave the Fermi sea unperturbed. While the implementation is experimentally challenging directly in the strongly interacting regime, it has been shown  \cite{cetina_2016} that such protocols can indeed be realized by quenching interactions using the optical resonance shifting technique developed in Ref.~\cite{cetina_decoherence_2015}.} 
A straightforward calculation shows that the Ramsey signal is given by
\begin{equation} \label{eq:ramsey}
 S(t)=\langle e^{i \hat H_0 t}e^{-i \hat H t} \rangle\equiv  \Tr [e^{i \hat H_0 t}e^{-i \hat H t}\hat \rho_{\text{FS}}],
\end{equation}
From this expression it is evident that the reverse RF absorption signal Eq.~\eqref{eq:InverseRF}  is determined by the Fourier transform of the Ramsey signal. Hence both signals contain the same physical information and can be used as complementary experimental tools (for more details see App.~\ref{SAAppendix}).

\paragraph*{\textbf{Spin-Echo interferometry}.---}
In addition to the Ramsey response,  more complicated interferometric protocols can be chosen which do not have simple conjugate observables in the frequency domain. One example is spin-echo interferometry, which defines  a more involved spin trajectory by  augmenting the Ramsey sequence by an additional \rtext{instantaneous} spin-flip at half of the evolution time. This leads to the  expression~\cite{knap2012} 
\begin{equation}
 S_\text{SE}(t)=\langle e^{i \hat H_0 t/2}e^{i \hat H t/2}  e^{-i \hat H_0 t/2}e^{-i \hat H t/2} \rangle.
 \label{eq:spinecho}
\end{equation}
In typical applications in the context of nuclear magnetic resonance (NMR) spectroscopy such  protocols are employed to echo-out external, quasi-static  perturbations. However, such protocols can also yield  additional information about the system dynamics and we will contrast the spin-echo and Ramsey response in the thermal and quantum regimes in Section~\ref{sec_decoh}.

\subsection{Functional determinant approach}
In order to calculate the absorption spectra $A(\omega)$ and the time-dependent many-body response such as the Ramsey and spin-echo signal, we use the functional determinant approach (FDA)~\cite{levitov_electron_1996,klich_03,schoenhammer07}. The FDA provides an exact numerical solution for systems which are described by fermionic bilinear Hamiltonians. Extensions to bosonic systems are possible \cite{klich_03} and have been recently \rtext{employed} for the description of Rydberg impurity systems \cite{SchmidtDem2016,SchmidtDemKillian2016,SchmidtDemKillian2017}.  The FDA reduces expectation values of many-body operators to determinants in  single-particle Hilbert space by virtue of the formula
\begin{equation}\label{eq:klich}
\langle e^{\hat{  Y}_1}\ldots e^{\hat{  Y}_N} \rangle=\det[1-\hat n+\hat ne^{\hat{  y}_1}\ldots e^{\hat{y}_N}].
\end{equation}
Here $\hat{  Y}_i$ are arbitrary, fermionic, bilinear many-body operators and $\hat y_i$ their single-particle representatives. Furthermore, $\hat n$ denotes the single-particle occupation number operator. The identity \eqref{eq:klich} allows us to solve the time-dependent many-body problem exactly by using the analytical solutions for the single-particle orbits obtained for the model Eq.~\eqref{eq:h}, cf.~App.~\ref{sec:solModel}.

\section{Universal many-body response: Radio-frequency spectra\label{rfsection}}

We now turn to RF absorption spectra. Our results are obtained by evaluating Eq.~\eqref{eq:StandardRF} and \eqref{eq:InverseRF}  using the FDA. While there has been much theoretical effort to predict \rtext{the properties of impurities immersed in Fermi gases} at zero temperature using approximate techniques \cite{Chevy2006,Lobo2006,Comescot2007,Comescot2008,prok2008,Veillette2008,Prokofiev2008b,Pilati2008,Punk2009,Mora2009,Combescot2009,Cui2010,schmidt_excitation_2011,massignan_repulsive_2011,Schmidt2012b,Yi2012,Levinsen2012,Massignan2012,Ngampruetikorn2012,Trefzger2013,Trefzger2013b,Massel2013,Edwards2013,Astrakharchik2013,knap_dissipative_2013,Vlietinck2013,Doggen2013,Massignan2013,Parish2013,Kroiss2014,Kantian2014,Trefzger2014,Massignan2014,Bour2015,Kroiss2015,Christensen2015,Christensen2015b,Chen2016,Roscher2015,Lan2015,Hui2016,Sarang2015,duncan2016,parish2016,Mao2016,Mitchison2016}, our calculations are to our knowledge the first to \rtext{predict corresponding spectra at arbitrary temperature, which are exact in the limit of infinitely heavy impurities} (for comparison to recent ultracold atom experiments see also \cite{cetina_2016}). Studying the frequency resolved response provides insight about the relevant many-body states and allows us to identify the excitation branches which will become of importance in finding the analytical solution of the many-body dynamics in Section \ref{sec_decoh}. 

In this section, we focus mostly on the low-frequency response which is universally determined by the scattering phase shift close to the Fermi surface. The high-frequency response \rtext{is discussed in more detail in Section}~\ref{sec:shorttime}.

\subsection{Reverse RF spectra} 

\begin{figure}[t] 
  \centering 
  \includegraphics[width=0.904\linewidth]{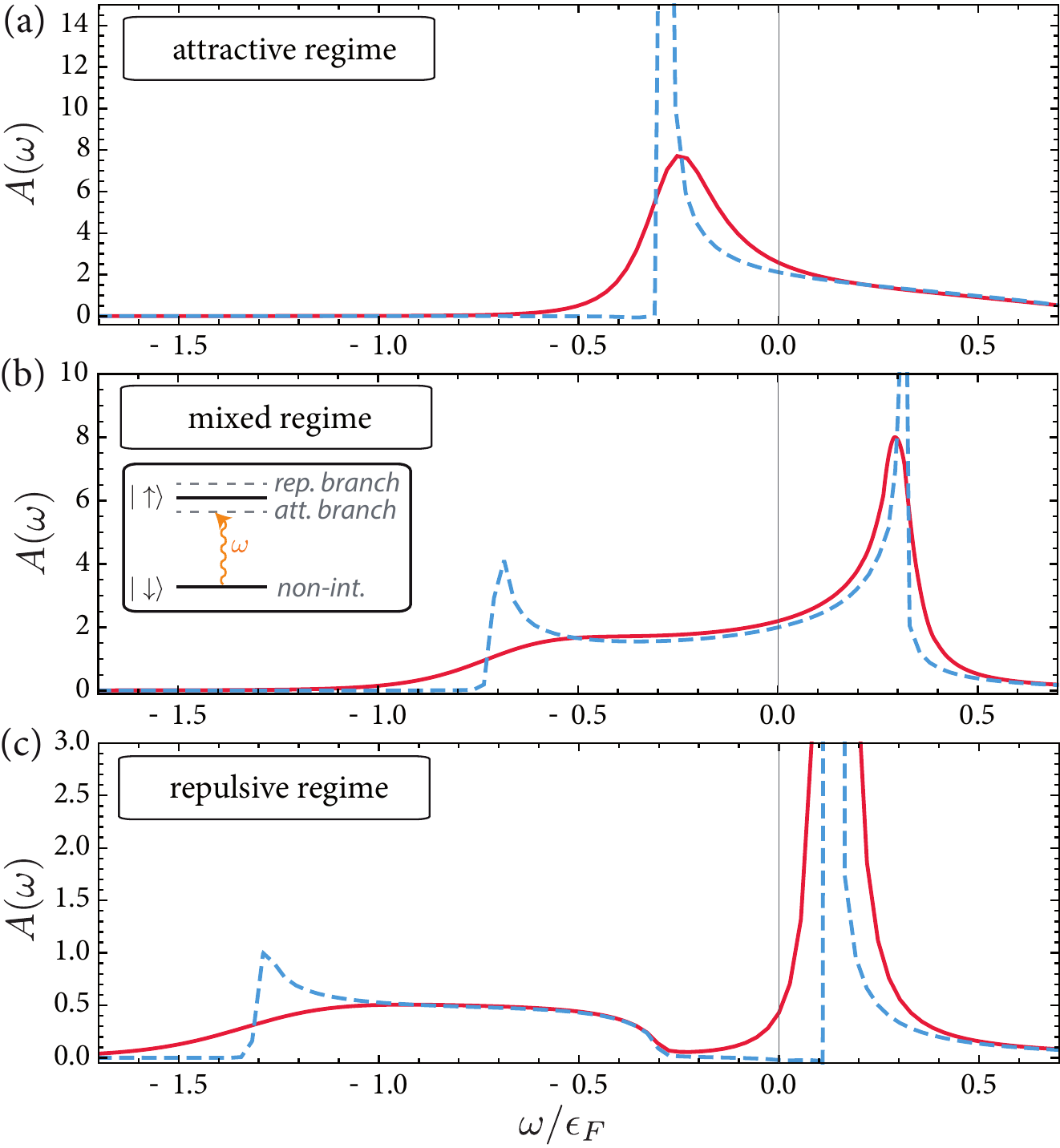}
  \caption{ \textbf{Finite temperature radio-frequency absorption spectra.} RF response in the reverse scheme where the impurity is initially in a hyperfine state non-interacting with the Fermi gas and driven to a state interacting with the gas (see inset in (b) for an illustration). The spectra are shown for temperatures $T/T_F=(0.01,\ 0.16)$ (dashed, solid), for interaction strengths $1/k_Fa=(-0.91,\ -0.1,\ 1.0)$ ranging from the attractive to the repulsive side of the Feshbach resonance, and for a fixed Feshbach range parameter $k_Fr^*=0.71$. \rtext{For $T/T_F=0$ the spectral onset of features will be replaced by sharp edges.}}
  \label{fig:linresponse}
\end{figure}

First we consider the reverse RF absorption spectrum, where the system is initially prepared in the non-interacting impurity spin state and then driven into an interacting  final  state. In the reverse RF scheme the measured response is identical to the impurity spectral function~\cite{schmidt_excitation_2011,kohstall_metastability_2012,cetina_2016,parish2016} and as such it reveals the spectrum of the Hamiltonian \eqref{eq:h}. In Fig.~\ref{fig.densityrfspec}(a) we show a density plot of the predicted absorption spectrum as a function of frequency $\omega/\epsilon_F$ and inverse interaction strength $1/k_Fa$ for the temperature $T/T_F=0.16$ and range parameter $k_Fr^*=0.71$. In Fig.~\ref{fig:linresponse} corresponding spectral cuts are shown for each of the interaction regimes realized by the specific interaction strengths $1/(k_Fa)=-0.91,\ -0.1,\ \text{and } 1$, respectively, which range from the `attractive' ($a<0$) to the `repulsive' side ($a>0$) of the Feshbach resonance. We show results for both low ($T/T_F=0.01$, dashed) and high temperatures ($T/T_F=0.16$, solid). \rtext{Note that the results are shown for the finite value $T/T_F=0.01$ as we discuss this case also when considering the real-time response in Section~\ref{sec_decoh}.}

The reverse RF spectra are dominated by two main excitations. The `attractive \rtext{excitation} branch'  appears as a pronounced response at negative detuning $\omega/\epsilon_F<0$. It is dominant in the attractive and \rtext{mixed} regime for $k_Fa<0$.  As the resonance is crossed to positive values of $1/k_Fa$ [Fig.~\ref{fig:linresponse}(a) to (c)] the attractive \rtext{branch} loses weight and the `repulsive \rtext{excitation} branch' emerges  at positive energies  and carries most of the  spectral weight. As can be seen from the spectral cuts in  Fig.~\ref{fig:linresponse}, at ultralow temperatures both excitation branches exhibit a non-analytical onset in frequency. These absorption edges are a key signature of the Anderson orthogonality catastrophe (OC)  \cite{anderson_infrared_1967,nozieres_singularities_1969,nozieres_recoil_94,knap2012}, and they are a consequence of the  absence of quasiparticles at zero temperature \rtext{for an infinitely heavy impurity}.

\begin{figure*}[t] 
  \centering 
    \includegraphics[width=\linewidth]{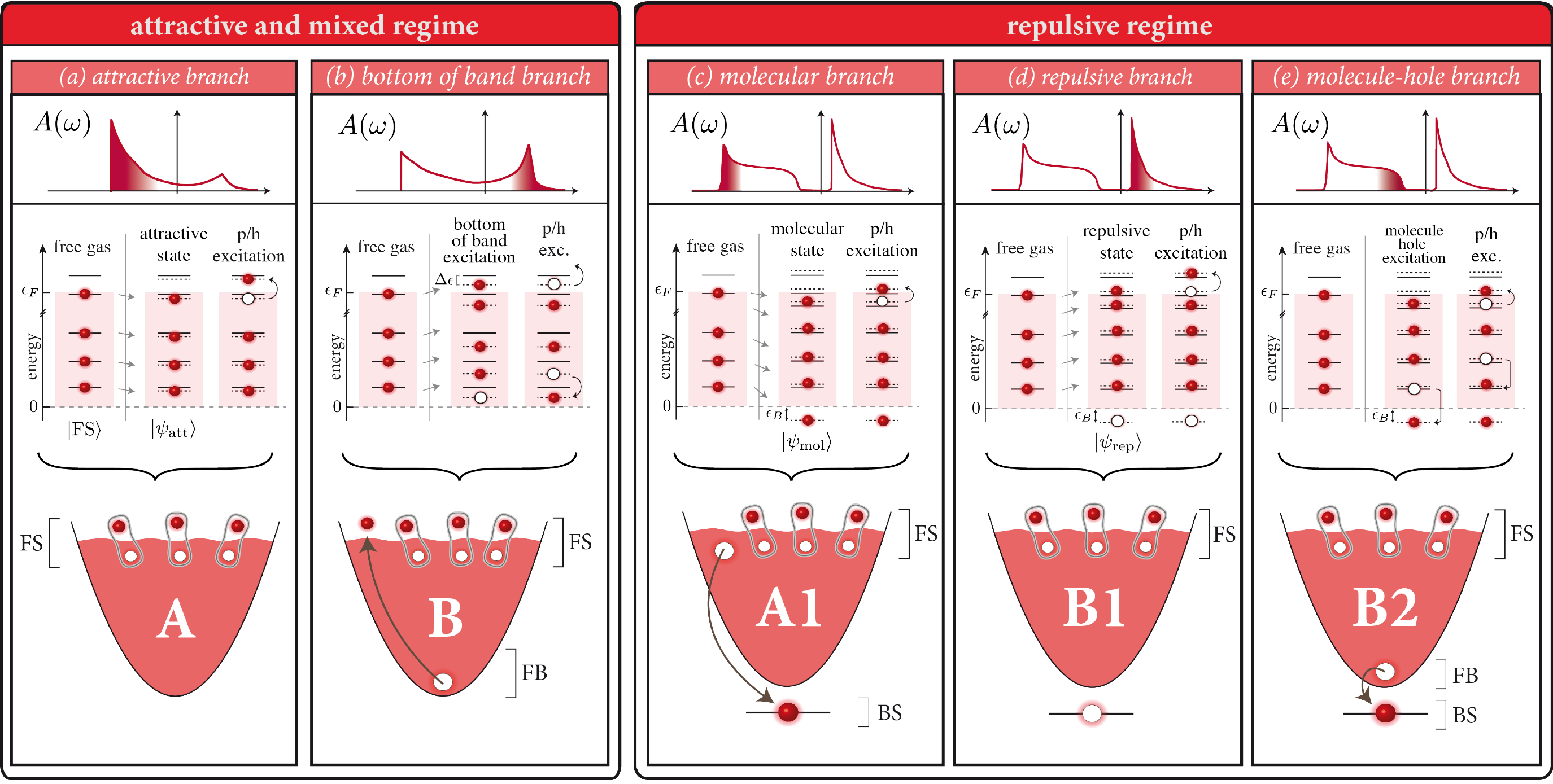}
    \caption{ \textbf{Illustration of the excitation branches dominating the OC dynamics. } In the upper panels we illustrate  typical absorption spectra realized in the respective interaction regimes.  The middle panel shows representatives of many-body states constituting the respective excitation branch.  In this panel we illustrate typical rearrangements of atoms (solid red spheres) from their non-interacting single-particle states (small solid, horizontal lines) into interacting states (small dashed, horizontal lines). Dependent on the interaction, those states are either shifted upwards or downwards in energy, and their wave function overlap with the non-interacting states is modified as well. \rtext{In this panel $\epsilon_F$ is the Fermi energy, $\epsilon_B$ the binding energy of the molecule present for $a>0$, $\Delta\epsilon$ the energy shift of single-particle levels due to interactions with the impurity, and $\ket{\text{FS}}$ represents the Fermi sea in its non-interacting ground state.} The \rtext{combined effect of the processes shown in this middle panel leads} to the characteristic features in the absorption spectra shown in the upper panels where shaded regions highlight the corresponding contributions. \rtext{The characteristic excitations of each branch are summarized in the lowest panel  where FS, FB, and BS are short-hand for Fermi-surface, bottom-of-the-Fermi-sea, and bound-state contributions, respectively.} }
  \label{fig.illustration}
\end{figure*}

\paragraph*{\textbf{From Fermi polarons to the orthogonality catastrophe---}} 
\rtext{Before we turn to the case of an infinitely heavy impurity, which is the focus of this report, let us briefly  consider the general case of an impurity of arbitrary mass $M$. The formation of a polaron is associated with the dressing of the impurity by fluctuations in the many-body bath. These fluctuations correspond to various ways in which the fermions reoccupy their single-particle energy levels due to the interaction with the impurity. This reoccupation becomes apparent when expanding the many-body eigenstates of the system in particle-hole fluctuations \cite{Chevy2006,Comescot2007}:
\begin{align}\label{expansion}
\ket{\psi_\alpha}&=\sqrt{Z_\alpha}\hat d^\dagger_\mathbf{0}\ket{\text{FS}} +\sum_{\veck\vecq}\alpha_{\veck\vecq} \hat d^\dagger_{\vecq-\veck}\hat c^\dagger_\veck\hat c_\vecq \ket{\text{FS}}+\ldots 
\end{align}
Here we consider the case of zero total momentum, the operators $d^\dagger_\vecp$ create an impurity in the momentum state $\vecp$, and $\ket{\text{FS}}$ represents the non-interacting Fermi sea in its ground state. The second term in this expression corresponds to the generation of a \textit{single} particle-hole fluctuation from the Fermi sea, but to obtain the exact eigenstates of the system the complete expansion (indicated by the dots) has to be considered. }

\rtext{While all eigenstates of the many-body Hamiltonian can be described by Eq.~\eqref{expansion}, of particular interest are  often those states $\ket{\psi_\alpha}$ that have finite overlap with the non-interacting state of the system $\ket{\psi_0}\equiv\hat d_\mathbf{0}^\dagger\ket{\text{FS}}$. States with such a finite \textit{quasiparticle weight} $Z_\alpha\equiv |\braket{\psi_0}{\psi_\alpha}|^2$ represent quasiparticles, called polarons, that bear close resemblance to their non-interacting counterparts.  As outlined in the introduction, for a mobile impurity in three dimensions two such polaron states dominate the absorption response: the `attractive Fermi polaron' $\ket{\psi_\text{att}^\text{pol}}$ exists at negative energy for sufficiently weak attractive interactions, while the `repulsive Fermi polaron' $\ket{\psi_\text{rep}^\text{pol}}$ appears at positive energy in the repulsive interaction regime (in addition to a dressed molecular state existing in this regime at negative frequencies).}

\rtext{Polaron states can only exist when the particle-hole fluctuations generated by the higher-order terms in Eq.~\eqref{expansion} are moderate enough to allow for a finite quasi-particle weight $Z_\alpha$.  Although particle-hole fluctuations close to the Fermi surface come with no energy cost, they  lead to a finite recoil energy experienced by an impurity of finite mass. For the leading term shown in Eq.~\eqref{expansion}, this energy is given by $E_\text{rec}=(\veck-\vecq)^2/2M$. In a simple picture this energy cost suppresses particle-hole fluctuations and leads to an efficient protection of the Fermi polaron's quasi-particle weight required for well-defined quasiparticles with a finite $Z>0$ to exist.}

\rtext{This is different for impurities of infinite mass and from our simple argument it is apparent that something remarkable must happen in this limit. Here the recoil energy vanishes and the energetic suppression of high-order particle-hole fluctuations is absent. As it turns out the fluctuations become indeed so dominant that they lead to the complete disintegration of the quasiparticle and the  weight $Z=0$ vanishes identically. Hence the many-body ground state $\ket{\psi_\text{GS}}$, which for a mobile impurity at weak interactions had been $\ket{\psi_\text{att}^\text{pol}}$, becomes now completely orthogonal to its non-interacting counterpart, $Z=|\braket{\psi_0}{\psi_\text{GS}}|^2=0$, and one encounters the `orthogonality catastrophe' (OC): no quasiparticles exist \cite{anderson_infrared_1967}.}

\rtext{Describing the OC requires the inclusion of the higher-order terms in the expansion \eqref{expansion} which poses a  challenge for theory that attracted intensive interest starting with the work of Anderson \cite{anderson_infrared_1967}. Among other approaches \cite{de_dominicis_stationary_1964,mahan_excitons_1967,nozieres_singularities_1969,yuval_exact_1970,combescot_1971,mahan_many_2000,hentschel_fermi_2005,pustilnik_dynamic_2006,pereira_exact_2008,pereira_spectral_2009,mkhitaryan_fermi-edge_2011,Pimenov2017}, for overviews we refer to \cite{ohtaka_theory_1990,rosch_quantum-coherent_1999,mahan_many_2000,nazarov_quantum_2009}, the FDA provides an efficient tool to address this problem and yields exact results for infinitely heavy impurities, also in the case of finite temperature where thermal averages have to be performed.}

\paragraph*{\textbf{Single-particle interpretation.---}}  \rtext{The fluctuations in Eq.~\eqref{expansion} correspond to the various ways in which  fermions can occupy their single-particle energy levels.  The analysis of this reoccupation of states becomes particularly simple in the limit of an infinitely heavy impurity, where it corresponds to a static scattering center and the exact single-particle states are thus easily calculated.}

\rtext{Our analysis is illustrated in Fig.~\ref{fig.illustration}. In the upper panels we show typical absorption spectra in the various interaction regimes, while in the middle panel we illustrate the single-particle energies of $\hat H$ and $\hat H_0$ as dashed and dotted lines, respectively.} \rtext{We show here only s-wave states; for short-range interactions, higher partial waves are not renormalized and hence irrelevant.} In the initial state where the impurity as a scattering center is  absent, the atoms fill up all these levels up to the Fermi energy $\epsilon_F$ and build a perfect Fermi sea $\ket{\text{FS}}$ (left subfigures). To the right, we show the single-particle energies when the impurity scattering center is present (dashed horizontal lines). We illustrate various occupations of those single-particle states which correspond to  dominant features in the absorption spectrum $A(\omega)$.

The specific structure of the absorption spectra can be understood in a simple single-particle picture when expressing the spectral function as
\begin{equation}\label{ASingleParticle}
A(\omega)=2\pi \sum_\alpha |\braket{\psi_\alpha}{\text{FS}}|^2\delta[\omega-(E_\alpha-E_{FS})].
\end{equation}
Here $\ket{\psi_\alpha}$ denotes the many-body eigenstates of the interacting Hamiltonian $\hat H$ with eigenenergies $E_\alpha$, and $E_{FS}$ is the energy of the initial non-interacting Fermi sea $\ket{\text{FS}}$. Eq.~\eqref{ASingleParticle} is given for $T=0$ and, as discussed in Section \ref{response}, it can straightforwardly be extended to finite temperature by an additional summation over initial states weighted by the thermal density matrix.

 A pronounced response in the spectrum is due to family of states $\{\ket{\psi_\alpha}\}$ which have a significant many-body Frank-Condon overlap $|\braket{\psi_\alpha}{\text{FS}}|^2$ with the initial state $\ket{\text{FS}}$. Together with the density of states the overlaps $|\braket{\psi_\alpha}{\text{FS}}|^2$ determine the specific shape of the absorption spectrum.  These dominant families of states are called excitation branches in the following.

\rtext{{\textit{Attractive excitation branch.}}} \rtext{When} the microscopic interaction is weakly attractive, i.e. $1/k_Fa\ll-1$, the dressing of the impurity by fermionic fluctuations leads to a reduction of energy.  \rtext{This leads to the formation of an attractive dressed impurity state $\ket{\psi_\text{att}}$  constituting  the new  ground state of the system}.  In the spectral cut shown in Fig.~\ref{fig:linresponse}(a) the attractive ground state can be identified as the pronounced \rtext{edge} feature  at negative detuning $\omega/\epsilon_F$. The  onset of this feature at negative frequency is determined by eigenenergy \rtext{$E_\text{att}$ of the attractive state $\ket{\psi_\text{att}}$}. This state is constructed by filling all interacting single-particle states up to the Fermi energy \rtext{and it has zero quasiparticle weight $Z=|\braket{\text{FS}}{\psi_\text{att}}|^2=0$ in the limit of infinite system size}. 

As illustrated by the horizontal dashed lines in Fig.~\ref{fig.illustration}, each single-particle level is subject to a small, negative energy shift $\Delta \epsilon$ which is determined by the scattering phase shift $\delta(E)\equiv\delta_{k=\sqrt{2mE}}$ evaluated at the respective single-particle energy $E$ (see Eq.~\eqref{eq:sol2} in Appendix~\ref{sec:solModel}). Each fermion occupying these single-particle states acquires this energy shift and the summation over all  $\Delta \epsilon$ determines the \rtext{energy $E_\text{att}$}, shown as a dashed line in Fig.~\ref{fig.densityrfspec}(a). This \rtext{summation} reflects Fumi's theorem \cite{Fumi1955}, which states that the ground state energy is determined by the sum over the single-particle phase shifts \cite{mahan_many_2000,giamarchi_quantum_2004,Comescot2007,massignan_repulsive_2011}\rtext{, see also Appendix \ref{FumiApp}.} 

\rtext{We emphasize once more that for $T=0$} the immobile impurity in its attractive ground state has zero quasiparticle weight $Z=|\braket{\text{FS}}{\psi_\text{att}}|^2=0$ in the thermodynamic limit\rtext{, signaling the breakdown of the polaron picture.} This is different for a mobile impurity in three dimensions \footnote{The case of a mobile impurity in a Fermi gas  in two dimension  is still not fully understood \cite{rosch_quantum-coherent_1999}; see also the recent work \cite{Pimenov2017}.}: here \rtext{$\ket{\psi_\text{att}}$ corresponds to the attractive polaron state $\ket{\psi_\text{pol}}$ with a finite quasiparticle weight $Z$ that leads to a delta-peak response in absorption spectroscopy at $T=0$. This is in contrast to the infinite-mass impurity where the finite weight in the delta function is fully redistributed into the asymmetric wing attached to the onset of the spectrum at $E_\text{att}$.} 

\rtext{This asymmetric continuum, visible on top of the attractive ground-state excitation in Fig.~\ref{fig:linresponse}, is due to the excitation of an arbitrary number of \textit{low-energy particle-hole excitations} close to the Fermi surface.}  One representative member of this family of states, massively contributing to the sum in Eq.~\eqref{ASingleParticle}, is illustrated in the third column in the center panel in Fig.~\ref{fig.illustration}(a). The whole family of states, including \rtext{$\ket{\psi_\text{att}}$}, contributing to the edge and the attached `wing' in the absorption spectrum (red shaded area in the spectrum in Fig.~\ref{fig.illustration}(a)) constitutes the \rtext{`attractive excitation branch'}.

For weak attraction each of the slightly renormalized \textit{single}-particle states has a large overlap with its non-interacting counter part (dotted versus solid lines in the center panels of Fig.~\ref{fig.illustration}) and hence, by moving all fermions `down' to the attractive \rtext{ground-state} configuration,  the largest many-body overlap $|\braket{\psi_\alpha}{\text{FS}}|^2$ can be achieved. This leads to the \rtext{attractive branch} as the dominant feature in the spectrum in the attractive interaction regime and it almost saturates the spectral sum rule.

This changes as attraction is increased. As illustrated in Fig.~\ref{fig.illustration}(b), for stronger attraction the single-particle states are  shifted further down in energy and the attractive \rtext{ground state} becomes more deeply bound. At the same time the overlaps of the single-particle states with their corresponding non-interacting counterparts become progressively smaller so that the spectral feature of the \rtext{attractive excitation branch} loses weight. 

\rtext{{\textit{Bottom-of-the-band excitation branch.}}} Related to this loss of spectral weight is the appearance of an additional feature in the absorption spectrum as unitary scattering close to the Fermi surface, i.e.~\rtext{$\delta_F=\pi/2$}, is approached. In this strong-coupling regime, the overlap of single-particle levels is such that a new pronounced excitation is favored at an energy $\epsilon_F$ above the attractive \rtext{ground-state excitation}.  As  illustrated in the second column of the center panel in Fig.~\ref{fig.illustration}(b), this feature corresponds to the distribution of the fermions into single-particle levels such that the lowest scattering state \rtext{at the bottom of the Fermi sea} remains empty. Hence the feature is termed  the `bottom-of-the-band excitation' \cite{knap2012}. 

First let us consider contact interactions. Here, at unitarity, where interactions are resonant, the phase shift is $\delta(E)=\pi/2$ for all scattering energies. From Eq.~\eqref{eq:sol2} of Appendix~\ref{sec:solModel} it then follows that the single-particle levels are located at half-way of their non-interacting counterparts (see second column in the middle panel of Fig.~\ref{fig.illustration}(b)). As a consequence of this symmetry, the same overlap $|\braket{\psi_\alpha}{\text{FS}}|^2$ is achieved for moving  all particles `up'  in energy to build the bottom-of-band excitation state or `down' into the attractive \rtext{ground} state \rtext{$\ket{\psi_\text{att}}$} where all low-energy levels are filled. Furthermore, the energies of the \rtext{attractive}  and the bottom-of-band feature are given  by \rtext{$-\epsilon_F/2$ and $+ \epsilon_F/2$}, respectively. 

In the \rtext{mixed} regime, realized for $k_Fr^*>0$, the phase shift at the Fermi surface $\delta_F$ exceeds $\pi/2$. We find that this leads to the bottom-of-the-band excitation acquiring a larger spectral weight  compared to the attractive \rtext{excitation branch} while both excitations remain separated by approximately the Fermi energy. This behavior makes the bottom-of-band excitation distinct from the repulsive \rtext{state} excitation to be discussed below.

\rtext{Similarly to} the attractive \rtext{ground state}, also the bottom-of-the-band feature can be dressed by particle-hole excitations. This effect leads to the characteristic enhancement of response on both sides of the bottom-of-the-band excitation (see  shaded area in the upper panel of Fig.~\ref{fig.illustration}(b)). This collection of states defines the `bottom-of-the-band excitation branch' and it dominates the spectrum in the \rtext{mixed interaction} regime, see Fig.~\ref{fig.illustration}(b). 
\rtext{In a contact interaction model ($k_Fr^*=0$), where the mixed interaction regime is absent, the bottom-of-the-band branch is still existent, but it does neither dominate the spectrum in weight, nor -- as discussed in the following section -- does it represent the so-called leading branch in the real-time dynamics.}

\rtext{{\textit{Molecular excitation branch.}}} When the attractive interaction becomes sufficiently strong, a single-particle bound state appears at zero energy as the Feshbach resonance at $1/k_Fa=0$ is crossed to the repulsive interaction regime \rtext{where $a>0$}. This bound state with binding energy $\epsilon_B=-\hbar^2/2ma^2$ close to unitarity is energetically separated from the scattering states. \rtext{By filling all particles into the low-lying states \textit{including the bound state} one constructs the dressed `molecular ground state' $\ket{\psi_\text{mol}}$ (see Fig.~\ref{fig.illustration}(c)).} \rtext{This state, which becomes the new ground state of the system after crossing the Feshbach resonance to $a>0$, is reflected in the sharp spectral onset at its negative eigenenergy $E_\text{mol}$. Additional particle-hole excitations lead to a continuum of states attached to it, together constituting the `molecular  excitation branch'}. \rtext{However,  the molecular excitation branch does not represent the dominant branch in the absorption response (see Fig.~\ref{fig:linresponse}(c)).} In fact, the bound state has a wave function which decays exponentially in space as $\sim e^{-r/a}$ and hence it has decreasing overlap with the low-energy scattering states as $a>0$ becomes smaller. In consequence, the weight of the  \rtext{molecular branch}  diminishes as one moves further away from the Feshbach resonance.

\rtext{{\textit{Repulsive excitation branch.}}} In contrast, a many-body state with a much larger many-body overlap $|\braket{\psi_\alpha}{\text{\text{FS}}}|^2$ can be constructed by leaving the bound state empty, and instead filling the lowest scattering states with all the atoms. This \rtext{excited `repulsive state'} $\ket{\psi_\text{rep}}$ is illustrated in the second column of the middle panel in Fig.~\ref{fig.illustration}(d) and leads to the spectral edge at positive energies observed on the `repulsive' side of the Feshbach resonance. 

\rtext{The existence of the repulsive state originates from the fact that the emergence of the bound state for $a>0$ is tied to a reassignment of  scattering states. While for $a<0$ the single-particle levels are shifted downwards in energy, for $a>0$ the scattering states are effectively shifted upwards in energy (see Fig. \ref{fig.illustration}(c-e)).} This reassignment of states becomes particular apparent when considering the limit $a\to 0^+$ far away from resonance. Here the positive energy shifts of the scattering states approach zero from above. \rtext{Indeed it is this upward shift of single-particle scattering states  that motivates the term `\textit{repulsive}' interaction regime although the microscopic interaction of course still remains attractive.}

The repulsive state can again be `dressed' by fluctuations around the Fermi surface \rtext{leading to the continuous spectrum directly attached to the repulsive state} (illustrated as red shaded area in Fig.~\ref{fig.illustration}(d)). In contrast to the bottom-of-the-band excitation, however, only dressing towards energies \rtext{above the edge of the repulsive state}  is efficient \rtext{due to the finite binding energy of the molecule}. The  \rtext{energy of the repulsive state} can again be calculated analytically by a summation over the single-particle energy shifts reflecting Fumi's theorem, here applied to an excited many-body state. The resulting energy $E_\text{rep}$ is shown as dashed-dotted line in spectrum in Fig.~\ref{fig.densityrfspec}. 

\rtext{While for an infinitely heavy impurity the repulsive  state has zero quasiparticle weight, for an impurity of finite mass the repulsive state becomes the repulsive Fermi polaron extensively discussed in literature} \cite{Cui2010,schmidt_excitation_2011,massignan_repulsive_2011,Schmidt2012b,knap2012,knap_dissipative_2013} and observed in cold atoms \cite{kohstall_metastability_2012,koschorreck_attractive_2012,Scazza2016} as well as two-dimensional semiconductors \cite{Sidler2016}. Its analog for impurities immersed in a Bose-Einstein condensate was recently observed as well \cite{jorgensen2016,hu2016} following intensive theoretical studies  \cite{RathSchmidt2013,Li2014,Yi2015,Schmidt2015ang,Levinsen2015,Christensen2015,DehZinner2015,Volosniev2015,Pena2015,shchadilova2016,SchmidtLem2016,Pena2016,Ashida2017,Sun2017}.

\rtext{{\textit{Molecule-hole excitation branch.}}} The bound state can be filled by any state from the Fermi sea which leads to the broad excitation band of width $\epsilon_F$ above the molecular ground state called the `molecule-hole continuum' \cite{schmidt_excitation_2011,massignan_repulsive_2011,kohstall_metastability_2012}.  The upper edge of that band at $E_\text{mol}+\epsilon_F$ is again dressed by particle-hole excitations. The collection of these states constitutes the `molecule-hole excitation branch' \rtext{and it represents the analog of the bottom-of-the-band branch for $a>0$.}  Note that at energies $\omega<0$ but above $E_\text{mol}+\epsilon_F$ we find that a spectral gap appears where the weight is exponentially suppressed, \rtext{a phenomenon that is reminiscent of the `dark continuum' found in diagrammatic Monte Carlo studies of mobile impurities by Goulko \etal \cite{goulko2016}.}

\paragraph*{\textbf{Comparison to experiments.---}}The reverse RF absorption spectrum of impurities has been measured in LiK mixtures~\cite{kohstall_metastability_2012, cetina_2016}. In these experiments the K impurity atoms are mobile. However, as shown in \cite{cetina_2016}, by appropriately identifying the scattering parameters in our model,  our theory can also be applied also in this case \rtext{to obtain an approximate solution} due to the large mass imbalance between K and Li atoms and the relatively high temperatures $T/T_F\gtrsim 0.1$.  Specifically,  our reduced mass differs by a factor $(40/46)$ from the experimental one \cite{kohstall_metastability_2012, cetina_decoherence_2015,cetina_2016}.  According to Eq.~\eqref{eq:parameters} this difference in the effective mass affects the coupling strength $g$ between the open-channel atoms and the closed-channel molecule. In order to achieve the same coupling $g$ as realized in experiments we have to choose a rescaled Feshbach range parameter $k_F r^*=k_Fr_\text{exp}^*(40/46)^2$ which is reduced with respect to the experimental value $k_Fr^*_\text{exp}$ (for a detailed discussion see~Ref.~\cite{cetina_2016}). 

In Fig.~\ref{fig.densityrfspec} we show the comparison of the FDA prediction of the reverse RF spectrum to the experimental data obtained in \cite{kohstall_metastability_2012} taken at low (b) and high (c) RF power for the experimentally realized range $k_Fr_\text{exp}^*=0.95$ and temperature $T/T_F=0.16$.  In all subfigures we also show as dashed and dot-dashed lines the \rtext{analytically calculated energies of the attractive and repulsive state obtained from Fumi's theorem as also discussed in Refs.~\cite{Comescot2007,massignan_repulsive_2011}} and App.~\ref{FumiApp}. Our finite temperature spectra are in excellent agreement with the experimental data not only in energy but also  spectral line shapes (see also Ref.~\cite{cetina_2016}) without any free parameters or artificial broadening of spectral lines. 

\rtext{The good agreement in spectral line shapes when applying our theory to the finite mass case, where it becomes an approximation, has two origins. First, although for the infinitely heavy impurity the ground state -- e.g. $\ket{\psi_\text{att}}$ with energy $E_\text{att}$ -- loses all its quasiparticle weight, this weight is predominately redistributed to a continuum of states that are energetically close to $E_\text{att}$. Hence the \textit{integrated weight} taken from a sufficiently large energy window around the attractive state can gain an integrated strength that is comparable to the finite mass case. Therefore, for experiments that are subject to broadening of spectral lines due to external factors (such as trap average, laser lines width, etc.),  observed line shapes and weights for the finite and infinite-mass case can appear quite similar and distinguishing both cases remains a challenge for frequency-domain measurements. Second, temperature  has the effect of broadening spectral lines. As we will show in the following Section \ref{sec_decoh}, for finite temperature one can assign a thermal weight to the dominant attractive and repulsive state even for impurities of infinite mass. This leads to an absorption profile of Lorentzian line shape that is  similar to the case of a mobile impurities and which further contributes to the good agreement between infinite-mass theory and experiment.}

\subsection{Standard RF spectra}

\begin{figure}[t!] 
  \centering 
  \includegraphics[width=\linewidth]{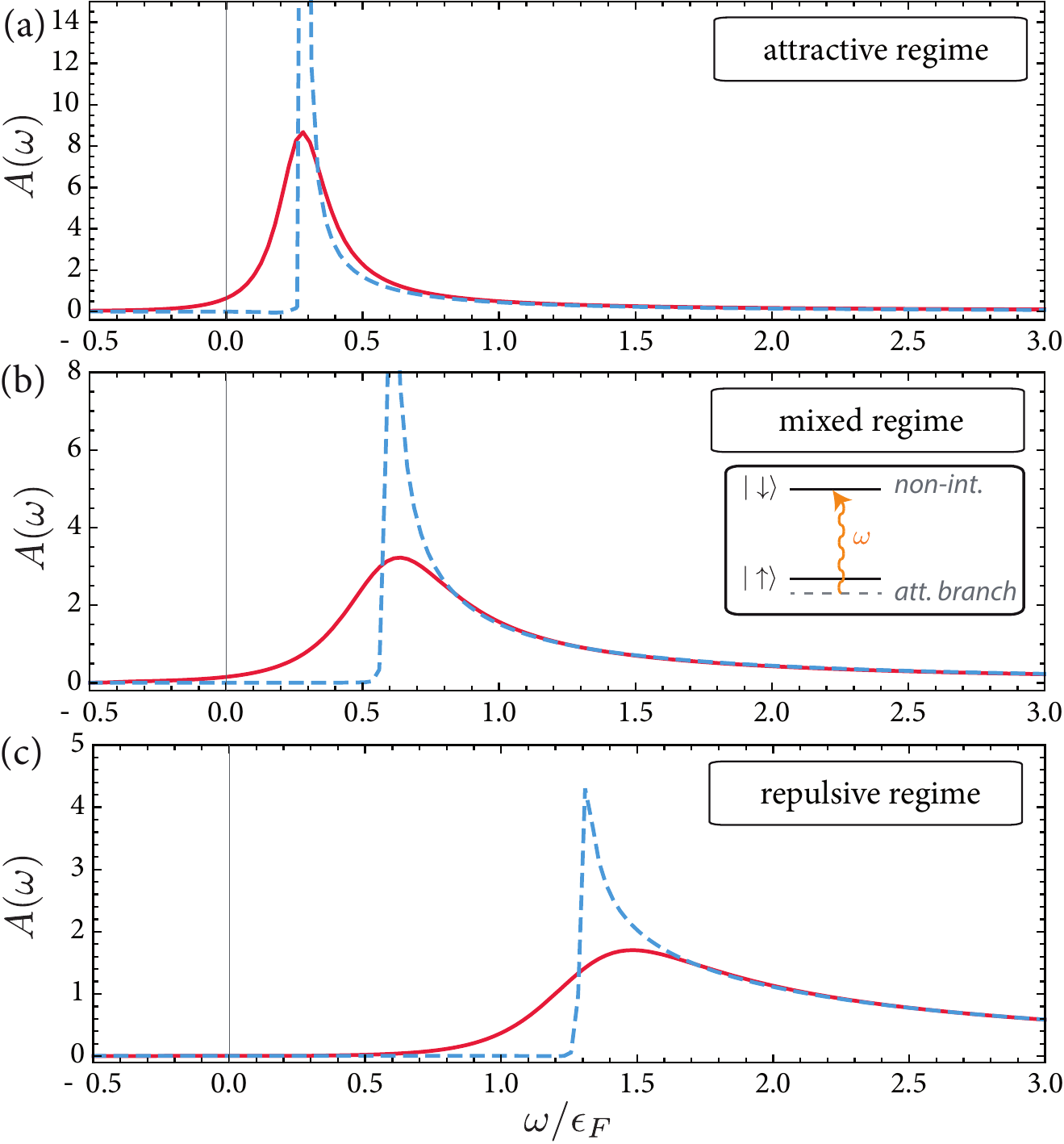}
  \caption{ \textbf{Radio-frequency absorption spectra in the standard RF scheme.} In the standard scheme, the \rtext{system is prepared in its many-body ground state where the impurity is initially interacting with the Fermi gas.} It is then driven to a state non-interacting with the gas (see inset in (b) for an illustration). The spectra are shown for the same parameters as in Fig.~\ref{fig:linresponse}, i.e.~$T/T_F=(0.01,\ 0.16)$ (dashed, solid), interaction strengths $1/k_Fa=(-0.91,\ -0.1,\ 1.0)$, and Feshbach range parameter $k_Fr^*=0.71$.}
  \label{fig:linresponsestandard}
\end{figure}

The attractive Fermi polaron has been observed first in a two-component $^6$Li Fermi gas using the `standard' RF scheme~\cite{schirotzek_observation_2009}, and its quasi-particle properties, such as its energy and residue were measured. To illustrate  the impact  of the different RF protocols we show in Fig.~\ref{fig:linresponsestandard} our prediction for the absorption spectrum as obtained in the standard RF scheme for the same parameters as in Fig.~\ref{fig:linresponse}.
As discussed in Section \ref{response}, in the standard RF scheme the system is initially prepared in the interacting \rtext{many-body} ground state and then driven to a non-interacting final state.  

The spectra are distinctly different from the reverse RF scheme. Since in the standard RF scheme the ground state, e.g. $\ket{\psi_\text{att}}$ for $a<0$, is prepared initially, one has always to pay its energy, e.g. $E_\text{att}$, to `break' this state. \rtext{This leads to a shift to positive energies in the absorption spectrum, which increases as the interactions are tuned across the Feshbach resonance. This shift} has been observed by Schirotzek \etal \cite{schirotzek_observation_2009}. In particular, as $k_F a>0$ and the bound state appears in the spectrum, the `break-up energy' involves the binding energy of the molecule. Note that the repulsive state \rtext{$\ket{\psi_\text{rep}}$, that corresponds to the repulsive Fermi polaron for finite-mass impurities,} cannot be revealed using the standard RF scheme, unless it is prepared as an excited, non-equilibrium initial state of the system \cite{koschorreck_attractive_2012}.

\section{Universal many-body response: Dephasing dynamics}\label{sec_decoh}

In typical condensed matter experiments, real-time observables such as the many-body overlap $S(t)$  in Eq.~\eqref{eq:ramsey} are challenging to measure due to the large size of the Fermi energy $\epsilon_F$. For instance, in typical solid state materials the corresponding  Fermi time $\tau_F=h/\epsilon_F$ is on the order of attoseconds \cite{Kittel1966}. 
In contrast, in ultracold atomic gases the Fermi energy is lower by many orders of magnitude due to the diluteness of ultracold atomic gases and  large atomic masses, which leads to  Fermi times on the order of $\mu$s to ms. The combination with interferometric techniques available in atomic experiments  makes it then possible to study many-body correlation functions in fermionic systems in real time and to observe striking far-from-equilibrium many-body dynamics.

\subsection{Exact numerical solution}

\begin{figure}[t] 
  \includegraphics[width=\linewidth]{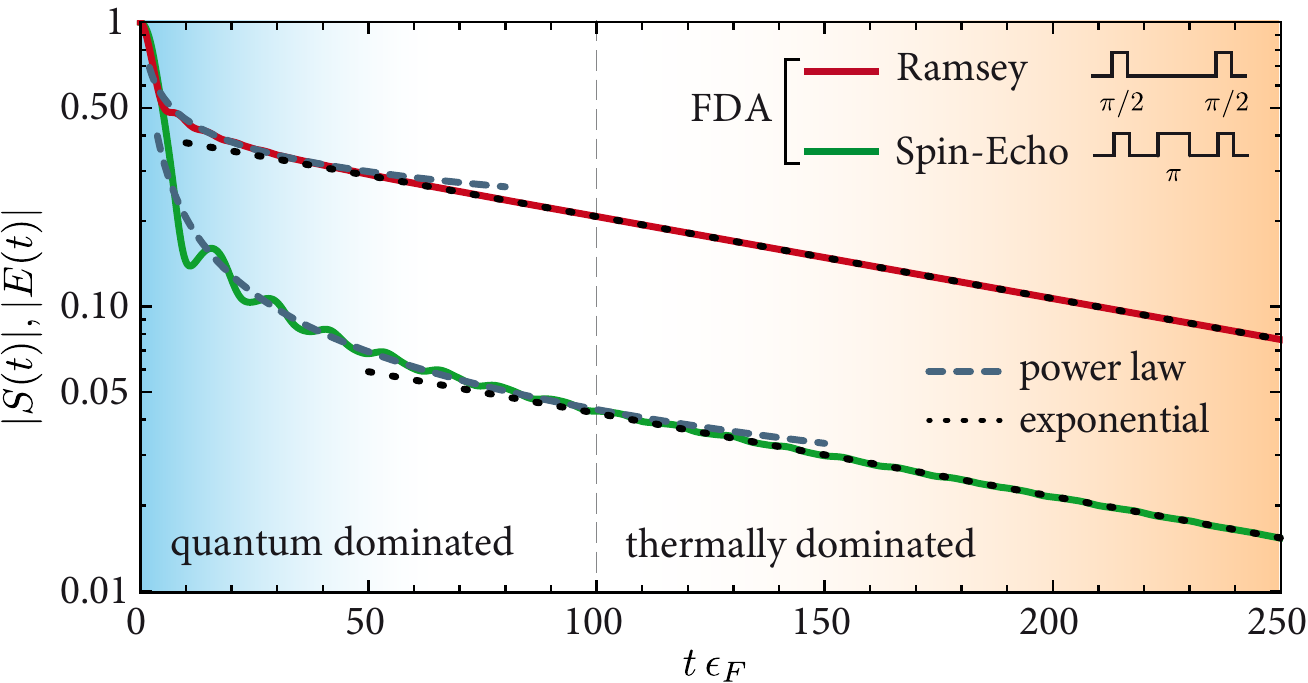}
  \caption{ \textbf{Real-time evolution of Ramsey and spin-echo response.} At finite temperature, the real-time Ramsey and spin-echo response of the impurity atom unveil a quantum-to-classical crossover from short to long times. Here, the effective range, temperature, and scattering length are given by $k_Fr^*=0.8$,  $T=0.01 T_F$, and  $k_F a=-1.1$, respectively. The vertical dashed line indicates the thermal time scale $\tau_\text{th}= \hbar/k_BT$. While for $t<\tau_\text{th}$ the spin-echo signal exhibits a three times faster power-law decay than the Ramsey signal, both signals are governed the same exponential decay rate in the thermal regime for $t>\tau_\text{th}$. \rtext{The dashed and dotted lines on top of the FDA data show the analytical result for the OC-characteristic power-law decay at early times and thermal exponential decay of coherence at long times, respectively.} }
  \label{fig.timeevolution}
\end{figure}

The time-dependent impurity problem represents an instance where  intriguing  dynamics can be observed in an exactly solvable many-body system. In Fig.~\ref{fig.timeevolution} we show the Ramsey signal (solid, red line) and the spin-echo response (solid, green line) for a system in the attractive interaction regime characterized by $k_Fr^*=0.8$, $T=0.01\, T_F$ and interaction strength $k_F a=-1.1$, as calculated by the FDA. \rtext{Again, we assume infinitely fast $\pi/2$ spin rotations both in the Ramsey and spin-echo sequence.} 

At times that are short compared to the inverse temperature, quantum mechanics governs the evolution: due to the sudden switch-on of interactions at $t=0$ the many-body wave function dephases, leading to a power-law decay of the Ramsey response with the exponent $(\delta_F/\pi)^2$, being the universal real-time signature of the Anderson orthogonality catastrophe~\cite{anderson_infrared_1967,nozieres_singularities_1969}.

\begin{figure*}[t] 
  \centering 
  \includegraphics[width=.95\textwidth]{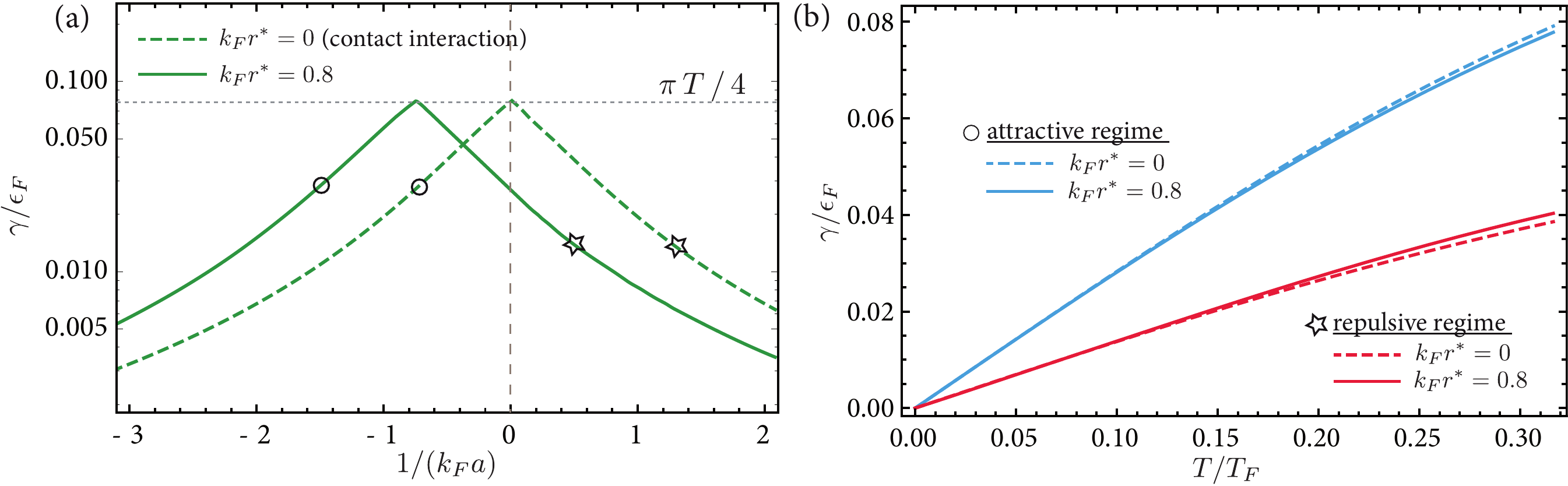}
  \caption{ \textbf{Finite temperature decoherence rate of Ramsey and spin-echo signal.} \fc{a} The long-time decoherence rate $\gamma$, as extracted from a fit to the long-time exponential decay of the Ramsey and spin-echo signal computed by FDA, is shown as a function of the inverse scattering length for $T/T_F=0.1$. The decoherence rate $\gamma$ of the two-channel model with $k_Fr^*=0.8$ (solid line) is compared with the results from a contact interaction model (dashed line). The decay rate of the Ramsey and spin-echo protocol are identical. \fc{b} Decoherence rate $\gamma$ as a function of temperature evaluated at the interactions specified as circle and stars in (a). \rtext{In (b)} the interaction parameters are chosen such that in the attractive and repulsive regime both zero and finite range give the same s-wave scattering phase shift $\delta_F$ at the Fermi energy. Specifically, in the attractive regime ($\delta_F=0.95$): $1/k_Fa=-0.72,\, k_Fr^*=0$ (dashed blue), and $1/k_Fa=-1.5,\, k_Fr^*=0.8$ (solid blue); repulsive regime ($\delta_F=-0.66$): $1/k_Fa=1.28,\, k_Fr^*=0$ (dashed red), and $1/k_Fa=0.5,\, k_Fr^*=0.8$ (solid red). 
  }
  \label{fig:deckfa}
\end{figure*}

When  time becomes comparable to the inverse temperature, $t\gtrsim \tau_\text{th}\sim \hbar/k_BT$, thermal fluctuations disrupt the coherent and time-reversal symmetric quantum many-body dynamics, and a crossover from quantum to predominantly thermal dynamics takes place, see Fig.~\ref{fig.timeevolution}.  The precise time scale for this crossover depends on the microscopic details, such as the scattering length and the effective range. However, this time scale depends also on the chosen observable: for instance, in the Ramsey signal a crossover to exponential dephasing sets in earlier as compared to the spin-echo signal.

One key signature of the asymptotic finite-temperature behavior is an exponential decay of coherence,  $|S(t)|\sim e^{-\gamma t}$. Remarkably, while in the quantum regime at early times\rtext{, before thermal decoherence sets in,} the dephasing rate is sensitive to the specific spin rotation protocol chosen (e.g., the spin-echo response decays faster than the Ramsey signal), using the theory of Toeplitz determinants \cite{martin_wave-packet_1992, levitov_electron_1996}, we  prove (see Appendix \ref{app:B}) that at finite temperature the long-time decoherence rate $\gamma$ is  identical for Ramsey and spin-echo interferometry --- even though in the spin-echo sequence the impurity spin is flipped in the middle of the time evolution. This fact is of particular relevance for recent experiments which inferred the long-time decay of the impurity Green's function from the spin-echo  and not the Ramsey signal \cite{cetina_decoherence_2015} (see also the discussion in Section~\ref{SectExpCom}). 

In \figc{fig:deckfa}{a}, we show the interaction dependence of the finite-temperature decoherence rates $\gamma$ at  $T/T_F=0.1$ which are identical for both Ramsey and spin-echo interferometry. The rates are extracted from a fit to the very long-time decay of the numerically evaluated $S(t)$ and $S_{\text{SE}}(t)$ response, such as shown in Fig.~\ref{fig.timeevolution}. We compare the decoherence rate for finite $k_F r^*=0.8$ and for zero range $k_F r^*=0$ corresponding to a model with  contact interactions.

As expected, for zero-range interactions, the maximum decoherence rate arises at $k_F a\to\infty$. The inclusion of a finite range has the  consequence of shifting this maximum of the decoherence rate away from the center of the Feshbach resonance. This shift can be understood from an analysis of the most relevant scattering processes which, due to Pauli blockade, take place  close to the Fermi surface.  The impurity-particle scattering rate is determined by the scattering amplitude $f(k)$ and hence the scattering phase shift $\delta_F$, cf.~Eq.~\eqref{scattampl},  evaluated at the Fermi energy $E=\epsilon_F$.

For finite range $r^*>0$, the magnitude of the scattering amplitude becomes maximal not at $k_Fa=\infty$ but when the phase shift $\delta_F$ equals $\pi/2$. The corresponding critical scattering length $a_\text{cr}$ marks the transition from the attractive to the \rtext{mixed} interaction regime.  The shift in the decoherence rate can be seen in Fig.~\ref{fig:deckfa}(a) where the maximum decoherence is reached at $1/k_F a_\text{cr}=-k_F r^*=-0.8$. The finite effective range in \eqw{scattampl} hence leads to the, at first sight, counterintuitive effect that, in the \rtext{mixed} regime, decoherence can slow down as the magnitude of the scattering length $a$ increases. Remarkably this also implies that the  position of the decoherence maximum can  be used to determine the few-body effective range from  thermal many-body observables.

In Fig.~\ref{fig:deckfa}(b), we show the temperature dependence of the exponential decay rate  obtained from  exact calculations of $S(t)$. In this figure, we compare the thermal decay rate $\gamma$ on both sides of the Feshbach resonance  for zero and finite-range interactions. To highlight the universality of our results,  we choose interaction parameters so that for both scenarios the phase shifts $\delta_F$ are equal. We find that only at temperatures $T/T_F\gtrsim0.15$ deviations become visible. This  signifies the highly universal character of the physics which is dominated by the low-energy scattering phase shift alone and does not depend on the microscopic details of interactions. 
For low temperatures, we prove below that the maximal thermal dephasing rate for heavy impurities interacting with a Fermi gas with short-range interactions is \rtext{in the long-time limit} universally given by
\begin{equation}\label{maxdec}
\gamma_\text{max} = \frac{\pi}{4}T.
\end{equation}
This value is saturated at the transition from the attractive to the \rtext{mixed} interaction regime.

As the temperature goes to zero, the thermal decoherence rate $\gamma$ vanishes. In this limit, the asymptotic exponential form of the decay breaks down and logarithmic corrections start to dominate. Those lead to the  power-law decay of $S(t)$ as the key signature of the Anderson orthogonality catastrophe with an exponent given by $(\delta_F/\pi)^2$. In contrast to the long-time exponential decay at finite temperature, at zero temperature the power law exponent \emph{does depend} on the trajectory of the impurity spin on the Bloch sphere which are different for Ramsey or spin-echo interferometry. In particular, quantum interference effects enhance the exponent at each spin flip in an interferometric sequence~\cite{knap2012}. Using the theoretical techniques of Ref.~\cite{dambrumenil_fermi_2005,knap2012}, we prove that the spin-echo exponent is enhanced by a factor three; details are given in App.~\ref{app:0T}.

\rtext{We note that for very narrow Feshbach resonances with $k_Fr^*\ll1$ the deviations from the  contact interaction model, visible in Fig.~\ref{fig:deckfa}(b), set in at  lower temperatures. This is due to the fact that for such resonances the phase shift $\delta(E)$ close to the Fermi surface becomes strongly energy dependent. The thermal softening of the Fermi distribution function allows fermions to probe this energy dependence which leads to the deviations of dephasing rate from the contact interaction model.}

\subsection{Analytical results}\label{analyticalsection}

While the previous subsection focused on an exact \textit{numerical} evaluation of the Ramsey signal at all times, we will now derive \textit{analytical} expressions at intermediate and long times. Analytical results for the short-time dynamics will be discussed in Sec.~\ref{sec:shorttime}. 

In this section, we are interested  in the many-body regime of  quantum dynamics where $t\epsilon_F\gg 1$ and study the crossover from the `low-T' regime (with $tT \ll 1$), where $S(t)$ exhibits power-law dephasing \cite{anderson_infrared_1967,mahan_excitons_1967,nozieres_singularities_1969,nazarov_quantum_2009},  to the `high-T' regime ($tT\gg 1$) showing exponential decay (Fig.~\ref{fig.timeevolution}) \cite{yuval_exact_1970}. While  analytic understanding of the Ramsey response has been developed in Ref.~\cite{knap2012} in the low-T regime for zero-range interactions, below we focus on the high-T regime.

\paragraph*{\textbf{Excitation branches.---}} 

\rtext{Similarly to} the absorption spectrum $A(\omega)$ discussed in Section~\ref{rfsection},   the time-dependent overlap $S(t)$ can be decomposed into the sum
\begin{equation}\label{SEigenstateDecomp}
S(t)= \sum_{\{\ket{\psi_\alpha}\} } |\braket{\psi_\alpha}{\text{FS}}|^2 e^{i E_{FS}t}e^{-i E_\alpha t},
\end{equation}
where $\{\ket{ \psi_\alpha} \}$ denotes the complete set of many-body eigenstates of the `interacting' Hamiltonian $\hat H$ with eigenenergies $E_\alpha$, and $\ket{\text{FS}}$ the non-interacting Fermi sea. 

Being \rtext{directly related by} Fourier transforms, $A(\omega)$ and $S(t)$ carry  the same physical information  (for details see App.~\ref{SAAppendix}). 
Depending on the interaction regime, different `excitation branches' $\{\ket{\psi_\alpha}\}$ dominate the dynamics of $S(t)$. 
In Section~\ref{rfsection} we identified five such branches (summarized in Fig.~\ref{fig.illustration}), and to make the discussion in this section self-contained we repeat here our main findings.

In the attractive and \rtext{mixed} regimes (a) and (b), the absorption spectra are dominated by the attractive  state together with particle-hole (p/h) excitations around the Fermi surface (FS); this collection of states constitutes the `attractive excitation branch' (branch A). Another relevant class of excitations in regimes (a) and (b) is the bottom-of-the-band contribution (FB, for `Fermi bottom'). Together with  p/h excitations it forms the `bottom-of-the-band branch' (branch B), see   Fig.~\ref{fig.illustration}(b). We find that these two branches describe accurately the long-time many-body dynamics and $S(t)$ can be approximated as
\begin{equation}\label{SRegimeAB}
S(t)\approx S_A(t) + S_B(t)\, ,
\end{equation}
where $S_A(t)$ accounts for the attractive branch A, and $S_B(t)$ accounts for the bottom-of-the-band branch B.

The two branch contributions can be written as
\begin{equation}\label{SAB}
S_A(t)=C_A\; S_0^{\rm (FS)}(t)\,, \quad
S_B(t)=C_B\; S_1^{\rm (FS)}(t)\; S_{-1}^{\rm (FB)}(t)\, .
\end{equation}
Here $S_n^{\rm (FS)}(t)$ and $S_n^{\rm (FB)}(t)$ represent excitations around the Fermi level and  the creation of a hole at the bottom of the Fermi sea, respectively. The subscript $n$ specifies the number of particles added or removed from the Fermi surface or bottom of the Fermi sea. Note that the total particle number is conserved in each of the branches. For instance, in $S_B(t)$ one particle is removed from the bottom of the band and inserted close to the Fermi surface. 

In the `repulsive regime' (c), where $\delta_F<0$ and a bound state (BS) is present, three relevant branches can be identified and  $S(t)$ approximated as 
\begin{equation}\label{SRegimeC}
S(t)\approx S_{A1}(t) + S_{B1}(t) + S_{B2}(t)\, .
\end{equation}
The three branches, denoted as `molecular excitation branch' A1, the `repulsive excitation branch' B1, and the `molecule-hole branch' B2 correspond to the most relevant spectral features discussed in Section \ref{rfsection}. As illustrated in Fig.~\ref{fig.illustration}(c), $S_{A1}(t)$ represents the attractive ground state `dressed' by  p/h excitations. In the spectral function those states correspond to the absorption edge at negative frequencies, while $S_{B1}(t)$ accounts for p/h excitations close to the Fermi surface on top of the repulsive state, cf.~Fig.~\ref{fig.illustration}(d). Finally, $S_{B2}(t)$ describes the edge of the molecule-hole continuum and represents the family of states where the bound state is filled by an atom from the bottom of the Fermi sea again dressed by p/h fluctuations at the Fermi surface, cf.~Fig.~\ref{fig.illustration}(e).
 Accordingly, the branch contributions can be expressed in terms of the processes (FS), (BS), and (FB),
\begin{eqnarray}\label{Region2}
S_{A1}(t) &=& C_{A1}\; S_{-1}^{\rm (FS)}(t)\; S_{1}^{\rm (BS)}(t)\, , \nonumber\\
S_{B1}(t) &=& C_{B1}\; S_0^{\rm (FS)}(t)\, ,\nonumber \\
S_{B2}(t) &=& C_{B2}\; S_0^{\rm (FS)}(t)\; S_{-1}^{\rm (FB)}(t)\; S_{1}^{\rm (BS)}(t)\,  . 
\end{eqnarray}

While the contributions $S_\alpha(t)$ can be analytically evaluated as we will now demonstrate, the coefficients $C_\alpha$ in Eqs.~(\ref{SAB},\ref{Region2}) depend on the microscopic details, and have to be determined  from a numerical evaluation of Eq.~\eqref{eq:klich}.

\paragraph*{\textbf{Temperature independent contributions.}---} 

First, we turn to the contributions from the bottom of the Fermi sea $S_{-1}^{\rm (FB)}(t)$ and the bound state  $S_{1}^{\rm (BS)}(t)$. Both involve states deep under the Fermi sea and hence yield temperature-independent contributions to the dynamics. The bound-state contribution is simply given by the phase accumulation of the form
\begin{equation}\label{eq.SBS}
S_{1}^{\rm (BS)}(t) \propto e^{-i\epsilon_{ B}t}\, ,
\end{equation}
where  $\epsilon_{B}<0$ is the energy of the bound state. 

For the bottom-of-the-band feature an analysis of the relevant many-body states yields (for details see Appendix \ref{App.FB})
\begin{equation}\label{eq.SBFS}
S_{-1}^{\rm (FB)}(t) \propto\int_0^\infty \frac{dE}{\sqrt{\epsilon_F E}} \sin^2\delta(E)\; e^{iEt}\, .
\end{equation}
Eq.~\eqref{eq.SBFS} has a power-law decay with time $t^{-1/2}$ at $1/\epsilon_F \ll t \ll 1/\epsilon_B^*$ and $t^{-3/2}$ at $1 \ll t \epsilon_B^*$, where we define the energy scale $\epsilon_B^*=1/(2m |a|^2)$ for both positive and negative $a$. A similar result for the bottom-of-the-band and bound-state contributions was reported in \cite{knap2012}.

\paragraph*{\textbf{Fermi-surface contributions.}---} 

In contrast to the (BS) and (FB) processes, p/h excitations around the Fermi surface (FS) involve arbitrarily low energies. Hence these processes are influenced by  finite temperature and they become responsible for the exponential dephasing of the Ramsey signal.  

The behavior of the contributions $S_n^{\rm (FS)}(t)$  can be understood analytically using two approaches. First, bosonization, valid at low temperatures, allows us to describe a crossover from  short-time power-law decay to  long-time exponential decoherence and allows us to reveal corrections to the temperature dependence of the decoherence rate. Second,  the theory of Toeplitz determinants provides analytical expressions for the decoherence rate of the various excitation branches at relatively high temperatures.

\subsubsection{Bosonization}\label{bosonization}

\begin{table*}[t]
\centering
\begin{tabular}{|c|c|c|c|c|c|c|}
\hline
 interaction regime & parameters & branch & leading & contribution to $S(t)$ & low-$T$ decay rate $\gamma$& high-$T$ correction \\
\hline
(a) `attractive' & $a<0$, $\delta(E)>0$, & A & yes & $S_0^{\rm (FS)}(t)$ & $\gamma_0=T\delta_F^2/\pi$ & 
    Eq.~(\ref{eq:integralomegagamma}) \\
\cline{3-7}
    & $\delta_F<\pi/2$     & B & no & $S_1^{\rm (FS)}(t)\; S_{-1}^{\rm (FB)}(t)$ & $\gamma_{1}=T(\delta_F-\pi)^2/\pi$ & 
    Eq.~(\ref{eq:sublead}) \\
\hline
(b)  `\rtext{mixed}' &  $a<0$, $\delta(E)>0$, & A & no & $S_0^{\rm (FS)}(t)$ & $\gamma_{0}=T\delta_F^2/\pi$ & 
    Eq.~(\ref{eq:sublead}) \\
\cline{3-7}
    & $\delta_F>\pi/2$     & B & yes & $S_1^{\rm (FS)}(t)\; S_{-1}^{\rm (FB)}(t)$ & $\gamma_{1}=T(\delta_F-\pi)^2/\pi$ & 
    Eq.~(\ref{eq:integralomegagamma})\\
\hline
(c)  `repulsive' &  $a>0$, $\delta(E)<0$ & A1 & no & $S_{-1}^{\rm (FS)}(t)\; S_{1}^{\rm (BS)}(t)$ & $\gamma_{-1}=T(\delta_F+\pi)^2/\pi$ & 
    Eq.~(\ref{eq:sublead}) \\
\cline{3-7}
    &  & B1 & yes & $S_0^{\rm (FS)}(t)$ & $\gamma_{0}=T\delta_F^2/\pi$ &
    Eq.~(\ref{eq:integralomegagamma}) \\
\cline{3-7}
    &  & B2 & yes & $S_0^{\rm (FS)}(t)\; S_{-1}^{\rm (FB)}(t)\; S_{1}^{\rm (BS)}(t)$ &   $\gamma_{0}=T\delta_F^2/\pi$ &
    Eq.~(\ref{eq:integralomegagamma}) \\
\hline
\end{tabular}
\caption{\textbf{Summary of the interaction regimes and branches.} The various branches contributing to the dephasing of $S(t)$. The leading branches, leading dephasing rates, as well as references to the high temperature corrections are indicated for the different interaction regimes (a), (b), and (c).}
\label{tab.branches}
\end{table*}

In the bosonization approach (for details see Appendix \ref{App.FB} and \cite{giamarchi_quantum_2004,gogolin_bosonization_2004}),  the energy dependence of the phase shift $\delta(E)$ is neglected and the dispersion relation is linearized around the Fermi surface. Hence, the approach is only applicable at low temperatures where the Fermi surface is sharply defined. Following standard bosonization techniques one can extract the power-law decay of coherence at $T=0$,
\begin{equation} \label{Bos2}
S_0^{\rm (FS)}(t)\propto e^{-i\Delta E t}\; t^{-(\delta_F/\pi)^2}\, ,
\end{equation}
which represents the well-known result for the Fermi-edge singularity \cite{nozieres_singularities_1969}. Here the energy $\Delta E$ determines the spectral onset of the attractive and repulsive polaron feature, which is given by Fumi's theorem \cite{mahan_many_2000,giamarchi_quantum_2004},
\begin{equation}\label{Fumi}
\Delta E = -\int_0^{\epsilon_F} \frac{dE}{\pi} \delta(E)\, ,
\end{equation}
as a sum over all phase shifts up to the Fermi energy (see App.~\ref{FumiApp}).

At finite temperature, a conformal mapping of complex time onto a cylinder with the periodicity $i/T$  leads to (see  Appendix \ref{App.FB} and Refs.~\cite{gogolin_bosonization_2004,Braunecker2006,Gutman2011}) 
\begin{equation}
S_0^{\rm (FS)}(t)\propto  e^{-i\Delta E t} \left(\frac{\pi T}{\sinh \pi T t}\right)^{(\delta_F/\pi)^2}\, ,
\end{equation}
which generalizes Eq.~\eqref{Bos2} to finite temperature.

Now we turn to the  Fermi-surface branch contributions $S^{(\text{\text{FS}})}_n(t)$ for $n\neq 0 $. These contributions  appear when atoms are transferred from the bottom of the Fermi sea to the Fermi surface ($n=1$) or from the Fermi surface to the bound state ($n=-1$). As derived in the theory of the Fermi-edge singularity \cite{combescot_1971}, and as also discussed in the context of full counting statistics (see, e.g., \cite{Ivanov2013,Ivanov2016} and references therein), the contributions $S_n(t)$ with $n\neq 0 $ are accounted for by a shift $\delta_F \to \delta_F \pm \pi$. Furthermore, as a particle is removed or added to the Fermi surface, the energy $\Delta E$ is modified as well, and $\Delta E\to \Delta E \pm \epsilon_F$. This leads to the general expression for the various Fermi-surface contributions,
\begin{equation}\label{FullBosonization}
S_n^{\rm (FS)}(t) \propto e^{-i(\Delta E + n \epsilon_F) t} 
\left(\frac{\pi T}{\sinh \pi T t}\right)^{\left(\frac{\delta_F}{\pi}-n\right)^2}\, .
\end{equation}
This equation describes the full crossover from the low-temperature regime with power-law behavior
\begin{equation}
S_n^{\rm (FS)}(t) \propto e^{-i(\Delta E + n \epsilon_F) t}\; t^{-\left(\frac{\delta_F}{\pi}-n\right)^2}
\end{equation}
to the finite temperature regime where, at sufficiently long times $t T\gg1$, the Fermi-surface contributions decay exponentially according to
\begin{equation}\label{TScaling}
S_n^{\rm (FS)}(t) \propto T^{\left(\frac{\delta_F}{\pi}-n\right)^2} e^{-\gamma_n t} e^{-i \omega_n t}\, .
\end{equation}
Here, we introduced the exponential decay rates and frequencies
\begin{eqnarray}\label{decayrates}
\gamma_n &=&T\frac{(\delta_F-n\pi)^2}{\pi},\nonumber\\
\omega_n&=&\Delta E + n \epsilon_F.
\end{eqnarray}
Remarkably, Eq.~\eqref{TScaling} predicts a \textit{prefactor} of the exponential decay which features a power-law dependence on temperature with an exponent governed by the OC $(\delta_F/\pi)^2$.

According to Eqs.~\eqref{SRegimeAB} and \eqref{SRegimeC}, $S(t)$ is given by a sum over  various branch contributions.  Out of those, the branch exhibiting the smallest decay coefficient $\gamma_n$, as determined by Eq.~\eqref{decayrates}, will survive at long evolution times. This branch, which we call the `leading' branch, determines the temperature scaling of the prefactor $\propto T^{\left(\frac{\delta_F}{\pi}-n\right)^2}$ in Eq.~\eqref{TScaling}. 

In Table \ref{tab.branches} we summarize our findings. In the column `low-T decay rate' we list the decay rates $\gamma_n$ which are applicable in the limit $T/\epsilon_F\ll1$ (provided $t T\gg 1$ and $t \epsilon_F\gg1$). Furthermore we show the definitions of the interaction regimes as well as the various contributions to the dephasing dynamics of the Ramsey signal $S(t)$.

\begin{figure}[b]
\includegraphics[width=\linewidth]{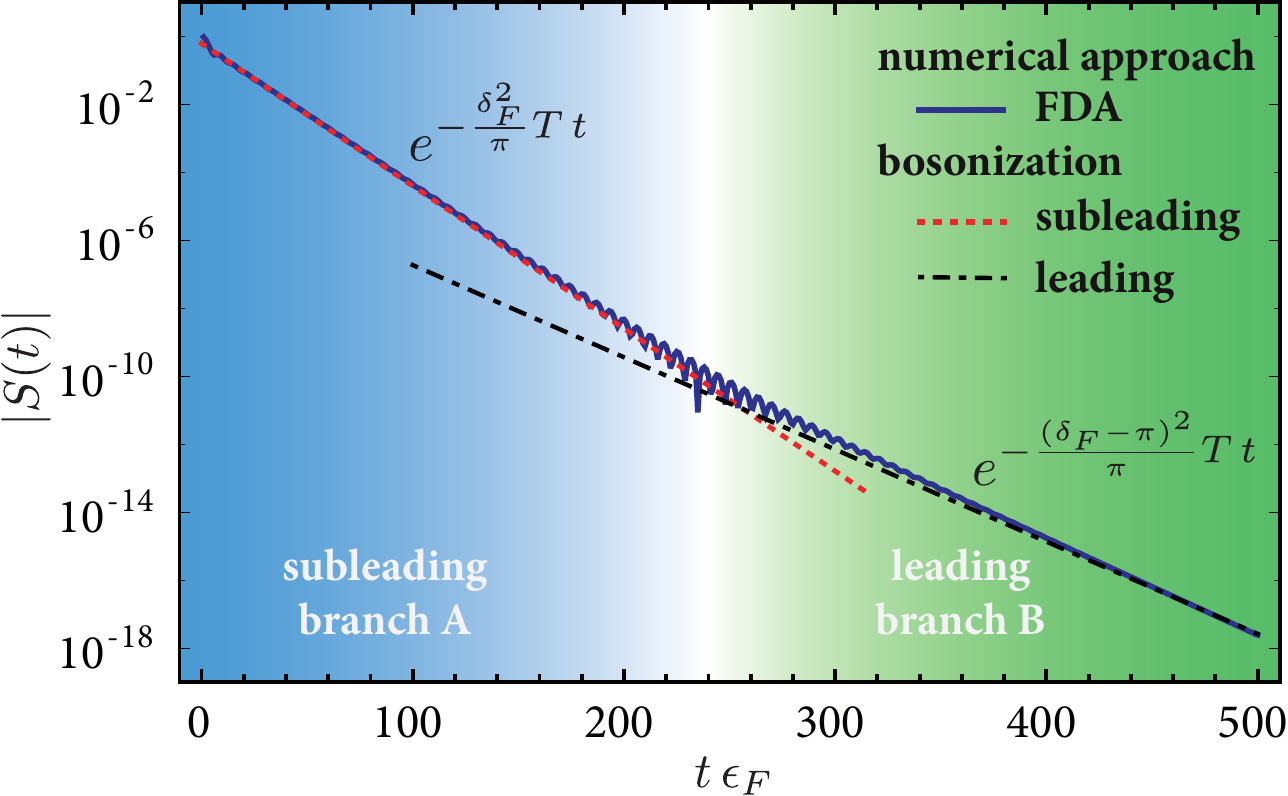}
\caption{\textbf{Crossover from subleading to leading branch dynamics.} Absolute value of the Ramsey contrast $S(t)$ as function of time. Interaction parameters are chosen to correspond to the \rtext{mixed} regime (b) with $T/T_F=0.1$, $1/k_Fa=-0.61 $, and $k_Fr^*=0.8$. The solid line shows the exact numerical evaluation of the dynamical overlap $S(t)$ using the FDA, while the dotted (dashed) line shows exponentials with exponents $\gamma_1$ ($\gamma_0$), respectively, which correspond to the leading (subleading) branch dynamics in this interaction regime. For the detailed dependence of the decoherence rates on $1/(k_Fa)$, see Fig.~\ref{fig.newdeccomp}.}
\label{fig.crossoversubleading}
\end{figure}

 \begin{figure*}[t]
\includegraphics[width=\linewidth]{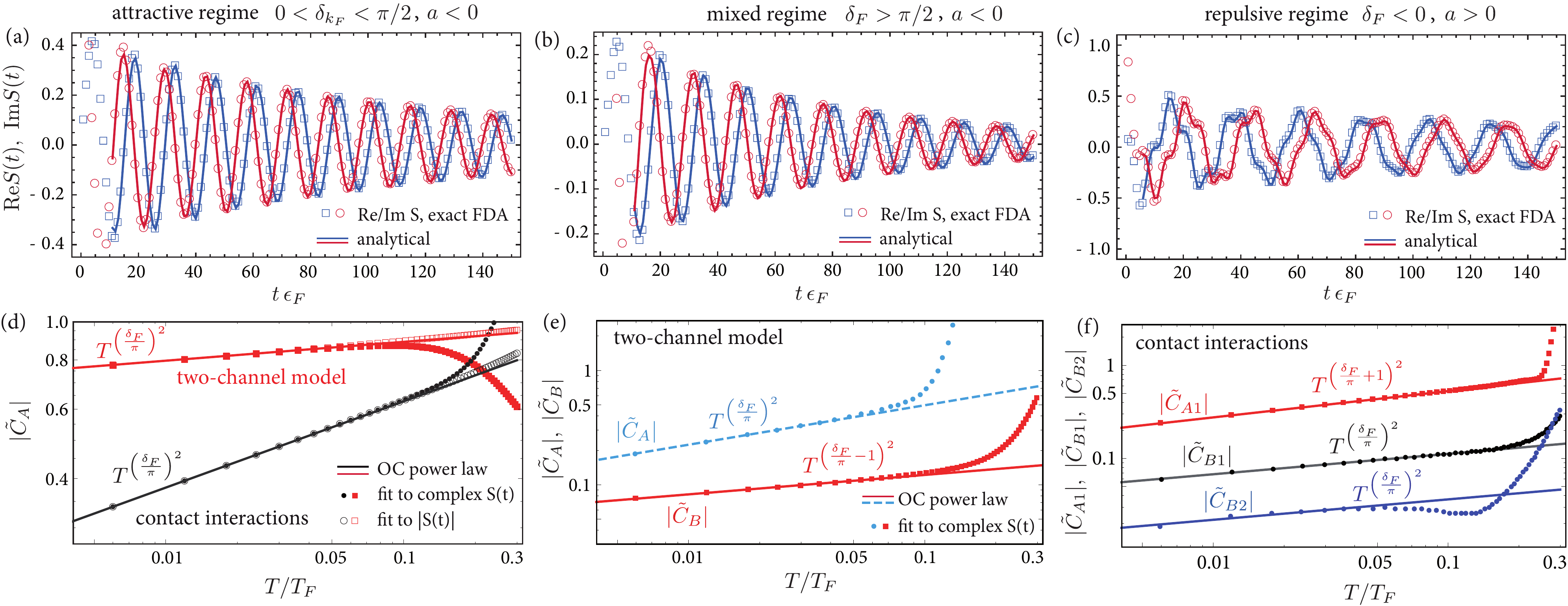}
\caption{\textbf{Power-low temperature dependence of dephasing dynamics.} The upper panels (a-c) show the real and imaginary  part of the Ramsey signal as calculated exactly using the FDA (symbols). The data, shown for the exemplary temperature $T/T_F=0.012$, is fitted by the analytical expressions~\eqref{SRegimeAB} and \eqref{SRegimeC} (solid lines). The results are shown for parameters (a) $1/k_F a= -0.1$, $k_Fr^*=0$, (b) $1/k_F a= -0.5$, $k_Fr^*=0.8$ (c) $1/k_F a= 2$,  $k_Fr^*=0$. In (d-f) the temperature dependence of the rescaled coefficients $\tilde C_\alpha(T)$ is shown where $\alpha=(\text{A,B,A1,B1,B2})$ specifies the excitation branch. The parameters are: (d) $1/k_F a= -0.1$, $k_Fr^*=0$ (black) and $1/k_F a= -2$, $k_Fr^*=0.8$ (red), (e) $1/k_F a= -0.5$, $k_Fr^*=0.8$,  (f)  $1/k_F a= 0.1$, $k_Fr^*=0$.  The open symbols in (d) correspond to an alternative fitting procedure, where the absolute value of $|S(t)|$ is fit by a single exponential decay function. }
\label{fig.Fitting1}
\end{figure*}

For instance, according to Eq.~\eqref{SAB}, in the attractive regime (a) both $S_0^{\rm (FS)}(t)$ and $S_1^{\rm (FS)}(t)$ contribute.  Since in this regime $\delta_F<\pi/2$, we find that $\gamma_0<\gamma_1$. Therefore $S_A(t)$, which represents the \rtext{attractive excitation branch}, is  leading. In contrast, in the \rtext{mixed} regime (b) $\gamma_1<\gamma_0$ and hence the bottom-of-the-band contribution $S_B(t)$ is the leading branch. Finally,  in the repulsive regime (c) both $S_{B1}(t)$ and $S_{B2}(t)$ are leading branches. They represent the dressed \rtext{repulsive state branch} and the edge of the molecule-hole continuum, respectively (cf.~Fig.~\ref{fig.illustration}).

The actual time scale from which on the leading branch observably \textit{dominates} the dynamics depends on the relative magnitudes of its coefficients $ C_\alpha(T) $. For instance, in the \rtext{mixed} regime (b) the `bottom-of-the-band branch' B  is  leading. However, its coefficient $C_B$ can be numerically so small that for experimentally observable times the attractive branch A actually dominates the dynamics. This effect is shown in Fig.~\ref{fig.crossoversubleading}. Here a crossover between two exponential decay rates can be seen in the Ramsey signal $S(t)$. For short and intermediate times, $S(t)$ exhibits an exponential decay with the `fast', subleading decay rate $\gamma_0$ (dotted line). Only at long times  $S(t)$  crosses over to the slower, leading decay  rate $\gamma_1$ (dot-dashed line).

The fact that the fast decay of the subleading branch can dominate the short-time behavior demonstrates that, although the observed dynamics may naively seem to have reached quasi-classical behavior, quantum effects can still lead to long-time interference effects.  The small signal at long times makes the experimental observation of this crossover between multiple exponential decoherence rates a challenge, similar to competing $T_2$ decay coefficients encountered in NMR spectroscopy \cite{Yeramian1987,wern1989}. We note that in Fig.~\ref{fig.crossoversubleading} we have  chosen interaction parameters that allow us to demonstrate the crossover at  long times. This choice leads to a very small Ramsey contrast in the crossover regime. In experiments more favorable interaction parameters can be chosen that lead to a substantially higher Ramsey contrast (as an example see Fig.~\ref{Fig.HighContrast} in App.~\ref{crossappendix}).

\paragraph*{\textbf{Power-law temperature dependence---}}

The power-law temperature dependence of the Fermi surface contributions given by Eq.~\eqref{TScaling} can be observed experimentally. One means to do so is to fit the asymptotic forms of $S(t)$ in Eq.~\eqref{SRegimeAB} and \eqref{SRegimeC} to the measured data of $S(t)$. To obtain unbiased results, the prefactor $T^{\left(\frac{\delta_F}{\pi}-n\right)^2}$ is absorbed in a rescaled definition of the coefficients
\begin{equation}\label{CScaling}
\tilde C_\alpha(T)=C_\alpha T^{\left(\frac{\delta_F}{\pi}-n\right)^2}
\end{equation}
where $C_\alpha$ is a temperature independent constant. Studying  the fit parameters $\tilde C_\alpha(T)$ as a function of temperature then allows one to reveal the intricate OC power-law dependence.

We  demonstrate  this fitting procedure for the three interaction regimes by numerically calculating the exact signal $S(t)$ shown in the upper panels of Fig.~\ref{fig.Fitting1}. This data is then fit by the respective asymptotic forms~\eqref{SRegimeAB} and \eqref{SRegimeC} using Eqs.~\eqref{eq.SBS}, \eqref{eq.SBFS}, and \eqref{FullBosonization} with rescaled coefficients $\tilde C_\alpha(T)$. The resulting fits of the complex $S(t)$  are shown as solid lines in Fig.~\ref{fig.Fitting1}(a-c) \rtext{and they compare remarkably well with the exact data (symbols) down to relatively short times.} 

The scaling of the extracted coefficients $\tilde C_\alpha$ with temperature is shown in the lower panels of Fig.~\ref{fig.Fitting1}. Specifically, in Fig.~\ref{fig.Fitting1}(d) we study the attractive regime where the attractive branch $A$ is leading. We find that the corresponding coefficient $\tilde C_A$ universally follows the predicted power-law dependence. Furthermore, our analysis shows that the subleading coefficient $|\tilde C_B(T)|$ is numerically smaller than $|\tilde C_A(T)|$  so that  the leading branch $A$ will also dominate the decay of $S(t)$. 

\begin{figure}[t] 
  \centering 
  \includegraphics[width=\columnwidth]{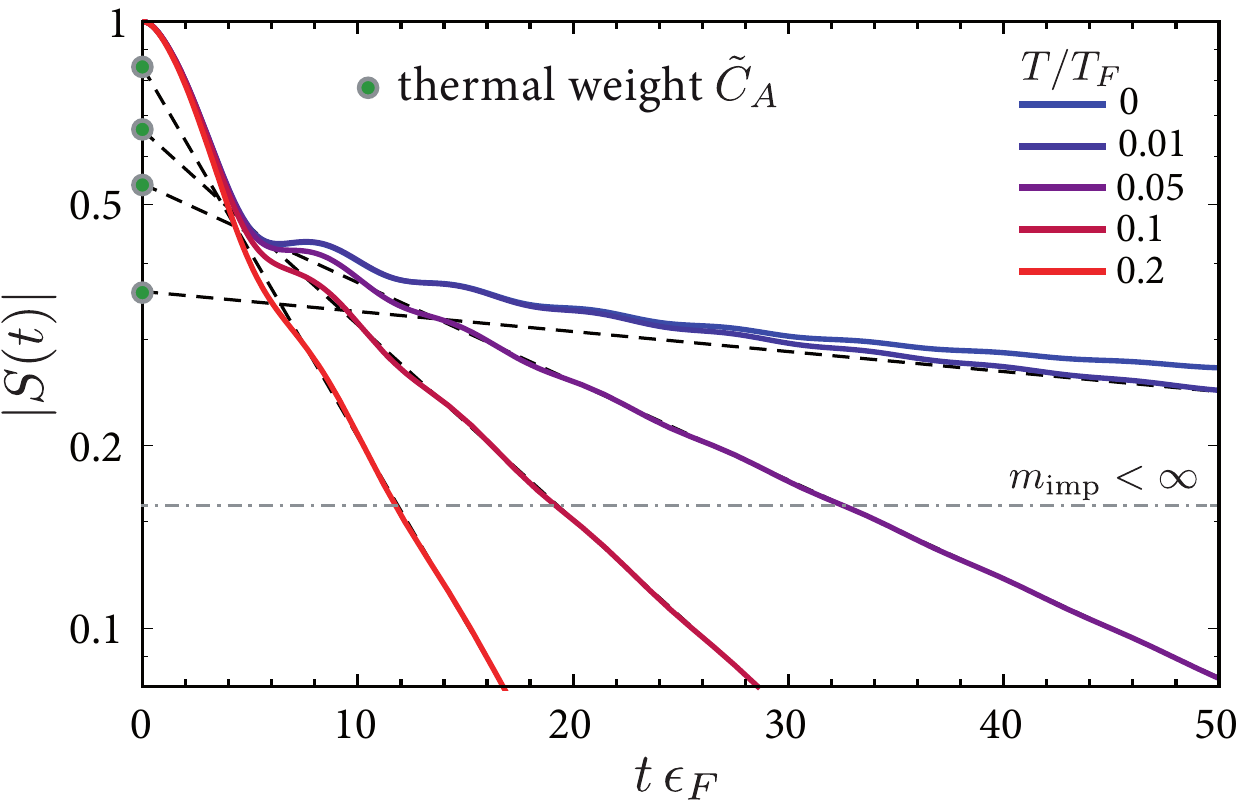}
  \caption{ \textbf{Thermal decay of the Ramsey signal and thermal weight.} The Ramsey contrast $|S(t)|$ is shown for various temperatures at fixed interaction parameters $k_F a =-1.1$ and $k_F r^*=0.8$. While at $T=0$ the contrast decays  to zero as a power-law, the decay is exponential at finite temperature.  The dashed lines are exponential fits to the data and the points  on the y-axis indicate the scaling behavior of the `thermal weights' $\tilde C_A$ (the fit for $T/T_F=0.01$ involves data at times not shown in the plot).  For an impurity of finite mass (e.g. $m_\text{imp}/m=40/6$ for a $^{40}$K impurity in a $^6$Li Fermi gas), the decay of the Ramsey contrast would saturate at the finite polaron quasi-particle weight $Z=|S(t\to\infty)|$ at zero temperature (illustrated by the dot-dashed line). 
  }
  \label{Fig.SEvolutions}
\end{figure}

As an alternative to a fit of the full complex signal of $S(t)$, one may also fit directly the absolute value $|S(t)|$ to obtain the temperature scaling of $\tilde C_\alpha$. In such a procedure, illustrated in Fig.~\ref{Fig.SEvolutions},  $|\tilde C_\alpha|$ as well as the decay exponent serve as fit parameters. The corresponding results are illustrated as open symbols in Fig.~\ref{fig.Fitting1}(d). This alternative procedure reveals that the temperature scaling of $\tilde C_A$, that is governed by the OC exponent $\left(\frac{\delta_F}{\pi}\right)^2$, can be observed up to temperatures as large as $T/T_F\approx 0.2$, making it accessible to current experimental technology.

Fig.~\ref{fig.Fitting1}(e) shows the results in the \rtext{mixed} regime ($k_Fr^*=0.8$, $1/k_Fa=-0.5$) for  both  coefficients $\tilde C_A(T)$ and $\tilde C_B(T)$. The dashed and solid curves show the predicted  asymptotic power law behavior $\sim T^{\left(\frac{\delta_F}{\pi}\right)^2}$ and $\sim T^{\left(\frac{\delta_F}{\pi}-1\right)^2}$, respectively. Although in this regime the bottom-of-the-band branch B is formally  leading, the corresponding coefficient $|\tilde C_B(T)|$ is numerically substantially smaller than $|\tilde C_A(T)|$. This finding reflects the crossover of exponential decay rates shown in  Fig.~\ref{fig.crossoversubleading}: for limited observation time the decay with the subleading, and hence larger decay rate $\gamma_{0}$ will dominate. This demonstrates that in experiments performed in the \rtext{mixed} interaction regime one has to be careful in assigning  thermal decoherence rates \rtext{from  early-time dynamics}.

In the repulsive interaction regime  the presence of the bound state leads to additional oscillations in the Ramsey signal. As can be seen in the upper panel Fig.~\ref{fig.Fitting1}(c) these oscillations are captured with remarkably high accuracy by the analytical expression Eq.~\eqref{SRegimeC} even at short times. In the lower panel Fig.~\ref{fig.Fitting1}(f) the temperature dependence of the corresponding rescaled Ramsey coefficients $\tilde C_\alpha(T)$ is shown. We find that up to  high temperatures $T/T_F=0.2$ the coefficient $\tilde C_{A1}$  follows the power-law prediction (solid line). Similarly to the \rtext{mixed} interaction regime, it is again not the leading branches B1 and B2 which have the numerically largest values  $|\tilde C_\alpha(T)|$. Instead, for the chosen interactions we find that the subleading bound-state branch $A_1$ dominates the dephasing at short and intermediate times.

 \paragraph*{\textbf{Relation to  ion mobility in $^3$He.}---} 
 
Our results on the scaling of the  coefficients $\tilde C_\alpha(T)$  allow us to draw connections to early work by Kondo and Soda \cite{Kondo1983} on ion mobility in $^3$He. In their work Kondo and Soda studied the renormalization of a heavy ion   due to its interaction with the fermionic quasiparticles in liquid $^3$He. In this case, the   ion Green's function  can be expressed as $G(\vecq,\omega)=Z(T)/[\omega-E_\vecq + i \Gamma(T)]$,
which in the time domain becomes
\begin{equation}\label{KondoG}
G(\vecq,t)=Z(T) e^{-\Gamma t} e^{i E_\vecq t}.
\end{equation}
Here $E_\vecp$ is the renormalized ion dispersion relation and $\Gamma(T)$ its quasiparticle lifetime. $Z(T)$ is the  quasiparticle weight that determines the ion mobility $\mu\propto Z(T)^2$ \cite{Kondo1983}. 

Using a perturbative expansion, Kondo and Soda predicted  the ion quasiparticle weight to scale as
\begin{equation}\label{KondoScaling}
Z(T)\propto\left(\frac{T}{T_F}\right)^{2V_0^2\rho^2}.
\end{equation}
Here $V_0$ is the microscopic contact coupling constant between the ion impurity and fermions of mass $m$, and $\rho$ is the density of states at the Fermi surface. This  scaling was predicted to be valid for temperatures $T_0\ll T \ll T_F$, where $T_0\equiv  (m/M)T_F $ approaches zero as the ion mass $M$ goes to infinity. 

The expression for the ion Green's function Eq.~\eqref{KondoG} suggests a comparison with our prediction for  the Fermi surface contribution to the impurity Green's function Eq.~\eqref{TScaling}. Remarkably, the scaling of the prefactor, cf.~Eq.~\eqref{CScaling} for  $n=0$, reproduces Kondo's scaling of the impurity quasiparticle weight in Eq.~\eqref{KondoScaling}, in our case derived from an exact calculation.

Having this relation in mind, one may interpret the coefficients $\tilde C_\alpha$ as wave-function renormalizations that determine the temperature-dependent spectral weight of the respective excitation branches. As temperature goes to zero we predict this thermal weight to vanish according to a power law  that is governed by the OC exponent, see  Fig.~\ref{fig.Fitting1}. This behavior is illustrated in Fig.~\ref{Fig.SEvolutions}, where we show Ramsey contrast curves at various temperatures. Exponential fits to the long-time decoherence data, shown as  dashed lines in Fig.~\ref{Fig.SEvolutions}, indicate that the weights of the exponential decays, shown as green dots, decrease with decreasing temperature. Simultaneously, the regime of quantum dephasing extends to longer times. In the limit of zero temperature, the thermal weight goes to zero and the thermal exponential decay of the Ramsey contrast is replaced by a power law with exponent $(\delta_F/\pi)^2$, characteristic of the Anderson OC. Remarkably, we find that each excitation branch is governed by its unique scaling exponent. While in liquid Helium the measurement of subleading excitation branches is difficult,  ultracold atoms  allow not only for the verification of  the power-law scaling Eq.~\eqref{CScaling} but also for the observation of subleading excitation branch dynamics.

We note that, while for an infinitely heavy impurity the thermal weight $\tilde C_A$ goes to zero at $T=0$, for a mobile impurity in three dimensions the Ramsey contrast will saturate at a finite quasi-particle weight $Z(T=0)= S(t\to\infty)$ (illustrated by horizontal dash-dotted line in Fig.~\ref{Fig.SEvolutions} \cite{schmidt_excitation_2011,massignan_repulsive_2011,cetina_2016,parish2016}). \rtext{Here, the time scale for the crossover to many-body dynamics that is greatly affected by the impurity mass is expected to be approximately given by the inverse impurity recoil energy $E_\text{rec}\sim (2k_F)^2/(2M)$. However, both a detailed study of the behavior of the quasiparticle weight as function of the inverse impurity mass $1/M$ as well as the related many-body dynamics remain open questions which are beyond the scope of this report.}

\subsubsection{Toeplitz-determinant approach}\label{sectoeplitz}

\begin{figure}[t]
\includegraphics[width=0.9\linewidth]{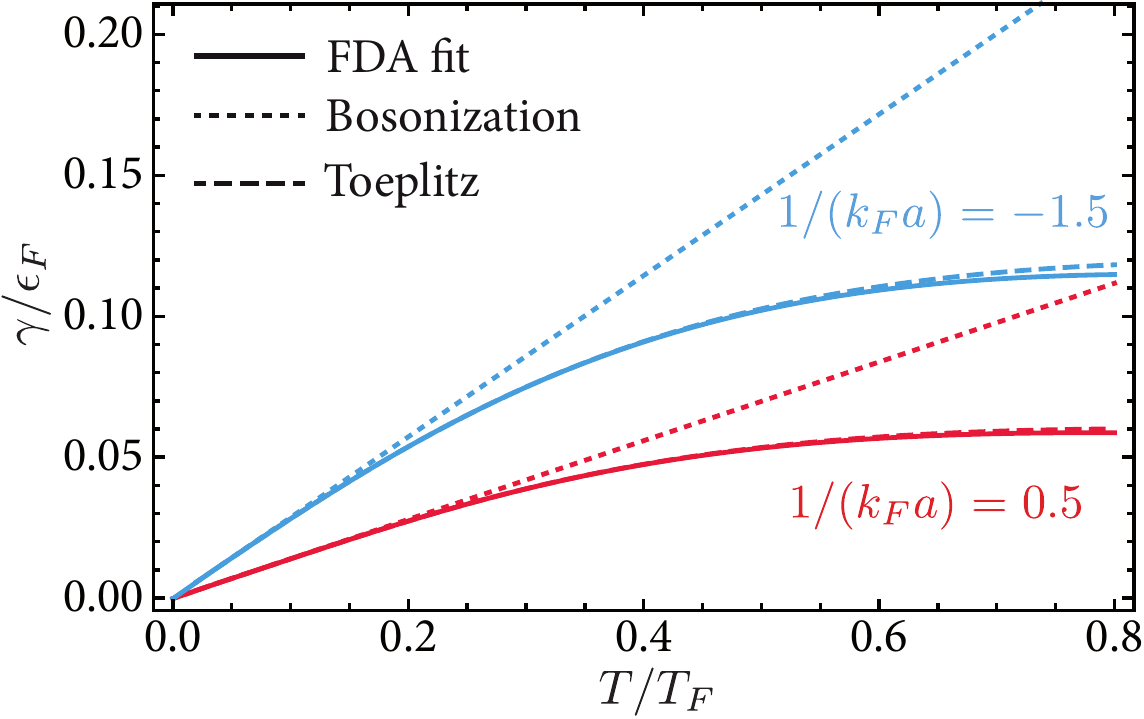}
\caption{\textbf{Decoherence rate at high temperatures.} We compare theoretical predictions from bosonization and the Toeplitz-determinant approach to the numerically exact long-time decoherence rate of the Ramsey signal as function of temperature in the attractive and repulsive regime for $k_Fr^*=0.8$. The Toeplitz determinant approach gives very accurate results up to high temperatures, while bosonization starts to fail at $T/T_F\approx 0.15$. 
}
\label{fig.newdec}
\end{figure}

In Fig.~\ref{fig.newdec} we compare the temperature dependence of the leading branch decoherence rate $\gamma_n$ as obtained from bosonization (dotted lines) with the exact numerical result (solid lines). We find that for sufficiently low temperatures both  agree. However, as $T/T_F$ is increased to values realized in current experiments, $T/T_F\approx0.2$,  deviations appear.  The reason for the failure of the bosonization approach lies in the fact that at high temperatures, on the one hand, the energy dependence of the phase shift $\delta(E)$ has to be taken into account, and, on the other hand, the assumption of a linearized dispersion relation becomes invalid.

These  effects can be taken into account in  a quasi-classical approach motivated by the theory of the Toeplitz determinants \cite{szego1952,Basor1991,levitov_electron_1996,gray2006,Hassler2008,Hassler2009}. In this approach, many-body overlaps such as Eq.~\eqref{eq:klich}  are evaluated in a quasi-classical basis of wave packets localized both in coordinate space and momentum; for details we refer to Appendix \ref{app:B}. In this basis, the evolution operator in Eq.~\eqref{eq:ramsey} is of a Toeplitz form, i.e. the kernel depends only on a time difference and, consequently, $S(t)$ can be expressed as a Toeplitz determinant. Such determinants were studied in various physical and mathematical context, and, applying these techniques, the  asymptotic behavior of the Fermi-surface contributions at long times can be obtained (see, e.g., \cite{Ivanov2013} for a related discussion in the context of full counting statistics).

 \paragraph*{\textbf{Leading branch decoherence---}} The leading asymptotic behavior of Toeplitz determinants is given by the so-called Szeg\H{o} formula, which allows us to map the calculation of the asymptotic Toeplitz determinant onto contour integrations, see e.g.~\cite{Hassler2009}. By going to the frequency representation, in which the kernel operator in Eq.~\eqref{eq:ramsey} is diagonal, one finds for the Fermi sea contributions (see Appendix \ref{app:B})
\begin{equation}\label{fullSEquationSzergo}
S(t)=\exp\left(t\int_0^\infty \frac{dE}{2\pi} \ln \left[1-n_F(E) + n_F(E) e^{2 i \delta(E)}\right]\right)\, ,
\end{equation}
where $n_F(E)$ is the Fermi occupation number.

\begin{figure}[t]
\includegraphics[width=\linewidth]{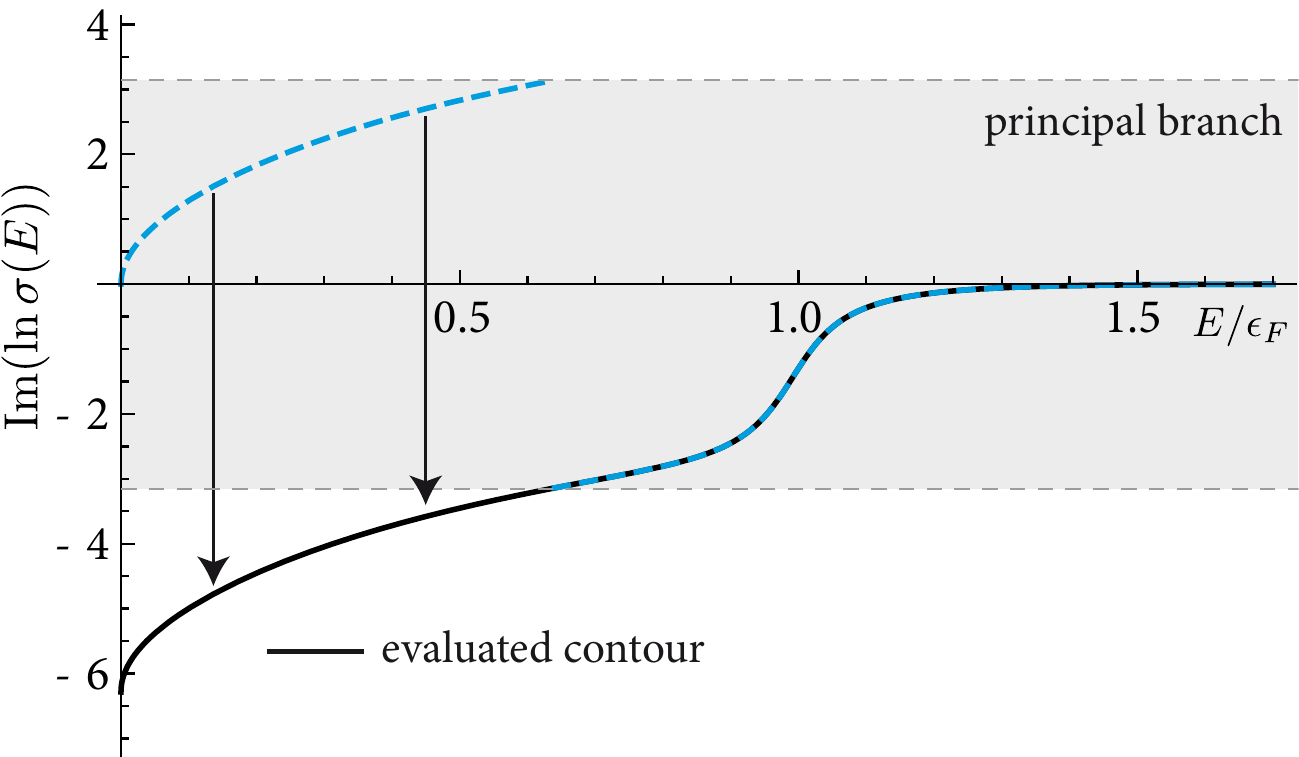}
\caption{\textbf{Evaluation of the Toeplitz determinant.} For the evaluation of the Toeplitz determinant according to Eq.~\eqref{fullSEquationSzergo} the branch of the logarithm has to be analytically continued as illustrated by the black solid line. Here the imaginary part  of the integrand $\ln \sigma(E)$ is shown as function of $E$ for interaction parameters in the \rtext{mixed} interaction regime (b) with $1/k_Fa=-0.5$, $k_Fr^*=0.8$, and $T/T_F=0.114$.}
\label{fig.C0Example}
\end{figure}

In the evaluation of Eq.~\eqref{fullSEquationSzergo} the branch of the logarithm must be chosen so that the integrand analytically continues along the integration contour and  tends to zero as $E\to\infty$. While in the attractive and repulsive interaction regime (a) and (c) one can remain in the principal branch of the logarithm, the \rtext{mixed} interaction regime (b) requires more care. Here the phase shift at the Fermi surface exceeds $\pi/2$ which requires to evaluate the integrand in Eq.~\eqref{fullSEquationSzergo} starting at low energies in the lower Riemann sheet and then analytically continue to the principal branch at high energies.  This is illustrated in Fig.~\ref{fig.C0Example} where we show the imaginary part of the integrand of Eq.~\eqref{fullSEquationSzergo} along the integration contour. 

Using this integration procedure, one obtains the oscillation frequency $\omega_L$ and the decay rate $\gamma_L$ from
\begin{equation}
i\omega_L + \gamma_L = -\int_0^\infty \frac{dE}{2\pi} \ln \left[\sigma(E)\right]\, ,
\label{eq:integralomegagamma}
\end{equation}
where we defined the so-called `symbol' of the Toeplitz determinant 
\begin{equation}\label{signum}
\sigma(E)=1-n_F(E) + n_F(E) e^{2 i \delta(E)}.
\end{equation}
However, the values of $\omega_L$ and $\gamma_L$ as obtained from Eq.~\eqref{eq:integralomegagamma} correspond only to the \textit{leading} branch of the respective interaction regime (see Fig.~\ref{fig.illustration}): in regime (a) this is branch A, in regime (b) it is branch  B, and in regime (c) these are the branches B1 and B2. Note that in the limit $T\to 0$, the integral \eqref{eq:integralomegagamma} is easily computable and reproduces the results derived in the bosonization approach, Eqs.~\eqref{decayrates}.

\paragraph*{\textbf{Subleading branch decoherence.---}} 

Calculating the frequencies and the decay rates for the subleading branch requires more effort. In the context of the theory of Toeplitz determinants, subleading branches were  studied in the case of singular symbols $\sigma(E)$. In that case, the theory of subleading branches is known under the name of the generalized Fisher-Hartwig conjecture and all branches decay as power laws \cite{Fisher2007,Basor1991,Deift2011}. However, in our finite-temperature case, the decay is exponential, and the theory of Fisher-Hartwig singularities does not directly apply. 

Still, the contribution of different excitation branches can be found from the following argument. Each branch contribution corresponds to a specific configuration of fermions (see Fig.~\ref{fig.illustration}) and should be an analytical functional of $\delta(E)$. Consequently, the subleading branch may be obtained as an analytical continuation from the regime where the corresponding contribution constitutes the leading branch and is given by Eq.~\eqref{fullSEquationSzergo}. This analytical continuation technically amounts to a continuous deformation of the leading integration contour in Eq.~\eqref{fullSEquationSzergo} into a new contour  $\mathcal C_{SL}$ in the complex energy plane so that it never crosses any singularities of the integrand.

As discussed in detail in Appendix \ref{app.toeplitz}, we find that for each interaction regime an integration contour $\mathcal{C}_{SL}$ can be chosen to give the desired subleading frequency and decay rate
\begin{equation}\label{contouromega}
i\omega_{SL} + \gamma_{SL} = -\int_{\mathcal{C}_{SL}} \frac{dE}{2\pi} \ln \left[\sigma(E)\right]\, .
\end{equation}
The specific contours are derived from an analysis of the analytical structure of the integrand $\ln \sigma(E)$ which is determined by the roots $E_r$ of the symbol $\sigma(E)$,
\begin{equation}\label{sigmaroot}
\sigma(E_r)=0\, ,
\end{equation}
in the complex energy plane. By choosing the appropriate integration contours  one can show that the subleading decay rate and frequency are given by  (see Appendix \ref{app.toeplitz}) 
\begin{eqnarray}\label{eq:sublead}
\gamma_{SL} &=& \gamma_L + | {\rm Im} E_r | \, \nonumber\\
\omega_{SL} &=& \omega_L - {\rm sign}({\rm Im} E_r) \; {\rm Re} E_r\, .
\end{eqnarray}
In the low temperature limit $T\to 0 $, $E_r=\epsilon_F+ i T (\pm \pi +2\delta_F)$ where the $-$ applies to the attractive and \rtext{mixed} regime (a) and (b) and $+$ holds for the repulsive regime $(c)$. Then the result Eq.~\eqref{eq:sublead} indeed reproduces Eq.~\eqref{decayrates}. 

\begin{figure}[t]
\includegraphics[width=\linewidth]{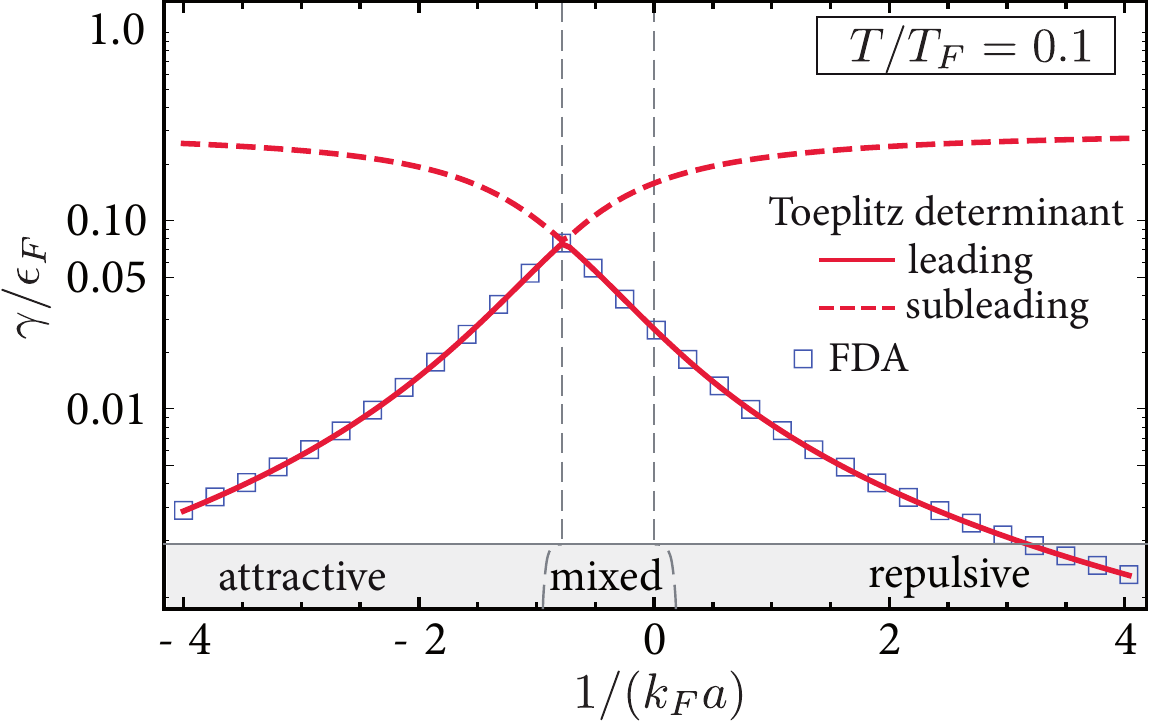}
\caption{\textbf{Dependence of decoherence rate on the interaction strength.} The interaction dependent decoherence rate of $S(t)$ obtained from the exact FDA and Toeplitz determinant approach at  $T/T_F=0.1$ and $k_Fr^*=0.8$. }
\label{fig.newdeccomp}
\end{figure}

The analytical Toeplitz approach is in remarkable agreement  with exact numerical results. In Fig.~\ref{fig.newdec} we compare the temperature dependence of decoherence rate as obtained from the Toeplitz approach and the exact FDA calculation. The predicted rates, shown for two interaction regimes, are in excellent agreement  up to  temperatures as high as $T/T_F\approx 1$, where bosonization completely fails. 

The virtue of the Toeplitz approach is further demonstrated in Fig.~\ref{fig.newdeccomp} where we show the interaction dependence of the decoherence rate. The rate of the leading (solid) and subleading branch (dashed) obtained from the Toeplitz approach compare remarkably well with the exact FDA results (symbols). But most importantly it also analytically predicts the rate of the subleading branch which can dominate the intermediate time evolution, cf.~Fig.~\ref{fig.crossoversubleading}. As discussed below, this becomes particularly relevant for experiments in which very long time scales are inaccessible due to small Ramsey contrast.

\subsection{Universal scaling relations}

\begin{figure}[t]
\includegraphics[width=\linewidth]{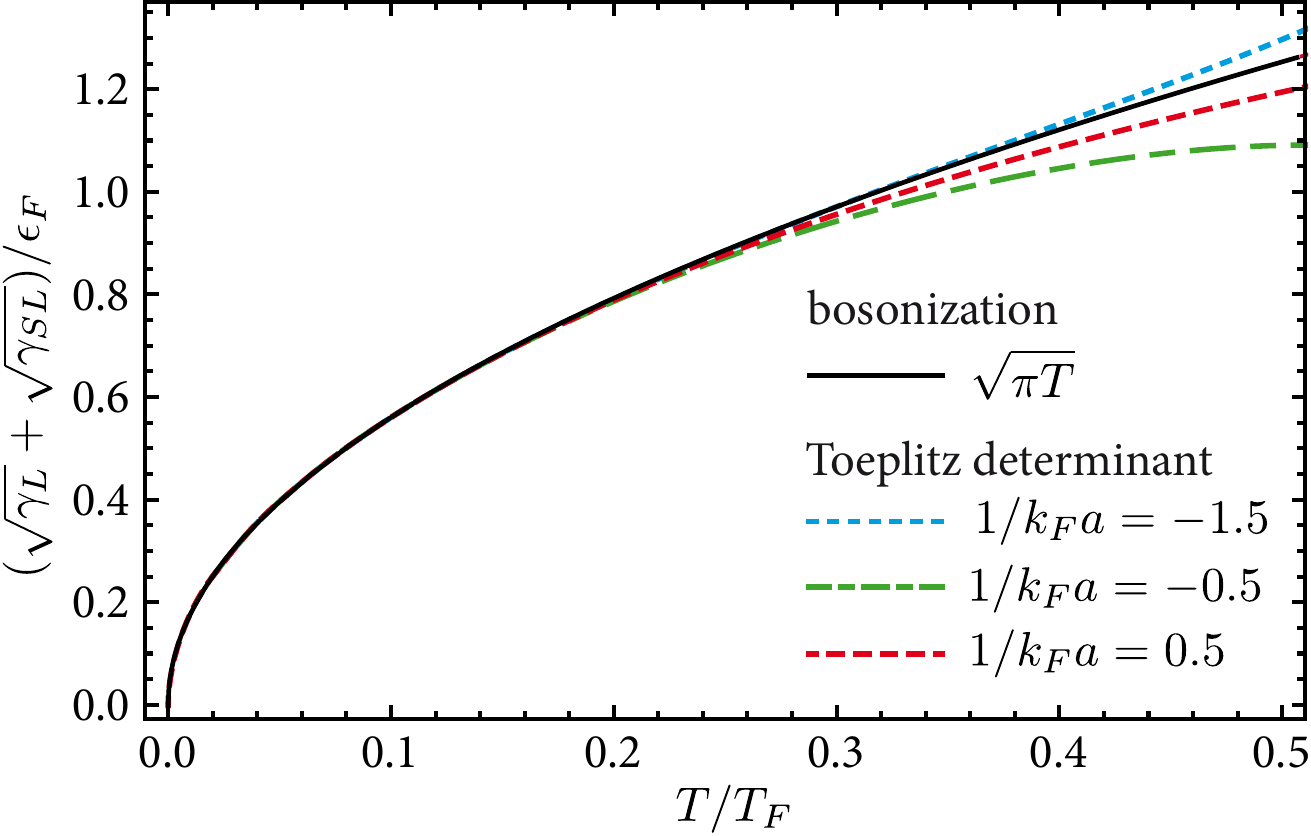}
\caption{\textbf{Universal scaling relations of decoherence rates.} At low temperatures the decoherence rates of the leading and subleading branch obey the universal scaling relations Eq.~\eqref{eq.scaling} (solid line) which are independent of interaction strength and hold up to temperature $T/T_F\approx0.2$ as can be seen from the comparison  to the result from the Toeplitz determinant theory for three different interaction strengths and fixed $k_Fr^*=0.8$.}
\label{fig.universalscaling}
\end{figure}

Within the bosonization approximation, the decay rates and frequencies for the leading and subleading branches obey simple  scaling relations,
\begin{eqnarray}\label{eq.scaling}
\sqrt{\gamma_L} + \sqrt{\gamma_{SL}} &=& \sqrt{\pi T} ,\\
|\omega_L-\omega_{SL}|=\epsilon_F \,.
\end{eqnarray}
These relations are universal and do not depend on the particular form of the phase shift $\delta(E)$. Being derived from bosonization, they are  valid in the limit of low temperature $T$, but remarkably, we find that
finite-$T$ corrections remain small up to rather high temperatures compared to $T_F$. This is illustrated in Fig.~\ref{fig.universalscaling}, where we compare the scaling of the decay rates obtained from the Toeplitz determinant approach to the prediction Eq.~\eqref{eq.scaling} for the various interaction regimes.

\subsection{Comparison to experiments}\label{SectExpCom}

\begin{figure}[t]
\includegraphics[width=0.85\linewidth]{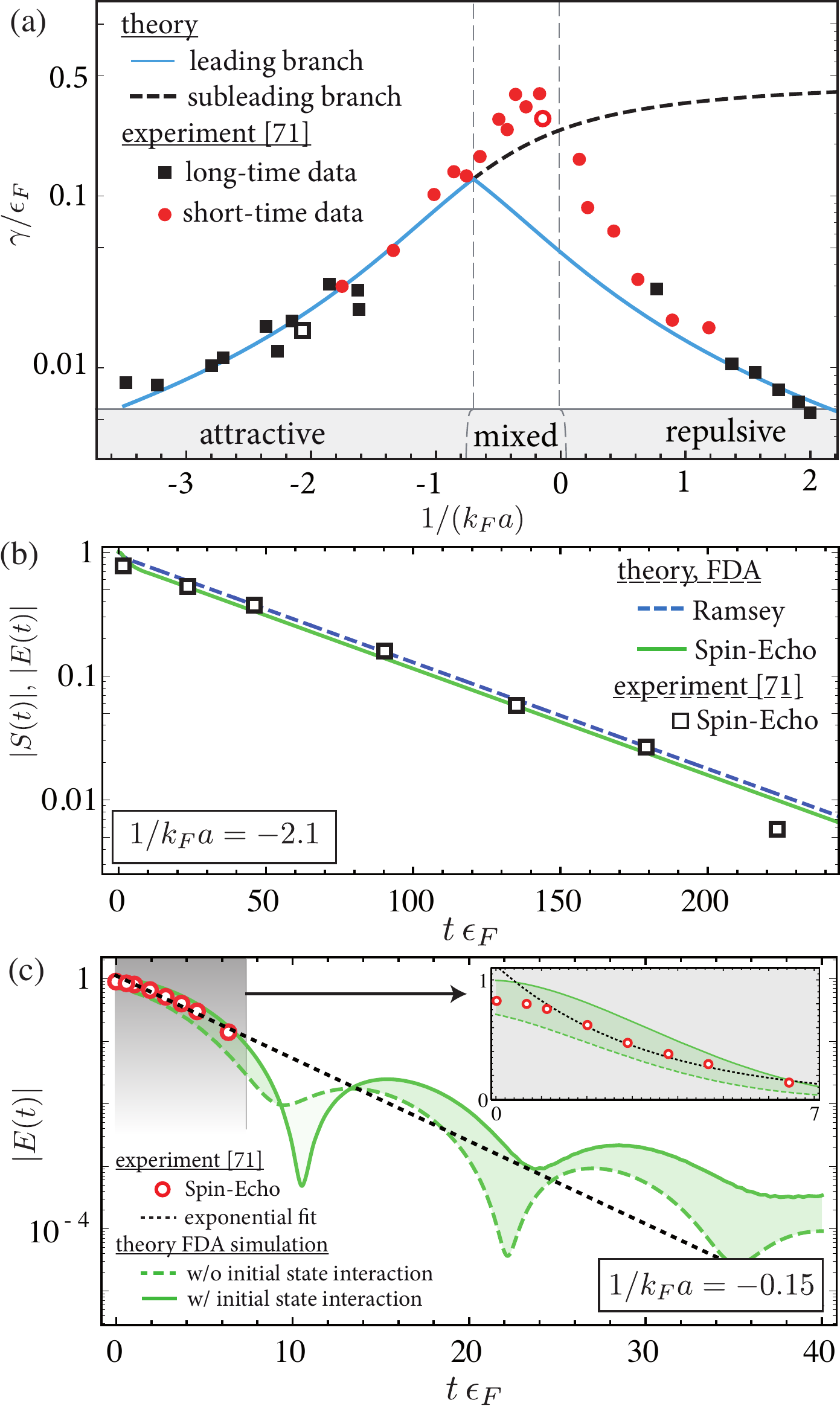}
\caption{\rtext{\textbf{Decoherence rate: theory and experiment \cite{cetina_decoherence_2015}.} (a) Decoherence rate as function of interaction strength $1/k_Fa$. The black symbols represent the experimentally measured decoherence rate of the spin-echo signal at long times while the red data points were extracted from exponential fits to the  short-time dynamics. The theoretical predictions from FDA and Toeplitz determinant theory are shown as solid and dashed lines. They are obtained for experimental temperature $T/T_F=0.16$ and rescaled resonance parameter $k_F r^*=(40/46)^2k_Fr^*_\text{exp}=0.7$ \cite{cetina_2016}. (b)  FDA simulation of the time-resolved responses (lines) and experimental data (symbols) for the interaction strength $1/k_Fa=-2.1$, indicated by the open square in (a). The Ramsey and spin-echo response lead to the same exponential decay rate. (b) Detailed FDA simulation  of the experimental spin-echo sequence at $T/T_F=0.2$ and $1/k_Fa=-0.15$, including finite pulse duration and initial state interaction. The dotted line is an exponential fit to the experimental data  that leads to the open red circle in (a). The inset shows a zoom into short evolution times on a linear scale. The experimental data is taken from Ref.~\cite{cetina_decoherence_2015}.}  }
\label{fig.compexp}
\end{figure}

The long-time thermal decoherence rate of impurities immersed in a Fermi gas  has recently been measured~\cite{cetina_decoherence_2015}. The experiment was performed using a dilute sample  of $^{40}$K impurities immersed  in  a $^{6}$Li Fermi gas, and the spin-echo decoherence rate was determined. The interaction between the impurities and the Fermi gas was  characterized by a narrow Feshbach resonance of range $k_F r^*_\text{exp}=0.93$ and the temperature was $T/T_F=0.16$. For relatively `weak' interactions with $1/k_Fa<-1.4$ and $1/k_Fa>0.8$ the loss in spin-echo contrast $|E(t)|$ was recorded up to long times $t \epsilon_F\approx220$ and  was fit to an exponential decay. In Fig.~\ref{fig.compexp}(a) the experimentally determined finite temperature decoherence rate is shown as black squares. In this figure we also show our prediction from the FDA with the appropriately rescaled resonance parameter $k_F r^*=(40/46)^2k_Fr^*_\text{exp}=0.7$ (see~Ref.~\cite{cetina_2016}) as the blue solid line. In this temperature regime the exact FDA agrees with the theory of Toeplitz determinants and we find full agreement with the experimental data (black squares).

\rtext{In a complementary approach} in Refs.~\cite{cetina_decoherence_2015,Christensen2015b}, the thermal decoherence  rate of the spin-echo signal was calculated using  Fermi liquid theory. In these works, which relied on an \textit{a priori} assumed equivalence of the spin-echo and Ramsey signal, good agreement with experimental observation was found  away from the Feshbach resonance. In the theoretical description, \rtext{which applies to the weakly interacting regime} \cite{cetina_decoherence_2015}, the inclusion of the decay of the repulsive polaron into the weakly bound molecular state was found to be important in the repulsive regime. This decay process was captured by adding a semi-phenomenological decay rate to the quasiparticle collision rate, obtained from Fermi liquid theory. Our results which are based on an exact solution, fully include all conversion processes between the repulsive polaron excitation branch and the molecular state. The excellent agreement of our prediction with the experimental data hence confirms the previous conjecture \cite{cetina_decoherence_2015} that the inclusion of  higher-order processes is crucial for an accurate description of the impurity dephasing dynamics.

Close to the Feshbach resonance, the experiment could not access the very long-time dynamics due to fast loss of spin-echo contrast. As a consequence, the data was measured only up to small times of $t\epsilon_F\approx 10$ and again fit to an exponential decay.  The resulting, experimentally measured rates are shown as the red dots in Fig.~\ref{fig.compexp}(a). We find that in the \rtext{mixed}, as well as in the repulsive regime at strong interactions, it deviates from the FDA prediction. 

Based on our previous discussion this discrepancy does, however, not come as a surprise, and we can identify two  possible explanations for the deviations. First, in the repulsive and \rtext{mixed} regime, very long times have to be reached to observe the leading long-time decoherence rate, as here the subleading branch  dominates the intermediate-time dynamics. In Fig.~\ref{fig.compexp}(a) we show the decoherence rate of the subleading branch as obtained from the theory of Toeplitz determinants as dashed line. We find that it agrees reasonably well with the observed enhanced decay rate in the \rtext{mixed} regime. This makes it plausible that the experiment may have observed the intermediate-time dynamics governed by the subleading branch. A second explanation for the deviations may be found in the fact that at the  times accessible in the experiment oscillations originating from bottom-of-the-band excitations still influence the dynamics of the Ramsey and spin-echo contrast. Indeed these oscillations  dominate the short and intermediate time dynamics in the strongly interacting regime, see e.g.~Fig.~\ref{Fig.HighContrast} in App.~\ref{crossappendix}. Only far beyond the thermal time scale $\tau_\text{th}=1/T$ a pure exponential decay can be expected,  and hence the experimental data may have been still strongly influenced by non-thermal quantum dynamics. Furthermore, \rtext{while the Ramsey- and spin-echo response lead to equivalent decoherence rates at very long times, their signals can show very distinct behavior at short and intermediate times.}

\rtext{To corroborate these arguments we have simulated the time-resolved signal for the experimental setup of Ref.~\cite{cetina_decoherence_2015} using the FDA and taking the experimental parameters as input. The result of our simulations for two representative interaction strengths (open symbols in Fig.~\ref{fig.compexp}(a)) are shown as curves in Fig.~\ref{fig.compexp}(b,c) where also the experimental, time-resolved data (symbols) for the spin-echo signal, as given in Ref.~\cite{cetina_decoherence_2015}, is shown. For weak interaction, see~Fig.~\ref{fig.compexp}(b), long evolution times were experimentally accessible. Here the theoretical simulation of the time-resolved spin-echo signal reveals that  the experimental data is well described by exponential, thermal decay with a decay rate in agreement with the experimental data. }

\rtext{This is in contrast to strong interactions, analyzed in Fig.~\ref{fig.compexp}(c). Here we simulate in detail the experimental sequence employed to obtain the red open data point in Fig.~\ref{fig.compexp}(a). To this  end, we include the finite pulse duration $\tau_{\pi/2}=\tau_\pi/2\approx 2.5/\epsilon_F$ of the spin-rotation pulses. Moreover, we simulate the fact that these pulses were performed at weak initial interaction $1/k_F a_\text{in}$ followed by a quench to the final interaction strength $1/k_Fa=-0.15$.  The details of the protocol are described in Ref.~\cite{cetina_2016}. The green shaded area in Fig.~\ref{fig.compexp}(c) shows the simulated response for varying  initial state interactions ranging from zero $1/k_F a_\text{in}=-\infty$ (solid), to intermediate interactions $1/k_F a_\text{in}=-2.25$ (dashed). The dotted black line is an exponential fit to the experimental data (symbols). From the simulation we find that independent of the initial state interaction, the dynamics is strongly influenced by bottom-of-the-band excitations that lead to pronounced oscillations in the signal. The comparison with the experimental data (symbols) in Fig.~\ref{fig.compexp}(c) makes it thus plausible that the experiment may have probed an intermediate time regime where dynamics cannot yet be modeled by a pure exponential decay, which will dominate only at much later times.}


\section{Universal short-time, high-frequency response \label{sec:shorttime}}

In the previous sections we have demonstrated that the combined use of the FDA, bosonization, and the theory of Toeplitz determinants allows us to obtain  a precise analytical understanding of the dephasing dynamics of heavy impurities immersed in a Fermi gas at intermediate and long times. The analytical approaches provided  an intuitive description of the universal long-time dynamics and hence low-frequency response in terms of a few relevant excitation branches. Here we turn to the  short-time behavior of the Ramsey signal which is more conveniently studied in the corresponding high-frequency response. To this end we  study the high-frequency absorption for a system where the impurities are interacting with the Fermi gas in both the initial and final state. Considering this scenario allows us to connect our predictions to universal analytical results obtained from a short-time operator product expansion \cite{Braaten2010}.

\subsection{RF response with finite initial and final state interactions}

For many experimentally accessible atomic species interactions between the impurity and the Fermi gas are present both in the  initial and final impurity spin state \cite{bartenstein2004}. In this case the absorption spectrum is given by
\begin{equation}\label{eq:MixedRF}
A(\omega)=2\text{Re}\int_0^\infty dt e^{i\omega t} \Tr [e^{i \hat H_i t}e^{-i \hat H_{f} t}\hat \rho_i].
\end{equation}
Here $\hat H_{i}$ ($\hat H_{f}$) is the initial (final) Hamiltonian of the system with initial (final) state scattering length $a_i$ ($a_f$) and range $r^*_i$ ($r^*_f$). 

\begin{figure}[t] 
  \centering 
  \includegraphics[width=\linewidth]{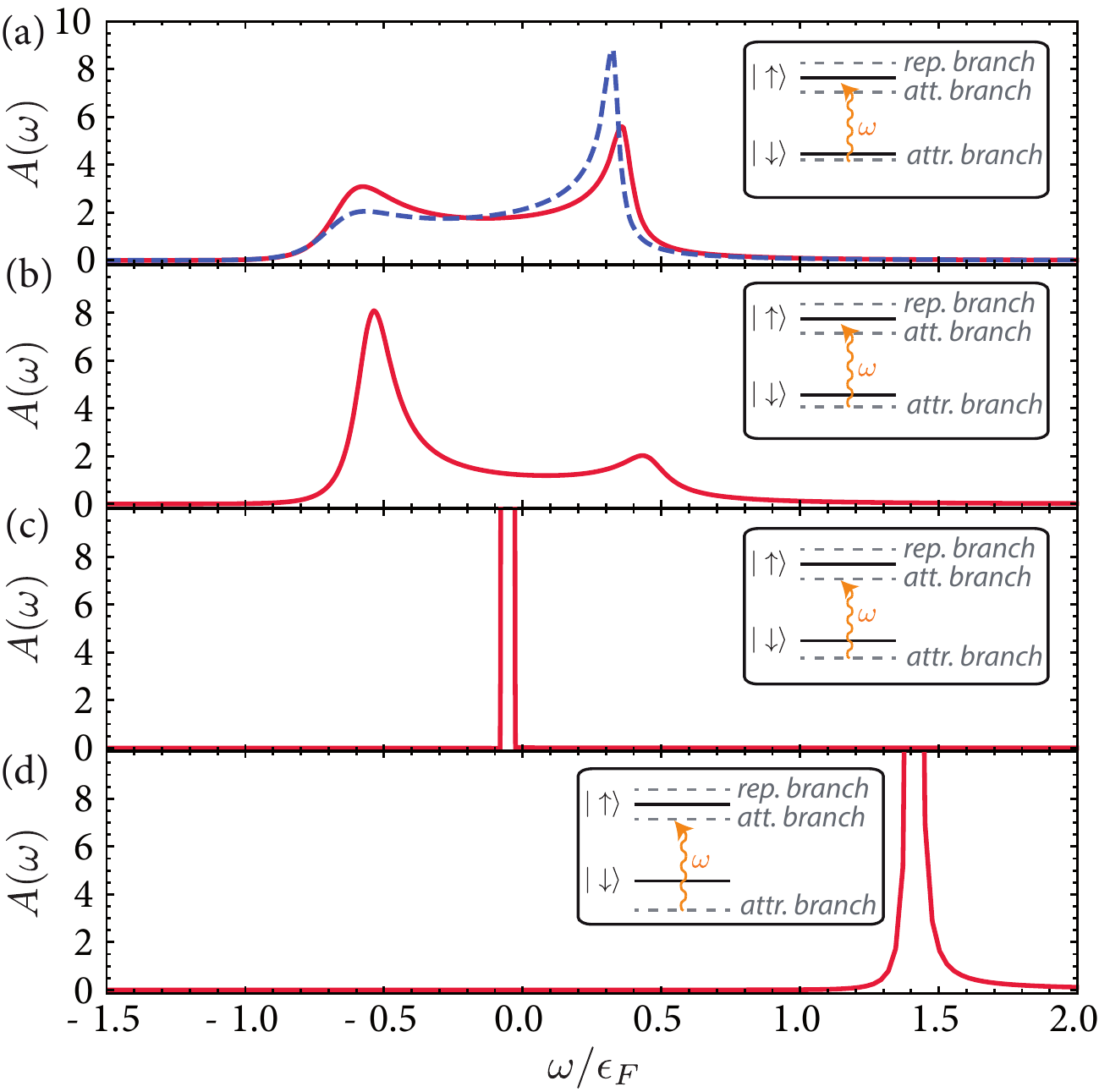}
    \caption{ \textbf{RF spectra for finite initial and final state interactions.} RF spectrum for fixed final state interaction $1/k_Fa_f=0.05$ while the initial state interaction between the \rtext{static} impurity and the Fermi gas is varied: (a) $1/k_Fa_i=-6$, (b) $1/k_Fa_i=-2$, (c) $1/k_Fa_i=-0.05$, (d) $1/k_Fa_i=2$. The temperature is given by $T/T_F=0.1$ and $k_F r_i^*=k_F r_f^*=0.8$ in both the initial and the final state.}
  \label{fig.mixedrf}
\end{figure}

Based on Eq.~\eqref{eq:MixedRF} we study the influence  of initial state interactions on the  RF response as shown in Fig.~\ref{fig.mixedrf}. In this figure, we keep the final state interactions fixed at $1/k_Fa_{f}=0.05$ while varying the interaction $1/k_F a_{i}$ in the initial state. For weak initial state interactions (solid line in Fig.~\ref{fig.mixedrf}(a)), the spectrum is only slightly shifted with respect to the perfect reverse RF response (dashed line). However,  as interactions in the initial state are increased, not only the spectrum is shifted further, but it also changes in shape, cf.~Fig.~\ref{fig.mixedrf}(b-d).

When interactions in the initial and final state become comparably strong, see~Fig.~\ref{fig.mixedrf}(c), the response approaches a $\delta$-peak due to the symmetry between the initial and final state. Using this feature---when the initial and final state have the same scattering length, $a_i=a_f$, yet differ in the Feshbach resonance width--- RF measurements on impurities provide a tool for the experimental determination of $r^*$.

\subsection{Analytical high-frequency response}

 We now turn to the absorption response at high frequencies. As has been shown recently \cite{Braaten2010}, the absorption behavior at high frequency can be related to seemingly unrelated quantities by so-called `Tan relations' \cite{Yu2006,Baym2007,Punk2007,Tan2008a,Tan2008b,Tan2008c,Braaten2008,Braaten2008b,Barth2011,Valiente2011,Valiente2012,Langmack2012,Hofmann2013}. These relations apply to arbitrary two-component Fermi gases (the case of an impurity immersed in a Fermi gas is a special case), where the two fermion species interact with short-range potentials. For such gases the momentum distribution decays as  \cite{lifshitz1980}
\begin{equation}\label{highmomtail}
n(\textbf{k})\to \frac{C}{k^4}
\end{equation}
at large momenta $k=|\textbf{k}|$. A decade ago, Tan discovered theoretically that the coefficient $C$, which has been termed the `contact', is related to various quantities by simple, universal relations \cite{Tan2008a,Tan2008b,Tan2008c}. For instance, Tan's adiabatic theorem states that the change of energy $E$ of an arbitrary two-component Fermi gas interacting with contact interactions, due to a change of the interspecies scattering length $a$ is determined by
\begin{equation}\label{Tanadiabatic}
\frac{dE}{d(1/a)}=-\frac{C}{8\pi\mu_\text{red}}
\end{equation}
where $\mu_\text{red}=m_1m_2/(m_1+m_2)$ is the reduced mass, and $m_{1,2}$ are the masses of the two fermion species. Similar formulas relate the contact $C$ to the pressure,  the density-density correlator, the virial theorem, and the inelastic two-body loss observed in such systems \cite{Tan2008a,Tan2008b,Tan2008c,Braaten2008,Braaten2008b,Barth2011,Langmack2012,Hofmann2013}. 

The contact $C$ appears also in the RF absorption spectrum when final and initial state interactions are present. Using an operator product expansion (OPE) Braaten~\etal~\cite{Braaten2010} predicted that at high frequencies the RF response follows
\begin{equation}\label{OPEtail}
A(\omega)\to\frac{1}{4\pi (2\mu_\text{red} \omega^3)^{\frac{1}{2}}}\left(\frac{1}{a_{i}}-\frac{1}{a_{f}}\right)^2 \frac{C}{a_{f}^{-2}+2\mu_\text{red} \omega}
\end{equation}
where we have set the Rabi coupling $\Omega=1$. The contact $C$ depends on the initial state of the system, and hence in particular on the initial state scattering length $a_i$.

 \begin{figure}[t] 
   \centering 
   \includegraphics[width=\linewidth]{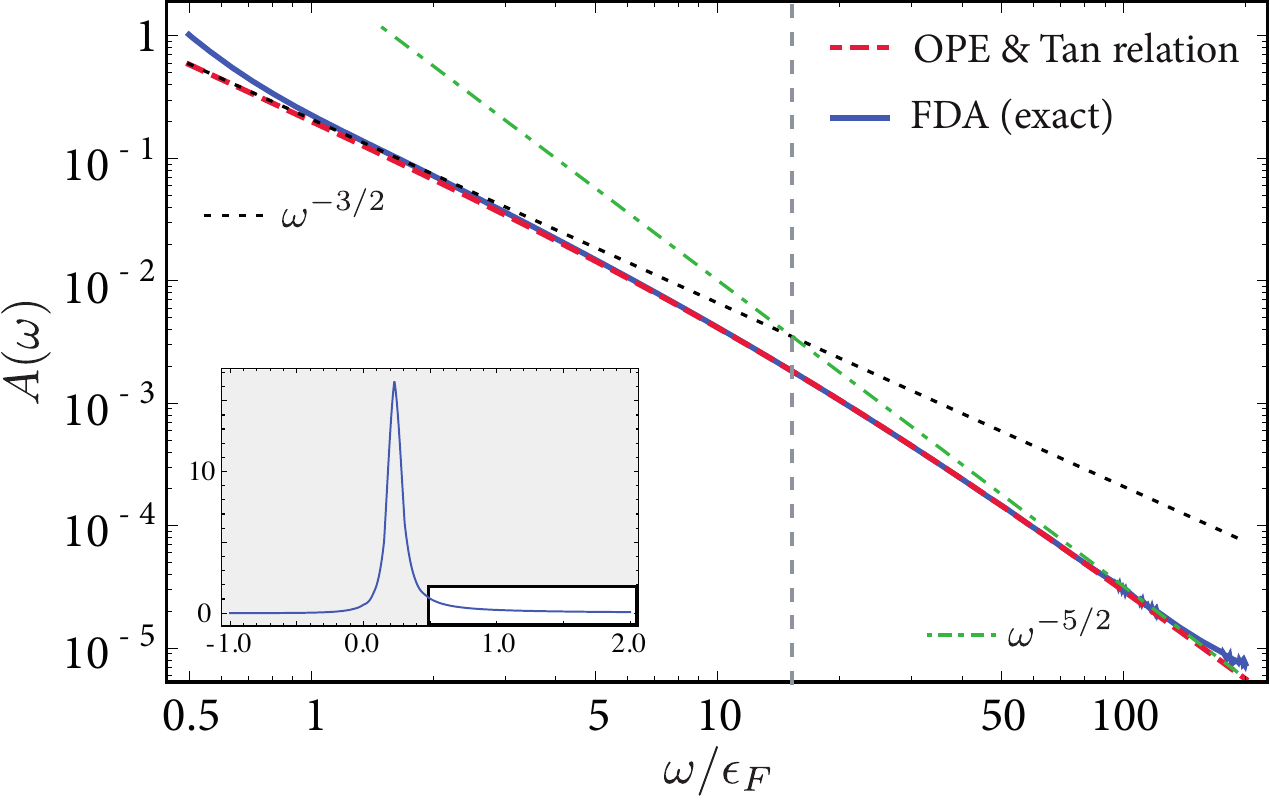}
   \caption{ \textbf{RF high-frequency tail.} RF spectrum on double-logarithmic scale as function of frequency as obtained from FDA  (blue solid) for contact interactions, $k_F r^*=0$, where both finite initial and final state interactions are present with $1/k_Fa_{i}=-0.5$, $1/k_Fa_f=-4$. The black dotted (green dot-dashed) line shows the limiting $\omega^{-3/2}$ $(\omega^{-5/2})$ behavior. The red dashed line gives the analytical prediction from the operator product expansion, Eq.~\eqref{OPEtail} \cite{Braaten2010}, with the contact $C$ calculated from the adiabatic theorem, Eq.~\eqref{Tanadiabatic}. The inset shows the rf response including low frequencies.}
   \label{fig:hightail}
 \end{figure}

The Tan relations can be derived from operator identities and as such hold  irrespectively of the state of the system. Only the contact $C$ itself then depends on the state and has either to be measured experimentally or determined from first principle calculations which are a challenge for generic many-body problems. However, being  valid for an arbitrary two-component Fermi gas, the Tan relations   apply also to our case of heavy impurities immersed in a  Fermi gas.

In Fig.~\ref{fig:hightail} we show the high-frequency response of an impurity on a double logarithmic scale as obtained from the FDA (solid line). In this figure, both initial and final state contact interactions are present ($k_Fa_{i}=-0.5$ and $k_Fa_{f}=-4$). The inset shows the response on a linear  scale including the polaron `peak' at low frequencies.  We  find that the signal undergoes  a crossover from a non-analytical $\omega^{-3/2}$ to  $\omega^{-5/2}$ behavior.  Having the exact  numerical FDA solution we may now turn to the verification of Eq.~\eqref{OPEtail}. To this end, we first calculate the ground-state energy $E\equiv \Delta E$ as function of the scattering length $a_i$ from Eq.~\eqref{Fumi}. Using the  adiabatic theorem, Eq.~\eqref{Tanadiabatic}, we then determine the contact as function of $a_i$. The resulting function $C(a_i)$ serves as input into the OPE prediction Eq.~\eqref{OPEtail} which has hence no free parameter (for an infinitely heavy impurity $\mu_\text{red}=m$, where $m$ is the mass of the atoms in the Fermi gas). In Fig.~\ref{fig:hightail} the resulting prediction is shown as red, dashed curve. We find that the analytical expression Eq.~\eqref{OPEtail}  describes with remarkable precision the exact RF response down to frequencies  of the order of the Fermi energy. \rtext{We note that we discussed here only the high-frequency response for contact interactions. The inclusion of a finite characteristic range $r^*$  leads to the crossover to  a modified power-law at frequencies $\omega\sim 1/(mr^{*2})$ \cite{Braaten2008b,Braaten2010}. For a detailed discussion of the role of finite range interactions we refer to recent work by Parish and Levinsen \cite{parish2016}.}

\paragraph*{\textbf{Relation to the long-time evolution.---}} 
The RF absorption response $A(\omega)$ is related to the Ramsey signal $S(t)$ by Fourier transformation (see also App.~\ref{SAAppendix}). Accordingly, the high-frequency behavior of $A(\omega)$ including the contact $C$ as given by Eq.~\eqref{OPEtail}, is naturally reflected in the \textit{short-time} dynamics of $S(t)$. The Tan relations connect the high-frequency behavior, however, also to the \textit{long-time} behavior of $S(t)$.  To illustrate this, we focus on low temperatures and the attractive interaction regime. 

As we have seen in Section~\ref{analyticalsection}, in the attractive interaction regime the attractive polaron excitation branch A, and hence the Fermi surface contribution $S_\text{FS}^{(0)}$,  dominates the long-time dynamics, cf.~Eqs.~\eqref{SRegimeAB}, \eqref{SAB}. Thus, at long times the phase evolution of the Ramsey signal follows $S(t)\sim e^{-i \Delta E t}$ (cf.~Eq.~\eqref{TScaling}). Here $\Delta E$ is given by the attractive `polaron' ground-state energy, which follows from Fumi's theorem, Eq.~\eqref{Fumi}. According to Eq.~\eqref{TScaling} the long-time phase evolution is approximately linear in time. Hence  $\Delta E$ can be extracted relatively easily as function of the scattering length from experimental long-time data \cite{cetina_2016}. We note that a similar relation of the phase evolution of the Ramsey signal to the Tan contact $C$ has recently been discussed also in an experimental study of the dephasing dynamics of the unitary Bose gas \cite{Fletcher2017}.

\rtext{The} ground-state energy of the system $E\equiv\Delta E$ can be used to calculate the contact $C$ from the adiabatic theorem \eqref{Tanadiabatic}. Since the contact $C$ is  the only free parameter that enters the high-frequency tail in Eq.~\eqref{OPEtail},  one can  predict the short-time (high-frequency) dephasing behavior of the corresponding Ramsey signal $S(t)$ from long-time dephasing dynamics.

\section{Summary and Outlook}\label{outlook}

We have studied the real-time and absorption response of impurities immersed in a Fermi sea. In cold atoms the interaction between the impurity and fermions can be controlled by Feshbach resonances and we showed that the system is described by an Anderson-Fano model. We found that three distinct interaction regimes can be identified that are universally determined by the phase shift at the Fermi surface. For each of these regimes we have computed the `standard' and `reverse' radio-frequency absorption spectra at both zero and finite temperature and calculated the real-time interferometric responses of the system exactly. 

In each interaction regime we identified analytically excitation branches in the many-body Hilbert space which dominate the dynamics. While in the attractive and repulsive regime  polaron-type excitations govern the dynamics, narrow Feshbach resonances allow one to additionally address a novel  `\rtext{mixed} regime' where `bottom-of-the-Fermi sea' excitations dominate the dynamics. This regime is special in exhibiting a quantum-interference-induced crossover in the decoherence dynamics  at times much larger than the thermal time scale $\hbar/k_BT$. 

We have analyzed the competition between quantum dephasing and thermal decoherence in the real-time response of the system. The former leads to a power-law decay of coherence up to a time scale set by the inverse temperature, while the latter gives rise to exponential decoherence. Using bosonization and Toeplitz determinant theory, we obtained analytic predictions for the intermediate to long-time response, which only depend on the scattering phase shift and temperature but not on the trajectory of the impurity spin. As a consequence, Ramsey and spin-echo interferometry exhibit the same decoherence rates at long times and we predicted that the maximum decoherence rate is given by $\pi k_B T/4$. 

Moreover, in analogy to earlier results on ion mobility in liquid $^3$He we find that at finite temperatures a finite spectral weight can be assigned to the various excitation branches that depends as a power-law on temperature. As temperature approaches zero, this weight vanishes as $\propto T^\alpha$, with $\alpha$ being related to the OC critical exponent.

In the present work we have considered the problem of an infinitely heavy impurity coupled to a Fermi gas. Extending on our exact solution, one may study the fate of the Anderson orthogonality catastrophe as the mass of the impurity becomes finite. This problem is still lacking a definite solution in low dimensions \cite{sols1987,prokofiev1993,Rosch1995,rosch_quantum-coherent_1999,jeblick2016}, and experiments with ultracold atoms, where the mass of the impurity can be tuned using state selective optical potentials \cite{Scazza2014,Hofer2015}, may shed light on this outstanding problem. To address this question theoretically, the functional determinant approach can be extended to account for Hamiltonians beyond bilinear order. In this way also interactions in the bath can be accounted for, as of relevance for a wide set of impurity problems in condensed matter physics. The inclusion of higher-order terms is expected to become of particular relevance when considering low dimensional systems  where enhanced quantum effects can lead to striking non-equilibrium dynamics \cite{mathy_quantum_2012,knap_quantum_2014,Gamayun2015,Gamayun2016,Meinert2016,Gamayun2017}, or when studying externally driven systems \cite{Kitagawa2010,Rechtsman2013,Knap2016,bordia2016,weidinger2016}. \rtext{Similarly higher-order effect will become of importance as the impurity concentration is increased. In this case bath-mediated interactions will become of relevance and modify the many-body dynamics of the system. The  detailed study of such effects  remains an open challenge.}

While the real-time Ramsey interferometry signal corresponds to the Fourier transform of the reverse radio-frequency response, more complicated interferometric measurements do not have a simple conjugate response~\cite{knap2012}. As an example we discussed spin-echo interferometry, yet, more complex protocols from nuclear-magnetic resonance can be envisioned, not only as a probe of many-body physics but also for controlling and manipulating many-body wave functions. Augmenting this technique with parameter ramps, one can adiabatically prepare many-body states and consequently probe them similar to pump-probe experiments in ultra-fast spectroscopy~\cite{orenstein_ultrafast_2012}. 

Moreover, in recent experiments  exotic states of matter have been created in  non-equilibrium transient regimes \cite{Kitagawa2010,fausti_light-induced_2011,polkovnikov_colloquium_2011,Rechtsman2013}. In these experiments, physics is probed on   scales ranging from long times to short times on the order of the Fermi time $\sim \hbar/\epsilon_F$. In light of these developments a detailed account of the short- to long-time dynamics becomes increasingly important for our theoretical understanding of new frontiers of condensed matter physics. In this respect  cold atomic systems provide a well-controlled starting point where dynamics can find a universal description on extended time scales as compared to traditional solid state systems. The description of dynamics in terms of excitation branches, put forward in this work, may provide a means to devise powerful variational wave functions that take into account most relevant parts of Hilbert space, and allow one to account for the interplay of  few- and many-body physics far from equilibrium.

\section*{Acknowledgments}%
We thank Michael Jag for providing the experimental data and Dima Abanin, Rudolf Grimm, Zoran Hadzibabic, Atac Imamoglu, Jesper Levinsen, Ivan Levkivskyi, Meera Parish, Yulia Shchadilova, Tarik Yefsah, Wilhelm Zwerger, and Martin Zwierlein for interesting and fruitful discussions. Parts of this work were inspired by Adilet Imambekov and Aditya Shashi. We acknowledge support from Harvard-MIT CUA, NSF Grant No. DMR-1308435, AFOSR Quantum Simulation MURI, AFOSR grant number FA9550-16-1-0323. R.~S. is supported by the NSF through a grant for the Institute for Theoretical Atomic, Molecular, and Optical Physics at Harvard University and the Smithsonian Astrophysical Observatory. M.~K. acknowledges support from the Technical University of Munich - Institute for Advanced Study, funded by the German Excellence Initiative and the European Union FP7 under grant agreement 291763, and the DFG grant No. KN 1254/1-1.  The research of D.A.I.\ was supported by the Swiss National Foundation through the NCCR QSIT. J.-S. Y. is supported by the Ministry of Science and Technology, Taiwan (Grant No. MOST 104-2917-I-564-054). M.~C. was supported by the Austrian Science Fund FWF within the SFB FoQuS (F40-P04).

\newpage

\onecolumngrid

\appendix

\setstretch{1.2}

\section{Exact single-particle wavefunctions in the two-channel model\label{sec:solModel}}

We calculate exactly the single-particle wavefunctions of model \eqw{eq:h}. We use the Ansatz $\ket{\Psi}=\alpha_m\ket{m}+\ket{\psi}$ to solve the Schr\"odinger equation $\hat H \ket{\Psi}=E\ket{\Psi}$ which gives (in units $\hbar=2m=1$) 
\begin{subequations}
\begin{align}
 \epsilon_m \alpha_m + g \int d^3 r \chi(r) \psi(r) = E \alpha_m \\
 g \chi(r) \alpha_m - \nabla^2 \psi(r) = E \psi(r) ,
\end{align}
\end{subequations}
where $\chi(r)=e^{-r/\rho}/4\pi\rho^2r$ is the form factor defined in the main text. These equations can be solved by choosing
\begin{subequations}\label{scattwavefcttwochannel}
\begin{align}
 \alpha_m &= \text{const.}\\
 \psi(r) &=  A\ \frac{\sin kr +\delta_k}{r}+ B\ \chi(r) ,
\end{align}
\end{subequations}
where the phase shift $\delta_k$ is a function of $k$. This leads to the following equations
\begin{subequations}
\label{eq:sol1}
\begin{align}
 E &= k^2 \\
 \alpha_m& =   \frac{A g}{\rho}\frac{k \rho \cos \delta_k + \sin \delta_k}{(k^2-\epsilon_m)(1+k^2 \rho^2)-\frac{g^2}{8\pi \rho}}
 \end{align}
\end{subequations}
while normalization requires
\begin{align}
 1 &= \alpha_m^2+2\pi A^2 R \Big(1+\frac{\sin 2\delta_k}{2kR}\Big) + \frac{g^2 \rho \alpha_m^2}{8\pi(1+k^2 \rho^2)^2}+2 A g\rho \alpha_m \frac{k \rho \cos \delta_k + \sin \delta_k}{(1+k^2 \rho^2)^2}.
\end{align}
Together these equations determine the unknown coefficients $A$, $B$, and $\alpha_m$.

Next, we calculate $k$ and $\delta_k$ from the boundary conditions set by constraining the atoms to a spherical box of radius $R$ which yields 
\begin{equation}\label{eq:sol2}
kR+\delta_k = n \pi
\end{equation}
This result is then compared to the scattering phase shift $\delta_k$ as obtained from  continuum scattering solution for the two-channel model  \eqw{eq:h} \cite{schmidt_efimov_2012}, 
\begin{equation}\label{eq:scattamp}
  f(k)=\frac{\mu_\text{red}}{2\pi\hbar^2}{g^2 \chi(\veck)^2}
  \left[-\frac{\hbar^2k^2}{2\mu_\text{red}}+\epsilon_m-
\frac{g^2\mu_\text{red} }{4\pi\hbar^2 \rho[1-ik\rho]^2}\right]^{-1},
\end{equation}
where we made factors of $\hbar$ and the reduced mass $\mu_\text{red}$ explicit. Using the low energy expansion of $f(k)$ one readily identifies the microscopic parameters of our model as discussed in \cite{schmidt_efimov_2012,cetina_2016} and given in Eq.~\eqref{eq:parameters}. Note within our model the effective range is given by $r_\text{e}=-2r^*+3\rho-4\rho^2/a$ which reduces to $r_e=-2 r^*$ for small values of $\rho$ as stipulated in the main text.

\section{Relation between $S(t)$ and $A(\omega)$}\label{SAAppendix}

\rtext{In this appendix we prove the simple statement that the time-dependent Ramsey signal $S(t)$ and the frequency-resolved absorption spectrum $A(\omega)$ are related by Fourier transformation. The absorption response is given by Fermi's golden rule Eq.~\eqref{Awresolved} (cf. also the $T=0$ limit in Eq.~\eqref{ASingleParticle}). After inserting the Fourier representation of the delta-function and using that $\ket{\psi_i}$ and $\ket{\psi_\alpha}$ are many-body eigenstates of $\hat H_0$ and $\hat H$, respectively (i.e.~$\hat H_0 \ket{\psi_i} = E_i \ket{\psi_i} $ and $\hat H \ket{\psi_\alpha} = E_\alpha \ket{\psi_\alpha}$) this equation can be rewritten as
\begin{align}\label{appeq1}
A(\omega)&=\int_{-\infty}^\infty dt \sum_{i,\alpha} \bra{\psi_i}\hat \rho_\text{FS} e^{i\hat H_0 t}\ket{\psi_\alpha}\bra{\psi_\alpha}e^{-i\hat H t}\ket{\psi_i} e^{i\omega t}=\int_{-\infty}^\infty dt \Tr\left[\hat \rho_\text{FS}e^{i\hat H_0 t}e^{-i\hat H t}\right] e^{i\omega t}.
\end{align}
This can be conveniently reexpressed as
\begin{align}\label{appeq2}
A(\omega)&=2 \text{Re}\int_0^\infty dt' \Tr\left[\hat \rho_\text{FS}e^{i\hat H_0 t'}e^{-i\hat H t'}\right] e^{i\omega t'}
\end{align}
where we used that $\hat \rho_\text{FS}$, $\hat H$, and $\hat H_0$ are hermitian. The integrand contains the overlap $S(t)$:
\begin{align}
S(t)=\Tr[\hat\rho_\text{FS} e^{i\hat H_0 t}e^{- i \hat H t}]
\end{align}
that is experimentally obtainable for positive times $t\geq0$ from the Ramsey signal. From this knowledge $S(t)$ at negative times can be inferred from $S(-t) = S^*(t)$. Hence the absorption spectrum can be inferred directly from the Ramsey signal. From Eq.~\eqref{appeq1} now simply follows also the reverse statement: multiplication of Eq.~\eqref{appeq1} by $e^{-i\omega t}/2\pi$ and integration over all frequencies yields
\begin{align}
\int_{-\infty}^\infty \frac{d\omega}{2\pi} A(\omega)e^{-i \omega t} &= \int_{-\infty}^\infty \frac{d\omega}{2\pi} \int_{-\infty}^\infty dt' e^{i \omega (t'-t)} \Tr[\hat\rho_\text{FS}e^{i\hat H_0 t'}e^{- i \hat H t'}]\nonumber\\
&=\int_{-\infty}^\infty dt' S(t')\delta(t'-t) = S(t)
\end{align}
which shows that the complex Ramsey signal $S(t)$ can be directly obtained from the Fourier transform of the absorption spectrum $A(\omega)$. }

\section{Fumi's theorem}\label{FumiApp}

\rtext{Here we give a short illustrative derivation of Fumi's theorem \cite{mahan_many_2000} for the case of an immobile impurity that interacts with contact interactions with a surrounding Fermi gas at zero temperature in three dimensions. We consider here the attractive ground state of the system at $a<0$, but the generalization to the repulsive state as well as the inclusion of the bound state for $a>0$ is straightforward.}

\rtext{Consider the impurity being localized at $\vecr=0$ in a spherical box of radius $R$ (cf. App. \ref{sec:solModel}). Only s-wave states have to be considered and the radial single-particle wave functions  in absence of the impurity potential are given by $u_n(r)\sim \sin(n\pi r/R)$ with $n$ the nodal quantum number. In presence of the scattering potential the wave functions acquire a scattering phase shift and are given by $v_n(r)\sim \sin(k_n r +\delta_{n})$ where the wave number $k_n$ is determined from the boundary condition Eq.~\eqref{eq:sol2} and we introduced $\delta_n \equiv \delta_{k_n}$. }

\rtext{The energy of the many-body state of interest is obtained by filling the single-particle states with fermions up to the Fermi energy. The single-particle energies in the non-interacting case are $\epsilon_n =  \left(\frac{n \pi}{R}\right)^2/2m$ while in presence of the scattering potential they are given by $\tilde \epsilon_n = \frac{1}{2m} \left(\frac{n \pi-\delta_{n}}{R}\right)^2$. In the attractive interaction regime this leads to a downward shift of single-particle levels. The `interacting' ground state energy of the attractive state (with respect to the non-interacting ground state) is given by summing over all single-particle energy shifts $\Delta\epsilon_n\equiv \tilde\epsilon_n - \epsilon_n$ up to the Fermi energy
\begin{align}\label{phasesum}
E = \sum_n \Delta \epsilon_n = -\frac{2\pi}{R^2}\sum_n n \,\delta_n\,.
\end{align}
In the second equation we used the fact that $\delta_n^2\ll 2 n \pi \delta_n$ is an increasingly good approximation as the box size $R$ is taken to infinity. In this limit $R\to\infty$ we may now replace the sum over $n$ by an energy integration using $\Delta E_n\equiv\epsilon_n - \epsilon_{n-1} = (2 n-1)(\pi/R)^2$. For large box sizes $R$, states with small $n$ lie at very small energies for which the phase shift is negligible. Thus we can take $\Delta E_n=2n  (\pi/R)^2$. From this follows now directly the continuum limit of Eq.~\eqref{phasesum}:
\begin{align}\label{phasesum2}
E =  -\frac{2\pi}{R^2}\sum_n n\, \delta_n &= -\frac{1}{\pi}\sum_n \underbracket[0.5pt]{2 \left(\frac{\pi}{R}\right)^2 n }_{\Delta E_n} \delta_n \nonumber\\
&\underset{R\to\infty}{=} -\int_0^{\epsilon_F} \frac{d E}{\pi} \delta(E)
\end{align}
In `repulsive regime' one has $a>0$ and a molecular bound state  with finite binding energy $\epsilon_B$ is present in the single-particle spectrum. If one is interested in the repulsive state the bound state remains unoccupied and Eq.~\eqref{phasesum2} directly applies. To obtain, however, the  ground state energy $E_\text{mol}$ in this regime  the molecule is occupied with a fermion and the molecular binding energy $\epsilon_B<0$ has to be added to Eq.~\eqref{phasesum2}. }

\section{Technical details on bosonization}\label{App.FB}

\subsection{Fermi-surface contribution}
The problem of a three-dimensional Fermi sea coupled to a static impurity scattering potential reduces for s-wave scattering effectively to a one-dimensional problem in a semi-infinite space ($r>0$). Furthermore, at sufficiently low temperatures one can linearize the spectrum near the Fermi surface so that one  obtains right and left moving fermions. If one unfolds the coordinate axis one can map left moving fermions at $r>0$ onto right moving fermions at $r<0$, thus arriving at the model of one-dimensional `chiral' fermions \cite{giamarchi_quantum_2004}.
Denoting the annihilation operator of the chiral fermions as $\psi(x)$, the effective Hamiltonian then can be expressed as
\begin{eqnarray}
H_0&=&-i v_F \int dx\; \psi^{\dagger}(x)\partial_x \psi(x)\, ,\\
H_{\text{int}}&=&\int dx\; V_0(x) \psi^{\dagger}(x) \psi(x)\, ,
\end{eqnarray}
where $V_0(x)$ is a scattering potential around $x=0$ (we assume that the impurity is infinitely heavy).
Using the standard bosonization approach, we rewrite the problem in terms of a bosonic field $\phi(x)$
\cite{levitov_electron_1996,giamarchi_quantum_2004,gogolin_bosonization_2004},
\begin{equation}
\psi(x)\propto e^{i \phi(x)}\, ,
\end{equation}
with the commutator
\begin{equation}
[\phi(x),\phi(x')]= i \pi\, \text{Sign}(x-x')\, .
\end{equation}
At long times, the short-range structure of the potential $V_0(x)$ may be ignored, with the effect of the
scattering incorporated in the scattering phase $\delta_F$:
\begin{eqnarray}
H_0&=& \frac{v_F}{4\pi} \int dx\; (\partial_x \phi(x))^2\, ,\\
H_{\text{int}}&=& - \frac{v_F \delta_F}{\pi} \partial_x \phi(0) + \Delta E\, ,
\end{eqnarray}
where
\begin{equation}
\Delta E=-\int^{\epsilon_F}_0 \frac{dE}{\pi} \delta(E)
\end{equation}
is the total energy shift due to the impurity (this relation is known as Fumi's theorem, see App.~\ref{FumiApp}).
Then  $S(t)$ can be calculated as
\begin{eqnarray}
S(t)&=& \langle e^{iH_0 t} e^{-i(H_0+H_{\text{int}})t}\rangle
= e^{- i \Delta E t} \left\langle \exp \left[ 
i \frac{v_F \delta_F}{\pi} \int_0^t dt' \partial_x \phi(0,t') 
\right] \right\rangle_{H_0} \nonumber\\
&=&  e^{- i \Delta E t} \left\langle \exp \left[ i \frac{\delta_F}{\pi} (\phi(0,t)-\phi(0,0)) 
\right] \right\rangle_{H_0} \propto
e^{- i \Delta E t} \left( \frac{\pi T}{\sinh\pi T t} \right)^{\left(\frac{\delta_F}{\pi}\right)^2}\, .
\end{eqnarray}
In the second step we used the linearized dispersion and replaced the time derivative by a spacial derivative $v_F dt' \to dx$, which allows us to easily evaluate the integral in the exponent.

\subsection{Bottom-of-the-band and bound-state contributions} 

The bottom-of-the-band and bound-state effects are associated with intermediate states
where a particle is moved from the bottom of the Fermi sea (or from the bound state) 
to the Fermi level, see Fig.~\ref{fig.illustration}.
Such contributions were discussed in Refs.~\cite{combescot_1971,knap2012} and here we extend those results. 
We start with the case of the bottom-of-the-band contribution.

The contribution to $S(t)$ from the intermediate states with one isolated hole deep inside the Fermi sea can be written as
\begin{equation}
S'(t) = \int\frac{dk}{2\pi} \, \sum_{\tilde m'} \,
\left|\left\langle FS | {\tilde\psi_k} | {\tilde m'} \right\rangle\right|^2 
e^{-iE_{\tilde m'} t +i (E_k - \epsilon_F) t} \, ,
\end{equation}
where $|FS\rangle$ is the ground state of the free Hamiltonian $H_0$,
$|{\tilde m'}\rangle$ are the eigenstates of the Hamiltonian {\em with} scattering $H_0+H_{\text{int}}$ 
{\em without} any holes deep under the Fermi surface (only
with particle-hole excitations around the Fermi level) and {\em with one extra particle} compared to the state $\ket{\text{FS}}$. Furthermore ${\tilde\psi_k}$ is the annihilation operator for the single-particle eigenstate of the scattering Hamiltonian with the momentum $k$ close to the bottom of the Fermi sea, $E_{\tilde m'}$ is the multi-particle energy of the state
$|{\tilde m'}\rangle$ (relative to the ground-state energy of $H_0$), $E_k$ is the single-particle
energy of ${\tilde\psi_k}$ measured from the bottom of the Fermi sea, and $\epsilon_F$ is the Fermi energy.

We can further expand ${\tilde\psi_k}$ in terms of the free-Hamiltonian states $\psi_k$:
\begin{equation}
\left\langle {\tilde m'} | {\tilde\psi_k^\dagger} | FS \right\rangle =
\int \frac{dk'}{2\pi} \left\langle {\tilde m'} | \psi_{k'}^\dagger | FS \right\rangle 
\left\langle \psi_{k'} | {\tilde \psi_{k}} \right\rangle\, .
\label{appendix-overlap}
\end{equation}
The overlap matrix elements can be computed as
\begin{equation}
\left\langle \psi_{k'} | {\tilde \psi_{k}} \right\rangle
= 4 \int_0^\infty \sin(k'x)\, \sin(kx+\delta(E_k))\, dx = \frac{4k'}{k'^2-k^2} \sin\delta(E_k)
\approx \frac{4}{k_F} \sin\delta(E_k)\, ,
\end{equation}
where $\delta(E_k)$ is the scattering phase at the energy $E_k$ and we have used $k\ll k_F \approx k'$.

Combining everything together and performing integration over $k'$ in Eq.~(\ref{appendix-overlap}), we
find
\begin{equation}\label{appequation1}
S'(t) = \frac{16}{k_F^2} \int\frac{dk}{2\pi} \sin^2\delta(E_k) \, e^{i E_k t}
\left\langle e^{iH_0 t} \psi(0) e^{-i(H_0+H_{\text{int}})t} \psi^\dagger(0) \right\rangle e^{-i\epsilon_F t}\, ,
\end{equation}
where the last average can now be understood in the linearized model of chiral fermions discussed in the
previous section. The important property of the above expression is that it factorizes into the bottom-of-the-Fermi-sea
and Fermi-surface contributions.

The first factor due to the bottom-of-the-Fermi-sea can be re-expressed, using the
quadratic dispersion relation $E_k=k^2/(2m)$ as
\begin{equation}\label{smin1app}
S_{-1}^{\rm (FB)}(t)=\frac{4}{k_F} \int_0^\infty \frac{dk}{2\pi} \sin^2 \delta(E_k) \, e^{i E_k t} 
= \frac{1}{\pi}\int_0^\infty \frac{dE}{\sqrt{E\epsilon_F}} \sin^2 \delta(E) \, e^{i E t}\, .
\end{equation}
Here we extended integration to infinity, assuming $T\ll \epsilon_F$ and $t\gg\epsilon_F^{-1}$. The time dependence of $S_{-1}^{\rm (FB)}(t)$ at large $t$ is determined by the behavior of the integrand around $E=0$. Since $\delta(E) \propto k \propto \sqrt{E}$ at $E\to0$, we find that
\begin{equation}
S_{-1}^{\rm (FB)}(t) \propto t^{-3/2}
\end{equation}
at very large $t$. Close to the resonance ($|\epsilon_B|=(2m|a|^2)^{-1} \ll \epsilon_F$), there is
an intermediate regime $\epsilon_F^{-1} \ll t \ll |\epsilon_B|^{-1}$. At such times, we may
approximate $\delta(E)\approx \pi/2$ in the integral, which gives
\begin{equation}
S_{-1}^{\rm (FB)}(t) \propto t^{-1/2}\, .
\end{equation}

The second factor in Eq.~\eqref{appequation1} is the Fermi-surface contribution. It can be calculated using the bosonization approach:
\begin{eqnarray}
S_{1}^{\rm (FS)}(t) &=& \frac{4}{k_F} \left\langle e^{iH_0 t} \psi(0) e^{-i(H_0+H_{\text{int}})t} \psi^\dagger(0) \right\rangle e^{-i\epsilon_F t}
= \frac{4}{k_F} \, e^{- i \Delta E t - i\epsilon_F t } 
\left\langle \exp \left[ i \left(\frac{\delta_F}{\pi}-1\right) 
\left(\phi(0,t)-\phi(0,0)\right)  \right] \right\rangle_{H_0} \nonumber\\
&\propto& e^{- i \Delta E t - i\epsilon_F t} \left( \frac{\pi T}{\sinh\pi T t} 
\right)^{\left(\frac{\delta_F}{\pi}-1\right)^2}\, .
\end{eqnarray}

The case of the bound-state contribution can be treated in a similar way. The only difference will be the
overlap matrix element $\left\langle \psi_{k'} | {\tilde \psi_{\text{BS}}} \right\rangle$ (here
${\tilde \psi_{\text{BS}}}$ is the wave function of the bound state), which will contribute to the overall
prefactor in the bound-state term (\ref{eq.SBS}).

\section{Asymptotic long-time response}\label{app:B}

\subsection{Finite temperature\label{app:ft}}

In this section, we derive the exponential decay of the Ramsey signal $S(t)$ at finite temperature at long times $t\epsilon_F\gg1$:
\begin{equation}\label{gammadef1}
S(t)\sim \exp(-\gamma t -i\omega t)\, .
\end{equation}
Furthermore, we will show that the decay rate is identical for the spin-echo and Ramsey protocols.

Our derivation is inspired by the Toeplitz-determinant technique used in full counting statistics for  linearized dispersion relations \cite{levitov_electron_1996,Hassler2008}
and resembles the Szeg\H{o} formula for Toeplitz determinants. However, here we go beyond the Toeplitz-determinant
approximation and take into account both the energy-dependence of the scattering phase and the nonlinear
dispersion relation.

The Ramsey overlap is given by
\begin{equation}\label{RamseyEq}
S(t)=\det \hat B \, ;\quad\quad \hat B=  1-\hat n +\hat n e^{i\hat h_0 t}e^{-i\hat h t} \, ,
\end{equation}
where $\hat h_0$ and $\hat h$ are the \textit{single-particle} Hamiltonians without and with scattering potential, respectively. 

Since the scattering potential is spherically symmetric, we can perform a partial wave expansion, and due to the low energies involved, we need to consider s-wave scattering only. We then use the standard approach to express the radial part of the scattering wave function $\Psi(r)$ by 
\begin{equation}
u(r)=r \Psi(r)
\end{equation}
where $r>0$ and $u(r)$ fulfills the radial boundary condition $u(0)=0$. Outside of the scattering potential  $u(r)$ takes the form
\begin{equation}
u(r) \propto \sin[kx+\delta(E)]\, ,
\end{equation}
where $\delta(E)$ is the scattering phase shift and $E=k^2/2m$ the scattering energy.

The key observation for calculating the determinant (\ref{RamseyEq}) is that the operator $\hat B$ acts nearly diagonally on quasiclassical wave packets localized both in momentum and coordinate space. For example we can use the Gaussian wave packets  
\begin{equation}\label{GaussianWavePacket}
u^{(0)}_{k_0,r_0}(r)={(2\pi)^{-1/4} \Delta_r^{-1/2}} \exp\left[{i k_0 r- \frac{(r-r_0)^2}{4\Delta_r^2} }\right]\, .
\end{equation}
where $\Delta_r$ determines the width of the wave packet in real space. From its Fourier transform it follows that the wave packet is localized in momentum space  around the momentum $k_0$ with width $\Delta_r^{-1}$.  

There are three time/energy scales in the problem: the inverse Fermi energy $\epsilon_F^{-1}$, the evolution time $t$, and the `collision time' $\tau_\text{col}=\partial\delta(E)/\partial E$ (this time is typically of the order of $\epsilon_F^{-1}$ or smaller, but becomes large at the bottom of the Fermi sea). Our further discussion assumes that the wave packets (\ref{GaussianWavePacket}) propagate quasiclassically and simply acquire an extra phase when scattering. This requires that the wave packets are sufficiently localized in coordinate space so that the  time $t$ fulfills 
\begin{equation}\label{wave-packet-limitation-1}
E^{-1}\ll\frac{\Delta_r}{v}\ll t
\end{equation}
where $v\equiv |v(k_0)|$ is given by the group velocity $v(k_0)=\left.\frac{\partial E}{\partial k}\right|_{k_0}$ of the wave packet.  We impose the condition (\ref{wave-packet-limitation-1}) at the Fermi surface (with $E=\epsilon_F$ and $v=v_F$), which implies $t\gg \epsilon_F^{-1}$. The condition (\ref{wave-packet-limitation-1}) would break down close to the bottom of the Fermi sea, but we can make this region arbitrarily small for large $t$.  Additionally, we require that the phase shift does not change much across the energy window of the wave packet, which  implies
\begin{equation}\label{wave-packet-limitation-2}
\tau_\text{col} \ll \frac{\Delta_r}{v}\, .
\end{equation}
For the same reason as above, we require this condition only at the Fermi energy (which, in turn, implies the applicability condition $t\gg \tau_\text{col}$).

To properly take into account the boundary conditions on the function $u(r)$, we consider anti-symmetrized superposition of wave packets 
\begin{equation}\label{AntisymmetrizedWavePacket}
u_{k_0,r_0}(r)=u^{(0)}_{k_0,r_0}(r)-u^{(0)}_{-k_0,-r_0}(r).
\end{equation}

These form an overcomplete set on $r>0$ with the completeness relation ($r,r'>0$)
\begin{equation}\label{decompositionOfUnity}
\int_{-\infty}^{+\infty} \frac{dk_0}{2\pi} \int_0^\infty dr_0\; u_{k_0,r_0}(r) u^*_{k_0,r_0}(r') =\delta(r-r').
\end{equation}
The two time evolution operators in $\hat B$ in Eq.~\eqref{RamseyEq} propagate the wave packets forward and backward in time. Neglecting residual diffusion of the wave packet due to the short range interaction potential (which is a good approximation for small collision times, cf.~\eqref{wave-packet-limitation-2}), $\hat B$ acts approximately diagonally on the wave packets:
\begin{equation}\label{quasiclassicalB}
\hat B \; u_{k_0,r_0}(r) \approx u_{k_0,r_0}(r)
\begin{cases}
1 &\quad \text{$-v(k_0)\,t<r_0$}\, , \\
1+n(E(k_0))(e^{2i\delta(E(k_0))}-1) & \quad\text{$-v(k_0)\,t>r_0$}\, .
\end{cases}
\end{equation}
Here we made use of the condition \eqref{wave-packet-limitation-2}, i.e. we assume the wave packet is sufficiently localized in momentum space around momentum $k_0$, so that phase shift is approximately constant across the wave packet's energy window.

Using the completeness relation \eqref{decompositionOfUnity} and  Eq.~\eqref{quasiclassicalB}, we can compute
\begin{multline}\label{AppCEq1}
\ln S(t) = \tr \ln \hat B = 
\int_{-\infty}^{+\infty} \frac{dk_0}{2\pi} \int_0^\infty dr_0\;
\theta(-r_0-v(k_0)t) \ln \left[1+n(E(k_0))(e^{2i\delta({E(k_0)})}-1)\right] \\
= \int_{-\infty}^0 \frac{dk_0}{2\pi}\; \frac{\partial E}{\partial k_0} t\;
\ln \left[1+n(E(k_0))(e^{2i\delta({E(k_0)})}-1)\right]
= t \int_0^\infty \frac{dE}{2\pi} \ln \left[1+n(E)(e^{2i\delta(E)}-1)\right]\, ,
\end{multline}
where  a small correction due to the normalization of Eq.~\eqref{AntisymmetrizedWavePacket} is assumed to be neglibible.
The result~\eqref{AppCEq1} is Eq.~(\ref{fullSEquationSzergo}) from the main text which determines $\gamma$ and $\omega$.

In particular,
\begin{equation}\label{eq:decayrate}
\gamma=- \re \int_0^\infty \frac{dE}{2\pi} \ln \left[1+n(E)(e^{2i\delta(E)}-1)\right]\, .
\end{equation}

Our derivation above generalizes the Toeplitz-determinant approach commonly used in full-counting statistics (see, e.g., \cite{Hassler2009} and references therein). The conventional Toeplitz-determinant approach usually assumes a linearized  spectrum, which results in the operator $\hat B$ being a Toeplitz matrix, and Eq.~\eqref{AppCEq1} resulting from the Szeg\H{o} theorem. In our derivation, we relax the assumption of a linear dispersion relation, however $\hat B$ may still be loosely thought of as a Toeplitz matrix, if we label states by their arrival time $\tau=r_0/v(k_0)$ at the scatterer. With this relation in mind, we continue to call our method the `Toeplitz-determinant approach', keeping the corresponding terminology of the Szeg\H{o} formula for Eq.~\eqref{AppCEq1} and `symbol' for the argument of the logarithm in it.

Repeating the same calculation for the case of the spin-echo response,
\begin{equation}\label{spinechoeq}
E(t)=\det \hat C; \quad \hat C=1-\hat n +\hat n e^{i\hat h_0 t/2}e^{i\hat h t/2}e^{-i\hat h_0 t/2}e^{-i\hat h t/2}\, ,
\end{equation}
we find
\begin{equation}\label{quasiclassicalC}
\hat C \; u_{k_0,r_0}(r) \approx u_{k_0,r_0}(r) 
\begin{cases}
1 & \quad\text{$-v(k_0)\,t<r_0$}\, , \\
1+n(E(k_0))(e^{-2i\delta(E(k_0))}-1) &\quad \text{$-v(k_0)\,t/2<r_0<-v(k_0)\,t$}\, , \\
1+n(E(k_0))(e^{2i\delta(E(k_0))}-1) &\quad \text{$-v(k_0)\,t/2>r_0$}\, ,
\end{cases}
\end{equation}
which leads to
\begin{equation}
\gamma_\text{SE} + i \omega_\text{SE} =-\frac{1}{2}\int_0^\infty \frac{dE}{2\pi}  \, \left\{ 
\ln [1+n(E)(e^{2i \delta(E)}-1)]+
\ln [1+n(E)(e^{-2i \delta(E)}-1)] \right\}
\end{equation}
It is straightforward to check that $\omega_\text{SE}=0$ and $\gamma_\text{SE}$ reduces to Eq.~(\ref{eq:decayrate}), i.e., the Ramsey and spin-echo decoherence rates are equal. 

Generalizing these derivations, we find that  the decoherence rate $\gamma$ does generally not depend on the trajectory of the impurity spin (on the Bloch sphere), i.e. arbitrarily many spin-echos with corresponding  time partitions as in Eq.~\eqref{spinechoeq} also yield the same result. Hence different trajectories give rise to the same exponential decay in the thermal long-time regime.

\subsection{Zero temperature\label{app:0T}}

At zero temperature, dephasing of the many-body wave function leads to a power-law decay of the dynamic Ramsey and spin-echo response~\cite{knap2012}. The power-law decay can be attributed to the creation of infinitely many particle-hole excitations at the Fermi level which gradually renders the many-body wave function orthogonal to the original Fermi sea. We note that the exponential decay rate $\gamma$ which we evaluated in the previous section for finite temperature tends asymptotically to zero with temperature, cf.~\eq{decayrates}. Approaching zero temperature logarithmic corrections to \eq{gammadef1} will emerge giving rise to the power-law decay.
Using the mapping onto a Riemann-Hilbert problem, introduced in Ref.~\cite{dambrumenil_fermi_2005} to solve generic time-dependent perturbations to the Fermi sea, we calculate the power law exponent for both Ramsey and spin echo protocols, see also Ref.~\cite{knap2012}.

We first study the Ramsey response \eq{eq:ramsey} and define $R(\lambda,\tau)$ as the time diagonal element of \eq{quasiclassicalB}, which dominates the asymptotic dynamics, to the power of an auxiliary parameter $\lambda$
\begin{equation}
 R(\lambda,\tau) = \begin{cases}
    \,\, 1,& \text{if } t< \tau\\
 \,\,  e^{2 i \lambda \delta_{F}} , & \text{otherwise}.
 \end{cases}
 \label{eq:RRamsey}
\end{equation}
The asymptotic behavior of the Ramsey signal can be obtained from~\cite{dambrumenil_fermi_2005}   
\begin{equation}
 \ln S(t) \sim \frac{i}{2\pi} \int_0^\lambda d\lambda \int_0^t d \tau \frac{d \ln Y(\tau+i0)}{d\tau} \frac{d \ln R(\lambda,\tau)}{d\lambda},
 \label{eq:SRH}
\end{equation}
where $Y(z)$ solves the Riemann-Hilbert problem
\begin{equation}
 Y(t-i0^+) Y(t+i0^+)^{-1} = R(\lambda, t).
\end{equation}
The function $Y(z)$ is analytic everywhere in the complex plane except for the interval $[0,t]$ and can be obtained from 
\begin{equation}
 \ln Y(z) = \frac{1}{2\pi i} \int \frac{\ln R(\lambda, z')}{z-z'} dz' .
\end{equation}
For $R(\lambda,\tau)$ given by \eq{eq:RRamsey}, we obtain
\begin{equation}
 \ln Y(z) = \frac{\lambda \delta_F}{\pi} \ln  \frac{z}{z-t} .
\end{equation}
In the vicinity of the branch points of $Y$ at 0 and $t$ we cut off the integral \eq{eq:SRH} at the Fermi energy $\epsilon_F$ of the problem which amounts to replacing $i0^+$ by $i \epsilon_F^{-1}$. Computing the integral gives
\begin{equation}
 S(t) \sim (i \epsilon_F t)^{-\delta_F^2/\pi^2},
\end{equation}
which is the well known result for the asymptotic behavior of $S(t)$~\cite{nozieres_singularities_1969,dambrumenil_fermi_2005}.

For the spin-echo response \eq{eq:spinecho}, $R_\text{SE}(\lambda, \tau)$ is obtained from \eq{quasiclassicalC}
\begin{equation}
R_\text{SE}(\lambda, \tau)= \begin{cases}
     \,\, e^{2i \lambda \delta_F}& \quad\quad t> 2\tau\\
     \,\, e^{-2i \lambda \delta_F}& \quad\quad 2\tau>t> \tau\\
 \,\,  1& \quad\quad\text{otherwise}
 \end{cases}
 \label{eq:RSE}
\end{equation}
yielding 
\begin{equation}
 \ln Y_\text{SE}(z) = \frac{\lambda \delta_F}{\pi} \ln  \frac{z(z-t)}{(z-t/2)^2} .
\end{equation}
With that we can evaluate \eq{eq:SRH}
\begin{equation}
 \ln S_\text{SE}(t) \sim - \frac{1}{2}\frac{\delta_F^2}{\pi^2}\left( \int_0^{t/2} d\tau - \int_{t/2}^t d\tau \right) \frac{d}{d\tau}  \ln  \frac{z(z-t)}{(z-t/2)^2} \sim -3 \frac{\delta_F^2}{\pi^2} \ln (i \epsilon_F t)
\end{equation}
and thus we find for the asymptotic behavior of the spin-echo response
\begin{equation}
 S_\text{SE}(t) \sim (i\epsilon_F t)^{-3\delta_F^2/\pi^2}.
\end{equation}
At zero temperature, the spin-echo exponent is therefore enhanced by a factor $3$ as compared to the Ramsey exponent, which demonstrates the importance of quantum interference effects.

\section{Bi-exponential crossover}\label{crossappendix}

The measurement of the crossover of exponential decay rates of the Ramsey signal $S(t)$ from subleading- to leading-branch dynamics can be an experimental challenge due to small Ramsey contrast.  In Fig.~\ref{fig.crossoversubleading} in the main text, we have shown an example where the crossover takes place at relatively long times $t$. There, the parameters were chosen so that  oscillations due to bottom-of-the band excitations at short times are damped out in the crossover regime.  At the correspondingly long times, the  Ramsey contrast $|S(t)|$ became inaccessibly small for experiments. 

\begin{figure}[t]
\includegraphics[width=0.5\linewidth]{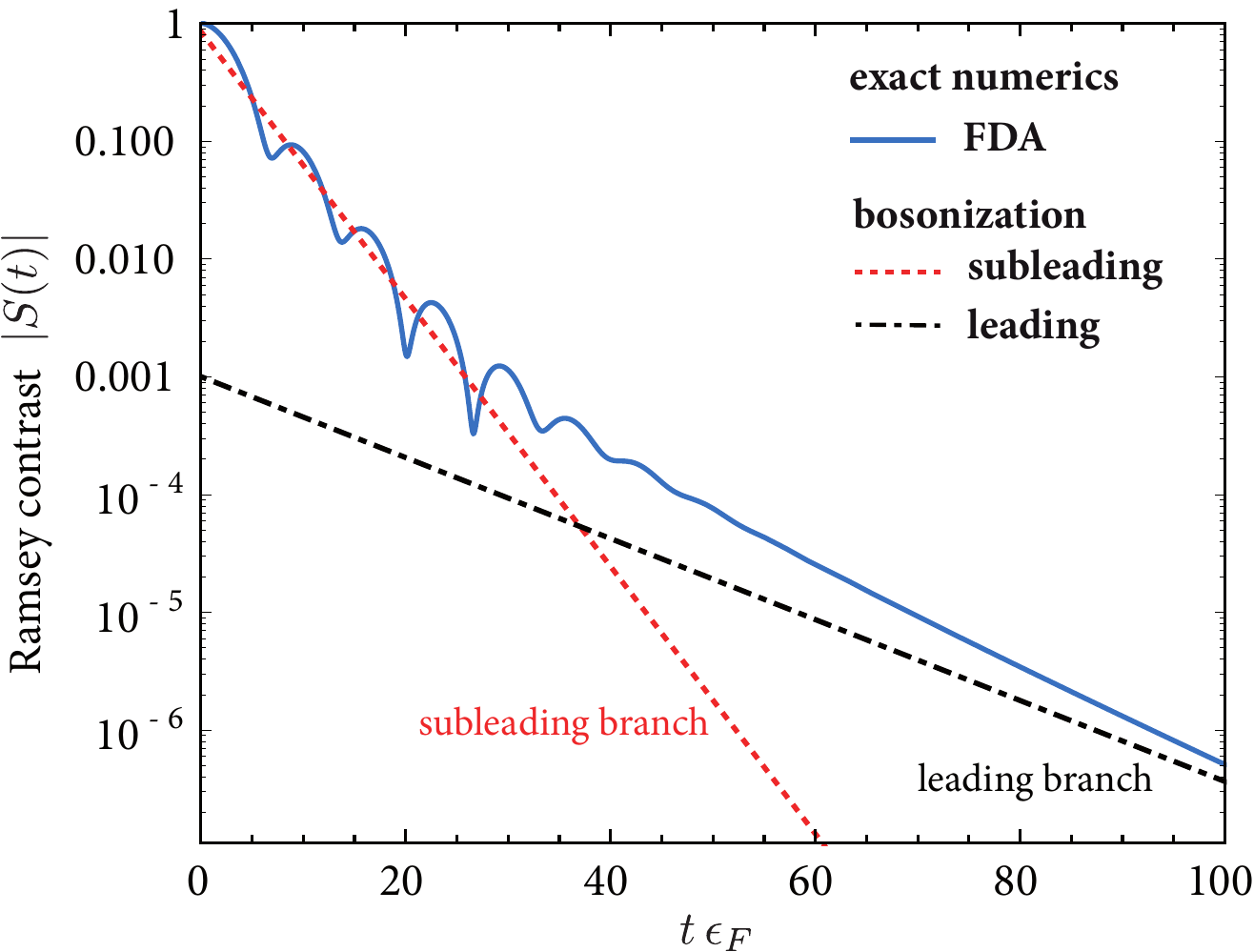}
\caption{\textbf{Crossover from subleading to leading branch dynamics.} Ramsey contrast $|S(t)|$ as function of time. Interaction parameters are chosen to correspond to the \rtext{mixed} regime (b). Compare to Fig.~\ref{fig.crossoversubleading}, we choose slightly different parameters,  $T/T_F=0.2$, $1/k_Fa=-0.61$, and $k_Fr^*=1.1$. The solid line shows the exact numerical evaluation of the dynamical overlap $S(t)$ using the FDA, while the dotted (dashed) line shows exponentials with the analytically predicted exponents $\gamma_1$ ($\gamma_0$), see Eq.~\eqref{decayrates}.}
\label{Fig.HighContrast}
\end{figure}

However, the precise time  at which the crossover takes place, as well as the corresponding magnitude of the Ramsey signal, is highly sensitive to the specific interaction parameters chosen. Hence, more favorable Ramsey contrast can easily be achieved by choosing only slightly modified parameters. For instance, in Fig.~\ref{Fig.HighContrast} we show the Ramsey contrast where, compared to Fig.~\ref{fig.crossoversubleading}, the temperature is increased from $T/T_F=0.1$ to $T/T_F=0.2$ and where $k_Fr^*$ is changed from $0.8$ to $1.1$. This slight variation  already yields an increase of the Ramsey contrast in the interference region by six orders of magnitude. This demonstrates that parameters can be  optimized  to  bring the observation of the subleading to leading branch dynamics within reach of experimental precision.

\section{Subleading branches of Toeplitz determinants}\label{app.toeplitz}

As outlined in Section \ref{sec_decoh}, analytical expressions for the different branches of the Fermi-surface contribution $S_n^{\rm (FS)}(t)$ may be obtained by choosing different integration contours $\mathcal{C}_{SL}$ in Eq.~(\ref{contouromega}), which, for convenience, we state here again,
\begin{equation}\label{contouromegaapp}
i\omega_{SL} + \gamma_{SL} = -\int_{\mathcal{C}_{SL}} \frac{dE}{2\pi} \ln \left[\sigma(E)\right]\, ,
\end{equation}
where, using the explicit form of Fermi distribution function, the symbol $\sigma(E)$ is given by
\begin{equation}\label{signumrepeat}
\sigma(E)=1-n(E)+n(E)e^{2 i \delta(E)}.
\end{equation}
The derivation of the  contours follows from an argument based on  analytic continuation: the leading branch is given by the Szeg\H{o} formula, leading to Eq.~\eqref{eq:integralomegagamma}, while the subleading branch can be obtained by analytically continuing contours from a neighboring interaction regime where the corresponding branch is leading.

\begin{figure*}[t]
\includegraphics[width=0.95\linewidth]{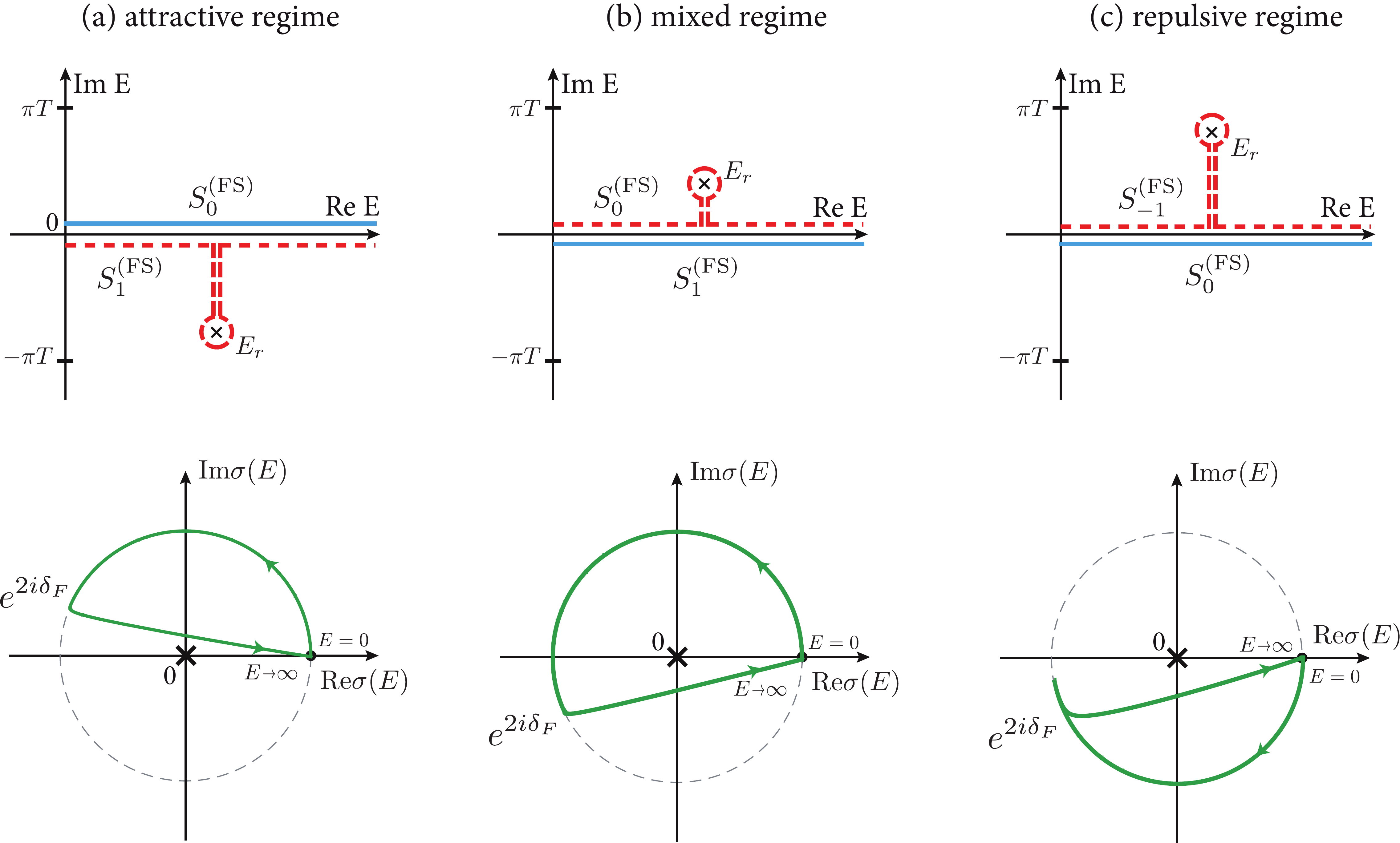}
\caption{\textbf{Integration contours for the evaluation of the Toeplitz determinant.}
The upper panel shows the integration contours in the complex variable $E$ for the leading (solid  lines) and subleading
(dashed lines) branches in Eqs.~(\ref{eq:integralomegagamma}) and (\ref{contouromega}) for the three interaction regimes in Table~\ref{tab.branches}. The cross marks the logarithmic branching point $E_r$ of the integrand. The lower panels schematically show the trajectories of $\sigma(E)$ in the complex plane (solid  lines) for the leading branch as $E$ spans the range from $0$ to $\infty$.  Low temperature $T\ll \epsilon_F$ is assumed. }
\label{fig.contours}
\end{figure*}

First, we consider  the \textit{leading} branches. In the upper panels of Fig.~\ref{fig.contours}, we show as solid lines the contours for the evaluation of the {leading} branch in the three interaction regimes (a), (b), (c) (according to Table \ref{tab.branches}). The corresponding trajectories of $\sigma(E)$ in the complex plane are shown in the lower panels of Fig.~\ref{fig.contours} as a function of the real variable $E$ ranging from $0$ to $\infty$ (assuming low temperature $T \ll \epsilon_F$). In the regimes (a) and (c), those trajectories do not encircle zero, and the asymptotics of the corresponding Toeplitz determinant follows from the conventional Szeg\H{o} formula. In other words, the leading branch $S_0^{\rm (FS)}(t)$ is given by Eq.~(\ref{eq:integralomegagamma}) with the usual integration along the real
axis.

At the transition between the regimes (b) and (c), only scattering deeply below the Fermi surface is modified. Therefore we deduce that the leading branch in the regime (b) is also given by an integration along the real axis. As discussed in Section \ref{sectoeplitz}, the branch of the logarithm is continued analytically along the contour, with the boundary condition $\log(\sigma(E{\to}\infty))=0$.  
In the interaction regime (b) we denote the leading branch as $S_1^{\rm (FS)}(t)$. This change of notation with respect to the leading branch contribution $S_0^{\rm (FS)}(t)$  in regimes (a) and (c) reflects that on passing from the regime (c) to the regime (b), we re-integrate the empty bound or bottom of the band state (see Fig.~\ref{fig.illustration}) into a hole near the bottom of the Fermi sea: as a consequence, the state without an extra particle at the Fermi surface in the regime (c) is interpreted as a state with one extra particle at the Fermi surface in the regime (b), hence the nomenclature $S_1^{\rm (FS)}(t)$. Note that the physical oscillation frequency is indeed continuous across the transition: while the shift of $\delta(E)$ by $\pi$ (see Fig.~\ref{fig.phaseshift}) leads to a decrease of $\omega_L$ by $\epsilon_F$, this energy   is  compensated by an additional contribution of $\epsilon_F$ from the hole at the bottom of the Fermi sea.

We now turn to the \textit{subleading} branches.
In order to obtain the corresponding integration contours, we examine first the transition between the regimes (a) and (b). At this transition, a zero $E_r$ of $\sigma(E)$ (and hence a logarithmic branching point of the integrand) crosses the real axis of $E$. The analytical continuity of the branches prescribes that they may be obtained by the same integrals (\ref{contouromega}), but with the contours deformed to accommodate for the shift of the zero of $\sigma(E)$, so that the integrand stays continuously on the same branch of the logarithm. In this way, we find the subleading branch in the regime (b) as a continuation of the leading branch in the regime (a) and the subleading branch in the regime (a) as a continuation of the leading branch in the regime (b). The corresponding integration contours for the subleading branches are shown in the upper panels of Fig.~\ref{fig.contours} as dashed lines: they deviate from the real axis to go around the zero of $\sigma(E)$ (and end up on a different branch of the logarithm, with respect to the integrand for the leading branch of their respective interaction regime). Finally, the subleading branch in the regime (c) is given by the same type of contour as in the regime (b) for the same reason as for the leading branch. 
For the calculation of integrals for the subleading branches, leading to the expressions Eq.~\eqref{eq:sublead}, the  contour deviations from the real axis can be chosen to be perpendicular to it, so that the contour retraces itself on the way back, as shown in Fig.~\ref{fig.contours}.  The universal jump of the logarithm of $2\pi i$ between the two branches then yields the result (\ref{eq:sublead}).

\setstretch{1}

\twocolumngrid


%

\end{document}